\documentclass[twocolumn]{aastex63}
\usepackage{hyperref}
\usepackage{color}
\usepackage{comment}
\usepackage{ifthen}
\usepackage{grffile}
\usepackage{xcolor}
\usepackage[utf8]{inputenc}
\usepackage{bm}
\usepackage{amsmath,mathtools}

\specialcomment{notes}{\begingroup\sffamily\color{red}}{\endgroup}
\includecomment{comment}

\begin{document}

\defcitealias{2019ApJ...877...37R}{R19}
\defcitealias{Hosek:2019kk}{H19}
\defcitealias{Kruijssen:2015fx}{KDL15}
\defcitealias{Sormani:2020my}{S20}

\title{Measuring the Orbits of the Arches and Quintuplet Clusters using HST and Gaia: Exploring Scenarios for Star Formation Near the Galactic Center}

\author[0000-0003-2874-1196]{Matthew W. Hosek Jr.}
\altaffiliation{Brinson Prize Fellow}
\affiliation{UCLA Department of Physics and Astronomy, Los Angeles, CA 90095}
\correspondingauthor{Matthew W. Hosek Jr.}
\email{mwhosek@astro.ucla.edu}

\author[0000-0001-9554-6062]{Tuan Do}
\affiliation{UCLA Department of Physics and Astronomy, Los Angeles, CA 90095}

\author[0000-0001-9611-0009]{Jessica R. Lu}
\affiliation{Department of Astronomy, 501 Campbell Hall, University of California, Berkeley, CA, 94720}

\author[0000-0002-6753-2066]{Mark R. Morris}
\affiliation{UCLA Department of Physics and Astronomy, Los Angeles, CA 90095}

\author[0000-0003-3230-5055]{Andrea M. Ghez}
\affiliation{UCLA Department of Physics and Astronomy, Los Angeles, CA 90095}

\author{Gregory D. Martinez}
\affiliation{UCLA Department of Physics and Astronomy, Los Angeles, CA 90095}

\author[0000-0003-2861-3995]{Jay Anderson}
\affiliation{Space Telescope Science Institute, 3700 San Martin Drive, Baltimore, MD 21218, USA}

\begin{abstract}
We present new absolute proper motion measurements for the
Arches and Quintuplet clusters, two young massive star clusters near the Galactic center.
Using multi-epoch HST observations, we construct proper motion catalogs for the Arches ($\sim$35,000 stars)
and Quintuplet ($\sim$40,000 stars) fields in ICRF coordinates established using stars in common with the Gaia EDR3 catalog.
The bulk proper motions of the clusters are measured to be
($\mu_{\alpha*}$, $\mu_{\delta}$) = (-0.80 $\pm$ 0.032, -1.89 $\pm$ 0.021) mas~yr$^{-1}$ for the Arches and
($\mu_{\alpha*}$, $\mu_{\delta}$) = (-0.96 $\pm$ 0.032, -2.29 $\pm$ 0.023) mas~yr$^{-1}$ for the Quintuplet,
achieving $\gtrsim$5x higher precision than past measurements.
We place the first constraints on the properties of the cluster orbits that
incorporate the uncertainty in their current line-of-sight distances.
The clusters will not approach closer than $\sim$25 pc to SgrA*, making
it unlikely that they will inspiral into the Nuclear Star Cluster within their lifetime.
Further, the cluster orbits are not consistent with being circular; the average value of
$r_{apo}$ / $r_{peri}$ is $\sim$1.9 (equivalent to eccentricity of $\sim$0.31) for both clusters.
Lastly, we find that the clusters do not share a common orbit, challenging one
proposed formation scenario in which the clusters formed from molecular clouds
on the open stream orbit derived by Kruijssen et al. (2015). Meanwhile,
our constraints on the birth location and velocity of the clusters offer mild support
for a scenario in which the clusters formed via collisions between gas clouds on the
$x_1$ and $x_2$ bar orbit families.
\end{abstract}


\section{Introduction}
The proximity of the Milky Way Galactic center (GC) offers a unique opportunity to
study star formation near a galactic nucleus.
The central $\sim$500 pc of the Galaxy, known as the Central Molecular Zone (CMZ),
hosts two of the most massive young clusters in the Galaxy:
the Arches and Quintuplet clusters.
With ages less than $\sim$5 Myr \citep{Najarro:2004ij, Martins:2008hl, Liermann:2012qq, Clark:2018qf, Clark:2018ij}
and masses of $\sim$10$^4$ M$_{\odot}$ \citep{Figer:1999lo, Figer:1999px, Clarkson:2012ty},
these clusters are the product of recent massive star formation events.
They are located at projected distances of $\sim$30 pc from the
central supermassive black hole (SgrA*), and, with the exception of the
Young Nuclear Cluster immediately surrounding SgrA*,
are the only known young clusters in the region.
Since the strong tidal shear in the CMZ is expected to
dissolve such clusters
on timescales of $\lesssim$20 Myr \citep{Kim:2000wd, Portegies-Zwart:2002hc},
the mechanism(s) that formed the Arches and Quintuplet clusters may
have produced additional unobserved clusters in the past
and thus play a significant (possibly dominant) role in star formation near the GC.

How the Arches and Quintuplet clusters
formed is under debate.
One possibility is that the clusters formed from
collisions between gas on the
``$x_1$'' and ``$x_2$'' families of orbits
found in a barred potential \citep[e.g.][]{Binney:1991cr}.
The $x_1$ orbits, which extend along the major-axis of
the galactic bar out to $\sim$1 kpc,
self-intersect as they approach the inner Lindblad resonance of the bar.
Gas clouds collide and shock at these locations,
lose angular momentum,
and begin to fall towards the GC on more radial trajectories.
Meanwhile, gas accumulates in the CMZ along the $x_2$ orbits,
which are elongated along the minor-axis of the bar
with radii of $\sim$0.1 -- 0.2 kpc.
The infalling $x_1$ gas collides with the
$x_2$ gas, enhancing gas densities and
possibly triggering star formation.
This behavior is seen in hydrodynamic simulations of large-scale gas flows
in the Galaxy \citep[e.g.][]{Sormani:2018pk, Armillotta:2019ip, Tress:2020zj, Armillotta:2020dm, Salas:2020cr}.

A second possibility is that the clusters formed
from molecular clouds on the open stream orbit derived
for dense gas in the CMZ by \citet[][hereafter KDL15]{Kruijssen:2015fx},
without the need for a gas collision.
In this ``open stream'' scenario, cloud collapse is triggered
by tidal compression as the natal clouds move
through pericenter passage \citep{Longmore:2013lq, Kruijssen:2015fx}
or transition into a tidally compressive regime in the
gravitational potential \citep{Kruijssen:2019kx}.
While this model has been found to be consistent with the observed
morphology and kinematics of the CMZ clouds \citep[e.g.][]{2016MNRAS.457.2675H, Langer:2017mv, Krieger:2017jt, Kruijssen:2019kx},
the actual location of several clouds
on the proposed orbit is debated \citep[e.g.][]{Butterfield:2018lb, Tress:2020zj}
and other configurations have been suggested \citep[e.g. two nuclear spirals;][]{Sofue:1995so, Ridley:2017hs}.

One way to test these formation scenarios is
to determine whether the observed motion
and orbital properties of the clusters
are consistent with these scenarios.
A major challenge in this approach is measuring the
\emph{absolute} proper motions of the clusters,
which is difficult due to the lack of observable distant background
sources needed to establish an absolute reference frame.
Past studies have used the \emph{relative}
proper motion of the clusters
compared to the field stars
as an approximation for their
absolute proper motions \citep[][]{Stolte:2008qy, Clarkson:2012ty, Stolte:2014qf}.
Recently, the situation has improved with
the success of the
European Space Agency's \emph{Gaia} mission,
which measures the absolute positions and proper motions
of billions of stars across the sky \citep{Gaia-Collaboration:2016cz}.
While the Arches and Quintuplet cannot be directly observed by
\emph{Gaia} (they are too extinguished at optical wavelengths to be detected),
there are several bright foreground stars near the clusters
with \emph{Gaia} measurements that can be
used to establish an absolute reference frame,
leading to the first
measurements of the absolute proper motions of the clusters by \citet[][]{Libralato:2020ev}.

A second challenge to this analysis is
that the present-day line-of-sight distances ($d_{los}$)
of the clusters are not well constrained.
As a result, there are many possible orbits
that can fit the on-sky position and three-dimensional
motion of each cluster \citep[][]{Stolte:2008qy, Stolte:2014qf, Libralato:2020ev}.
This uncertainty must be taken into account in order
to place statistical constraints on their orbits,
birth locations, and birth velocities.
The uncertainties in the cluster ages
must be accounted for, as well.

We use \emph{Hubble Space Telescope} (\emph{HST})
Wide-Field Camera 3 Infrared Camera  (WFC3-IR) observations
and the \emph{Gaia} Early Data Release 3  (EDR3) catalog to
measure the absolute proper motions of the
Arches and Quintuplet clusters to significantly higher
precision then previously achieved.
Combining these measurements with radial velocities
from the literature,
we forward model the orbits of the clusters
while taking the uncertainties in $d_{los}$,
cluster age, and position/motion measurements into account.
We calculate probability distributions for
the orbital properties, birth positions, and
birth velocities of the clusters,
and compare them to the predictions
of the $x_1$ -- $x_2$ gas collision
and open stream scenarios
to evaluate whether they are viable formation
mechanisms for the the Arches and Quintuplet clusters.

The paper is organized as follows:
in $\mathsection$\ref{sec:obs}
we describe the \emph{HST} observations and
astrometric measurements while in
$\mathsection$\ref{sec:astrometry}
we explain our methodology for
transforming the \emph{HST} astrometry
into the \emph{Gaia} reference frame
and present absolute proper motion catalogs
for the cluster fields.
Our measurements of the bulk absolute proper motions of the
clusters are reported in $\mathsection$\ref{sec:motion}.
We detail our approach to forward-model
the orbits of the clusters in $\mathsection$\ref{sec:orb_sims},
and present the corresponding constraints on their
orbital properties, birth velocities, and birth velocities in $\mathsection$\ref{sec:orbit}.
We compare our measurements to past work and place
them in the context of the proposed cluster formation scenarios in $\mathsection$\ref{sec:discussion}.
Finally, our conclusions are summarized in $\mathsection$\ref{sec:conclusions}.

\section{Observations and Measurements}
\label{sec:obs}
\begin{figure*}
\begin{center}
\includegraphics[scale=0.4]{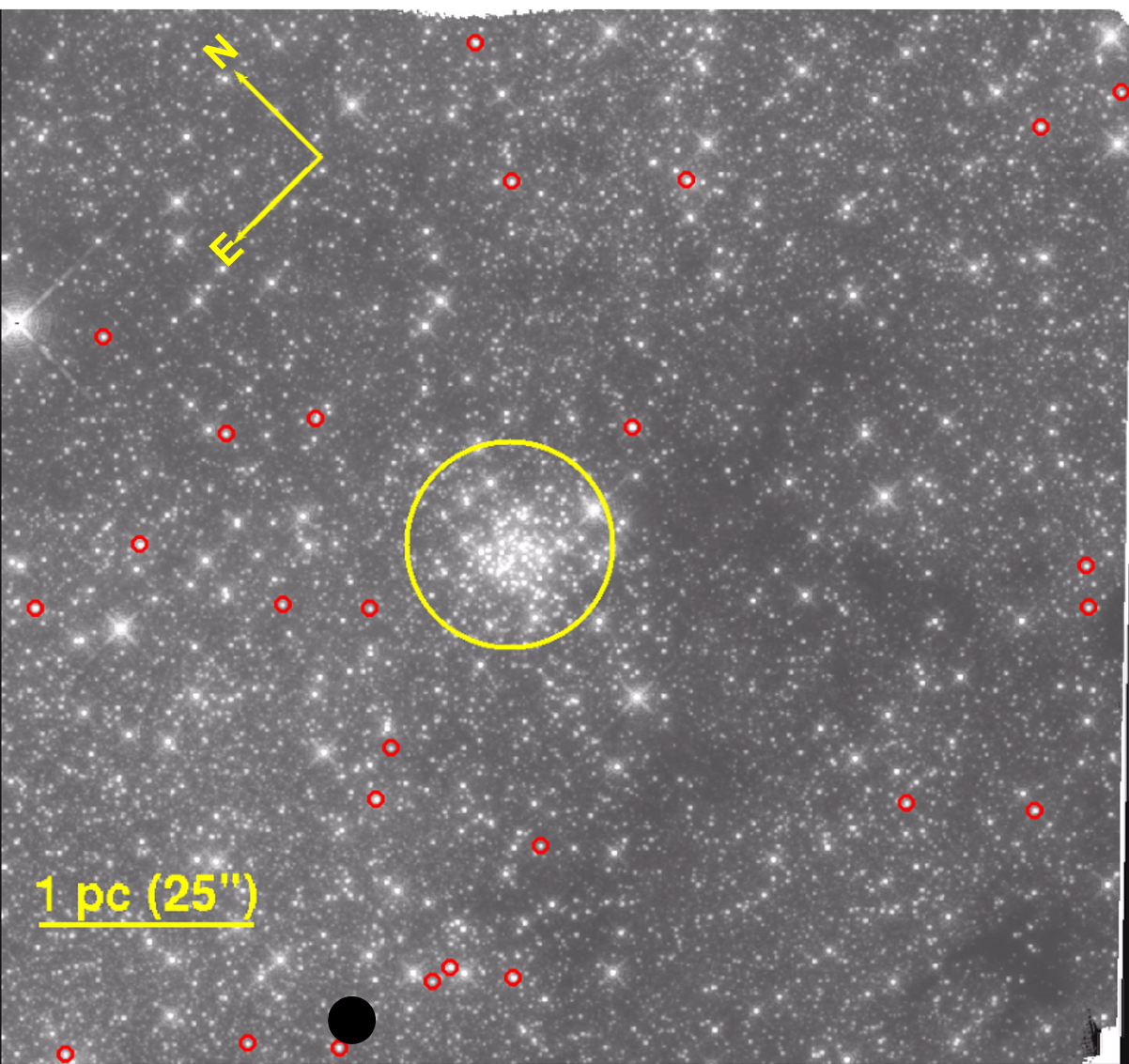}
\includegraphics[scale=0.4]{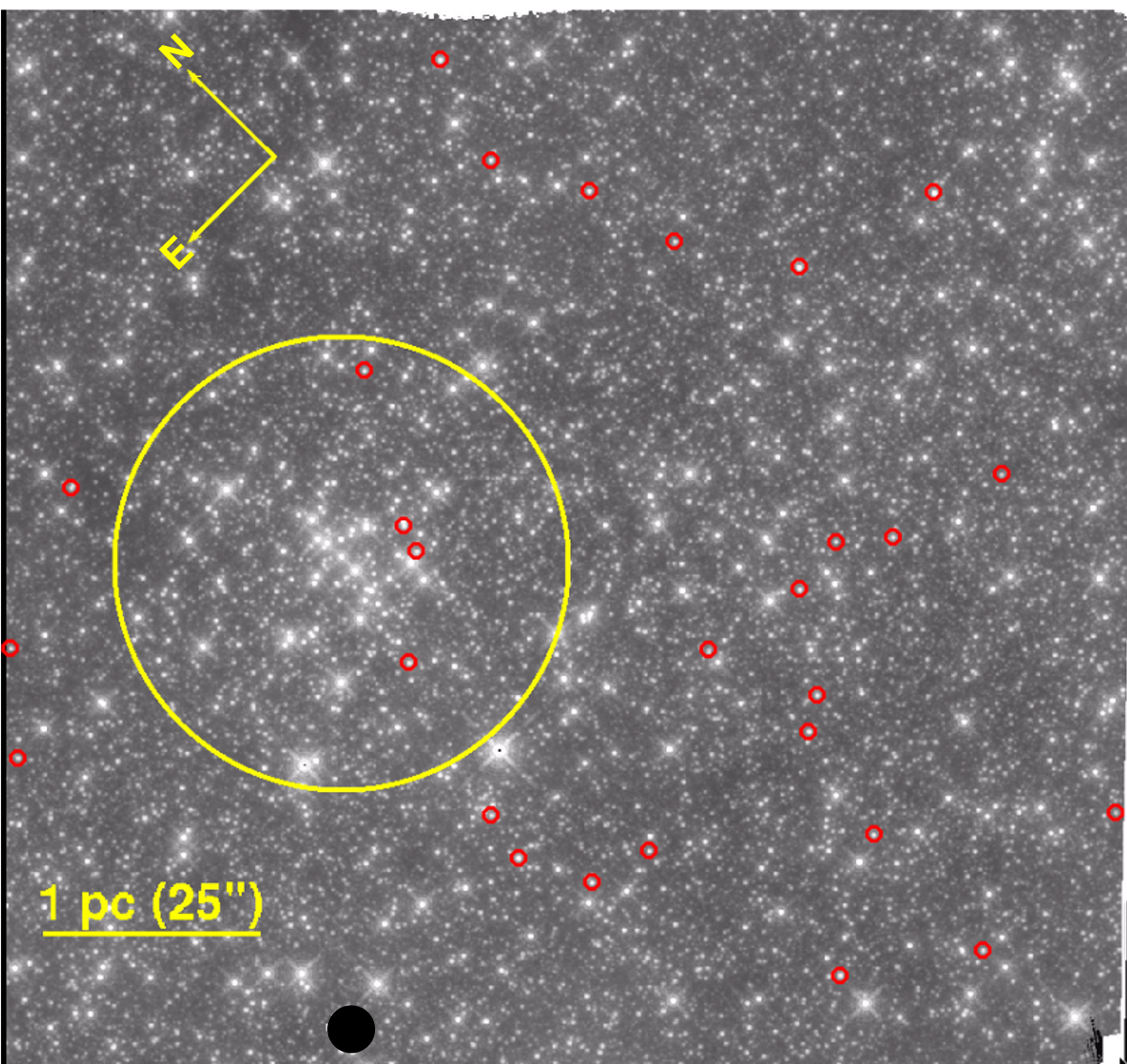}
\caption{ \emph{HST} WFC3-IR images of the Arches (left) and Quintuplet (right) clusters in the F153M filter with a log stretch. The yellow circle represents the half-light radius of the clusters. We use foreground stars from the Gaia EDR3 catalog (red circles) to establish an absolute reference frame for the \emph{HST} astrometry.}
\label{fig:clust}
\end{center}
\end{figure*}

The Arches and Quintuplet clusters were observed with \emph{HST}
WFC3-IR in 2010, 2011, and 2012 as part of a multi-cycle
GO program, with a fourth epoch of additional data obtained in 2016 (Table \ref{tab:obs})\footnote{All data used in this paper can also be found in MAST \dataset[10.17909/rgcy-2n46]{http://dx.doi.org/10.17909/rgcy-2n46}}.
The 2010 observations included images in the F127M, F139M,
and F153M filters, while additional F153M images were
obtained in 2011, 2012, and 2016.
The F153M observations were designed to maximize
astrometric performance, employing a
21-point sub-pixel dither pattern in order to
fully sample the point-spread function (PSF).
In addition, each epoch was observed at the same position angle in order
to reduce optical distortion between epochs.
Meanwhile, the F127M and F139M observations were designed
to obtain stellar photometry to approximately the same depth as the F153M data.
These observations provide a field of view of 132"~x~124" for each cluster\footnote{Note that
this field of view is only slightly larger than that of a single WFC3-IR field, due to the
compact dither pattern.}
with a plate scale of 0$''$.121 pix$^{-1}$.
The Arches field is centered at
($\alpha$(J2000), $\delta$(J2000)) = (17\textsuperscript{h}45\textsuperscript{m}50\textsuperscript{s}.49,  = -28$^{\circ}$49$'$19$''$.92)
while the Quintuplet field is centered at
($\alpha$(J2000), $\delta$(J2000)) = (17\textsuperscript{h}45\textsuperscript{m}50\textsuperscript{s}.49, -28$^{\circ}$49$'$19$''$.92).
A typical F153M image of each cluster is shown in Figure \ref{fig:clust}.

\begin{deluxetable}{c c c c c c }
\tablecaption{HST WFC3-IR Observations}
\label{tab:obs}
\tablehead{
\colhead{Cluster} & \colhead{Date} & \colhead{GO/PI} & \colhead{Filter} & \colhead{$N_{img}$} & \colhead{$t_{img}$}  \\
}
\startdata
Arches & 2010.6150 & 11671/Ghez & F127M & 12 & 599 \\
Arches & 2010.6148 & 11671/Ghez & F139M & 10 & 349 \\
Arches & 2010.6043 & 11671/Ghez & F153M & 21 & 349 \\
Arches & 2011.6829 & 12318/Ghez & F153M & 21 & 349 \\
Arches & 2012.6156 & 12667/Ghez & F153M & 21 & 349 \\
Arches & 2016.8009 & 14613/Lu & F153M & 21 & 349 \\
& & & & & \\
Quintuplet &  2010.6070 & 11671/Ghez & F127M & 12 & 599 \\
Quintuplet &  2010.6060 & 11671/Ghez & F139M & 10 & 349 \\
Quintuplet &  2010.6230 & 11671/Ghez & F153M & 21 & 349 \\
Quintuplet &  2011.6880 & 12318/Ghez & F153M & 21 & 349 \\
Quintuplet &  2012.6130 & 12667/Ghez & F153M & 21 & 349 \\
Quintuplet &  2016.8090 & 14613/Lu & F153M & 21 & 349 \\
\enddata
\tablecomments{Description of columns: \emph{Cluster}: cluster observed, \emph{Date}: Date observed, \emph{GO/PI}: HST GO number and PI of observations, \emph{Filter}: WFC3-IR filter used,
\emph{N$_{img}$}: number of images, \emph{t$_{img}$}: integration time per image, in seconds}
\end{deluxetable}

Stellar astrometry and photometry are extracted in the same manner as described by
\citet[][hereafter H19]{Hosek:2019kk} and \citet[][hereafter R19]{2019ApJ...877...37R}.
Briefly, initial measurements are obtained using the \texttt{FORTRAN} code \texttt{img2xym\_wfc3ir},
a version of the \texttt{img2xym\_WFC} package developed for WFC3-IR \citep{Anderson:2006il}.
Stars are iteratively detected and measured using a library of spatially-variable PSFs arranged in a 3 x 3 grid across the field.
After the first iteration, a uniform perturbation is applied to the PSF library in order to minimize the residuals of the
PSF fit.
Subsequent iterations use the perturbed PSF models to extract improved measurements.
Within a given epoch, common stars across the images are
matched and a first-order polynomial transformation (6 free parameters)
is used to transform the stellar positions into a
common reference frame for that epoch.

Next, a master starlist for each epoch is produced using \texttt{KS2}, a \texttt{FORTRAN} code that
uses the \texttt{img2xym\_wfc3ir} output to detect fainter stars in the field \citep[see][]{Anderson:2008qy, Bellini:2017xy, Bellini:2018ow}.
\texttt{KS2} uses the transformations to stack the images into a common reference frame.
Stars are iteratively detected in the image stack,
where the PSF of the detected sources are subtracted in each iteration
to allow the detection of fainter stars.
Final astrometric and photometric measurements of the stars
are made in each individual image at their detected location in the image stack.
The astrometric error for each source, $\sigma_{HST}$, is calculated as the standard error of
the stellar position across all the images in the epoch (e.g., $\sigma_{HST}$~=~$\sigma$~/~$\sqrt{N_{frames}}$,
where $\sigma$ is the standard deviation of the positions and $N_{frames}$ is the number of images the star is detected in).
Extracted magnitudes are converted from instrumental to Vega magnitudes using the KS2 photometric zeropoints derived in \citet{Hosek:2018lr}.
These master starlists contain $\sim$46,000 stars for the Arches cluster field and $\sim$56,000 stars for
the Quintuplet cluster field, with a typical depth of F153M $\sim$23 mag (95th percentile of the detected magnitudes).

\section{HST Absolute Proper Motions}
\label{sec:astrometry}
We use the \emph{Gaia} EDR3 catalog \citep{Gaia-Collaboration:2021bx} to transform the \emph{HST} astrometry
into an absolute reference frame.
The \emph{Gaia} catalog is aligned to the International Celestial Reference Frame (ICRF)
using the positions of over 2000 quasars.
The resulting \emph{Gaia} reference frame is found to be consistent with the
ICRF at the level of $\sim$0.01 mas with a relative rotation of
$<$$\sim$0.01 mas yr$^{-1}$ \citep{Gaia-Collaboration:2021bx}.
We ignore this uncertainty in the following analysis since this is well below the best astrometric precision
of the \emph{HST} measurements ($\sim$0.3 mas).

\subsection{Selecting Gaia Reference Stars}
\label{sec:gaia_ref}
There are a limited number of sources in common between the \emph{HST}
observations and the \emph{Gaia} EDR3 catalog that are suitable to use to
establish an absolute astrometric reference frame (Figure \ref{fig:clust}).
There are several reasons for this:
(1) the extremely high extinction towards the GC means that most of the sources detected in the
near-infrared \emph{HST} observations are too faint to be detected in the optical \emph{Gaia} observations,
and so common sources must come from the foreground stellar population;
(2) these foreground stars are among the brightest in the \emph{HST} images and thus may be saturated,
reducing their astrometric accuracy;
and (3) the fields have significant stellar crowding that can bias astrometric measurements.
With this in mind, we make the following set of quality cuts on the \emph{Gaia} catalog sources
in order to identify possible reference stars:

\begin{itemize}
\item \texttt{astrometric\_params\_solved} = 31, indicating that the source has a five-parameter astrometric solution (position, parallax, and proper motion) and a good measurement of the effective wavenumber that is used for the color-dependent PSF correction terms \citep[][]{Lindegren:2021ae}

\item  \texttt{duplicated\_source} = False, ensuring that no other \emph{Gaia} sources are within 0.$''$18 \citep[][]{Lindegren:2021ae}

\item \texttt{parallax\_over\_error} $\geq$ -10, to eliminate sources with significantly negative (e.g., unphysical) parallaxes \citep[e.g.][]{Arenou:2018dz}

\item \texttt{astrometric\_excess\_noise\_sig} $\leq$ 2, meaning that no significant excess noise is present in the astrometric solution beyond the statistical errors \citep[][]{Lindegren:2021ae}

\item \emph{Gaia} G $>$ 13 mag, as the systematic errors in sources brighter than this is not yet well characterized in the EDR3 catalog \citep[][]{Lindegren:2021ae}

\item F153M $\geq$ 13.8 mag for the \emph{HST} source matched to the \emph{Gaia} star, avoiding the regime where the \emph{HST} astrometric error begins to increase due to saturation.

\end{itemize}

After these cuts, 30 potential reference stars remain in the Arches field and 41 potential
reference stars remain in the Quintuplet field.
As a final check of the astrometric quality of the reference stars,
initial transformations are calculated using the procedure outlined in $\mathsection$\ref{sec:transforms}
to convert the \emph{HST} astrometry into the \emph{Gaia} reference frame.
Reference stars that have transformed \emph{HST} positions or proper motions that are
discrepant by $\geq$3$\sigma$ from
their \emph{Gaia} EDR3 values (where $\sigma$ is defined as the quadratic sum of the astrometric errors
from \emph{HST} and \emph{Gaia})
are iteratively removed from the sample,
where the largest outlier is removed
in each iteration and
the transformations are recalculated
to determine if the remaining outliers are still discrepant.
This process is repeated until no more 3$\sigma$ outliers exist.
Stars removed by this procedure
are often found in close proximity to the diffraction spikes of bright stars or
near the edge of the \emph{HST} field.
In total, 4 stars are removed from the Arches reference star sample
and 13 reference sources are
removed from the Quintuplet reference star sample via this process.

Ultimately, 26 \emph{Gaia} sources are used
as reference stars in the Arches field
and 28 \emph{Gaia} sources are used as reference stars in the Quintuplet field.
The locations of these reference
stars are shown in Figure \ref{fig:clust}.

\subsection{Transforming HST Astrometry into the Gaia Reference Frame}
\label{sec:transforms}
We transform the \emph{HST} master starlists into the \emph{Gaia} EDR3 reference frame on an epoch-by-epoch basis
using the reference stars identified in $\mathsection$\ref{sec:gaia_ref}.
First, the \emph{Gaia} EDR3 proper motions are used to calculate the expected positions of the reference stars
at a given \emph{HST} epoch, with the astrometric uncertainty propagated accordingly.
Next, 5 of the reference stars are matched to their \emph{HST} counterparts by-eye, and an initial transformation
from the \emph{HST} pixel coordinates into the \emph{Gaia} coordinates is calculated via a first-order polynomial (6 free parameters).
This transformation is applied to the entire \emph{HST} starlist to provide an initial estimate of their positions in the \emph{Gaia} frame.
The full list of reference stars is then compared to the transformed \emph{HST} positions, and matches are identified
as sources with positions that are consistent within 80 mas.

The set of matched reference stars is used to calculate the
final transformation from \emph{HST} pixels into \emph{Gaia} coordinates
via a second-order polynomial transformation (12 free parameters\footnote{Higher-order polynomial transformations
did not yield significant improvement compared to the second-order polynomial transformations, and thus were deemed to be unnecessary.}).
Each reference star is weighted by 1 / $\sigma_{tot}^2$, where $\sigma_{tot}$ is the quadratic sum of the intrinsic \emph{HST} and \emph{Gaia}
astrometric errors for that source.
We refer to the transformed \emph{HST} positions as the ``HST-Gaia reference frame,''
which is aligned with the ICRF via the \emph{Gaia} EDR3 reference frame.

The uncertainty of the transformation into the HST-Gaia reference frame
is calculated via a full sample bootstrap over the reference stars.
We resample the reference stars (with replacement) 100 times
and recalculate the second-order polynomial transformation for each sample.
The resulting transformations are then applied to the original \emph{HST} master starlists.
The transformation error ($\sigma_{trans}$) for each star is calculated as the
standard deviation of its transformed position across all the bootstrap iterations.
This is combined with the intrinsic astrometric error ($\sigma_{HST}$) to get the
total astrometric error $\sigma_{ast}$:

\begin{equation}
\sigma_{ast} = \sqrt{\sigma_{HST}^2 + \sigma_{trans}^2}
\end{equation}

\begin{figure*}
\begin{center}
\includegraphics[scale=0.3]{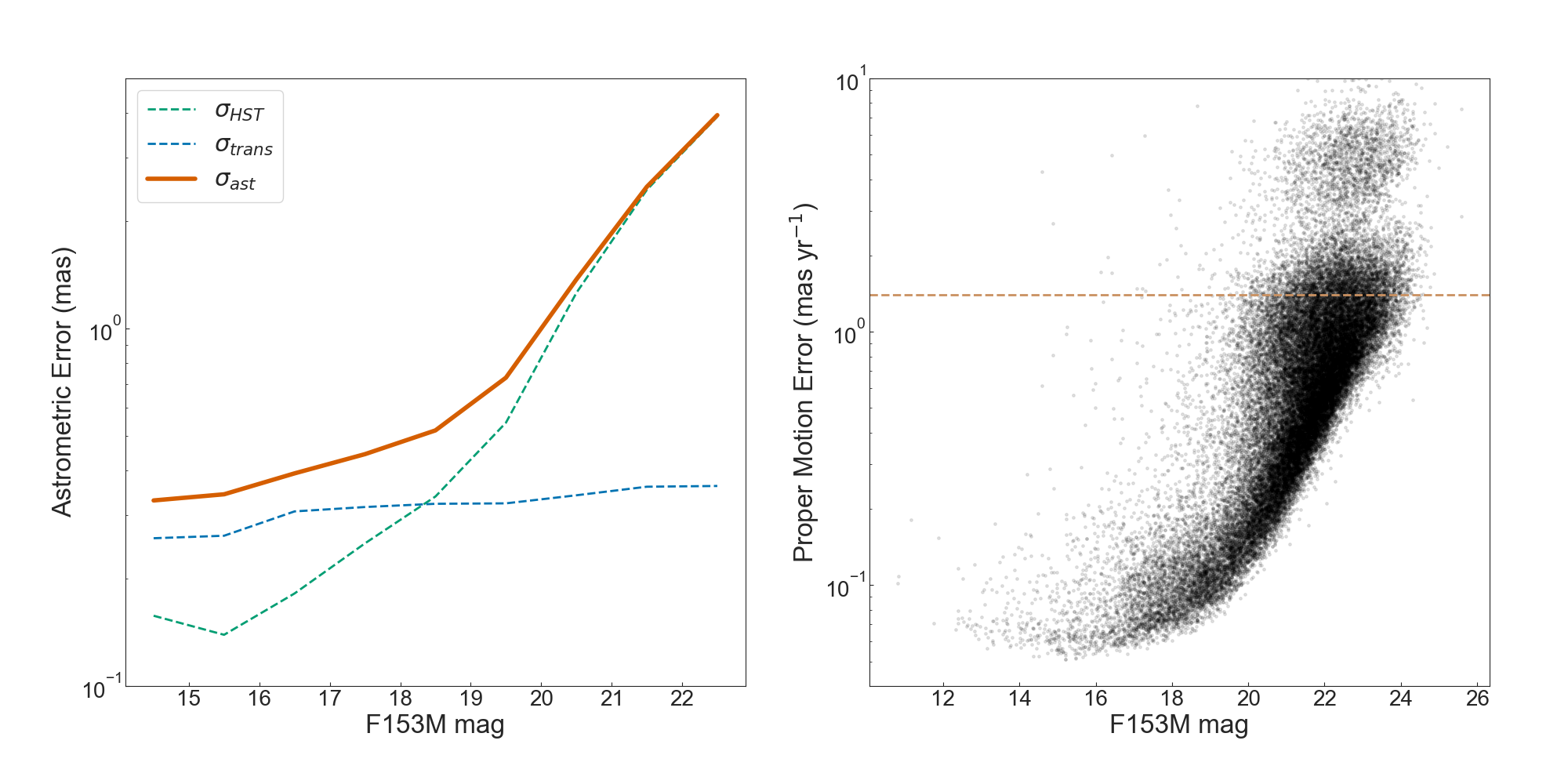}
\caption{\emph{Left:} The total astrometric error ($\sigma_{ast}$; red solid line), intrinsic \emph{HST} astrometric error ($\sigma_{HST}$; green dotted line), and transformation error ($\sigma_{trans}$; blue dotted line) as a function of magnitude for the Arches cluster field in the 2012 epoch. The best-measured stars have typical $\sigma_{ast}$ values of $\sim$0.3 mas, which is dominated by $\sigma_{trans}$. \emph{Right:} Proper motion error as a function of magnitude for the Arches cluster field. Stars with errors larger than 1.4 mas yr$^{-1}$ (dashed line) are excluded from the analysis. We achieve a precision of $\sim$0.07 mas yr$^{-1}$ for the brightest non-saturated stars. We achieve similar astrometric performance for the Quintuplet cluster field.}
\label{fig:ast_err}
\end{center}
\end{figure*}

Figure \ref{fig:ast_err} shows $\sigma_{ast}$ as a function
of F153M magnitude for the Arches cluster in the 2012 epoch,
which is representative of the other epochs.
Similar results are obtained for the Quintuplet cluster.
We achieve an error floor of $\sim$0.3 mas for the
brightest stars, which is dominated by the transformation errors.
Note that this is $\gtrsim$2x larger than the errors
presented in \citetalias{Hosek:2019kk} and \citetalias{2019ApJ...877...37R},
who use the same datasets.
This is because \citetalias{Hosek:2019kk} and \citetalias{2019ApJ...877...37R} measure \emph{relative} astrometry,
e.g., the positions and proper motions of stars relative to each other,
while we present \emph{absolute} astrometry tied to the \emph{Gaia}
reference frame.
Because of this, our astrometric transformations are limited to
the \emph{Gaia} reference stars in the field,
while the transformations for relative astrometry can use
thousands of reference stars among the general field population.
As a result, our transformation errors are much larger than those in \citetalias{Hosek:2019kk}
and \citetalias{2019ApJ...877...37R}.

\begin{figure*}
\begin{center}
\includegraphics[scale=0.3]{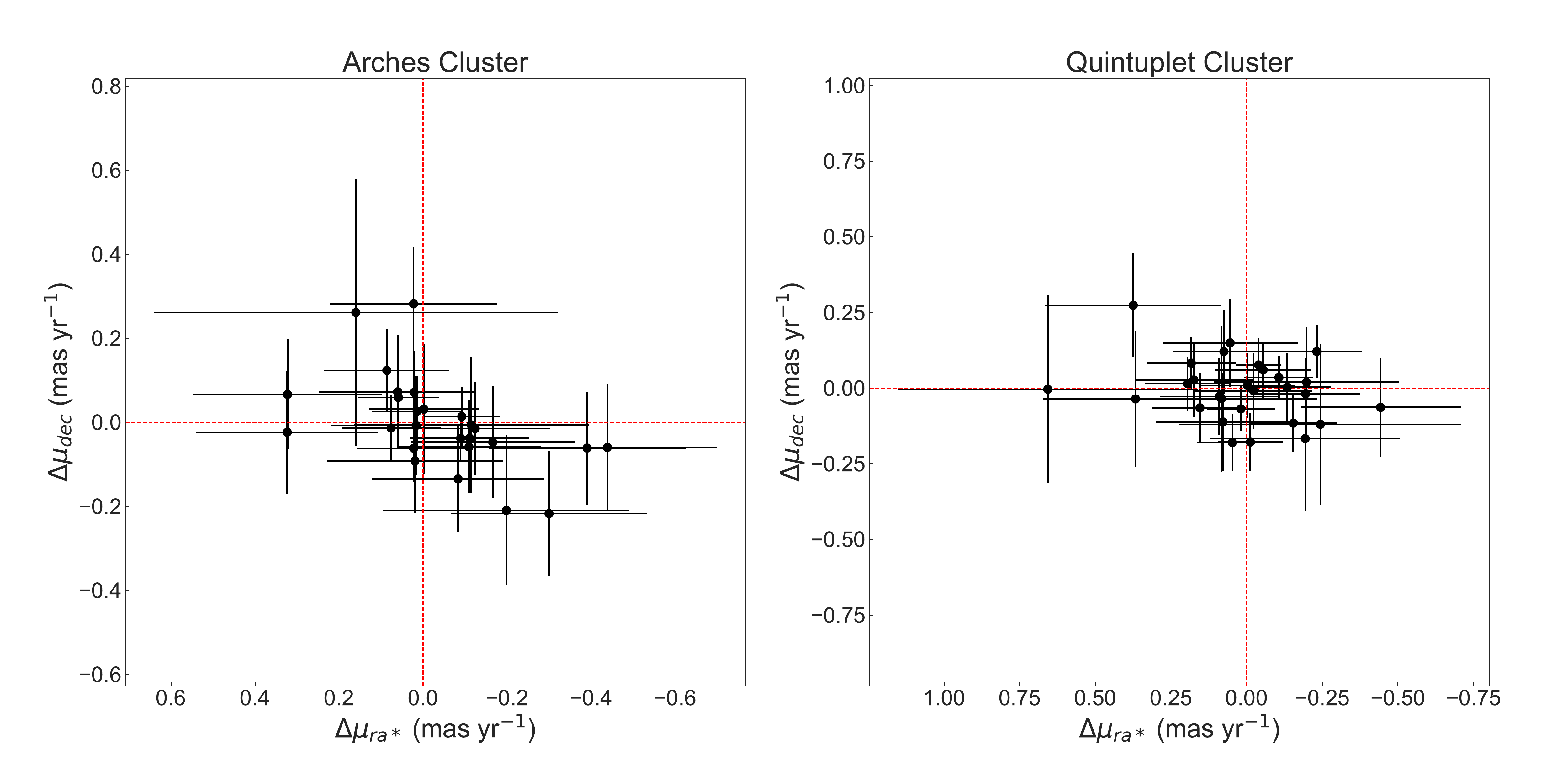}
\caption{The proper motion differences of the reference stars between the HST-Gaia and \emph{Gaia} EDR3 reference frames for the Arches (left) and Quintuplet (right) fields.
We find that the HST-Gaia reference frame is consistent with \emph{Gaia} EDR3 reference frame to $\leq$0.03 mas yr$^{-1}$.
This is the dominant source of error in our absolute proper motion measurements of the Arches and Quintuplet clusters.}
\label{fig:hst_gaia}
\end{center}
\end{figure*}

To assess the performance of the HST-Gaia reference frame,
we compare the proper motions of the reference stars derived from the
transformed \emph{HST} astrometry to their corresponding values in the
\emph{Gaia} EDR3 catalog (Figure \ref{fig:hst_gaia}).
The overall accuracy and uncertainty of the HST-Gaia reference frame
is defined as the error-weighted average of the proper motion differences
($\Delta$$\mu_{\alpha^*}$, $\Delta$$\mu_{\delta}$)
and the corresponding error in the weighted average.
We obtain ($\Delta$$\mu_{\alpha^*}$, $\Delta$$\mu_{\delta}$) = (-0.027 $\pm$ 0.03, -0.004 $\pm$ 0.02) mas yr$^{-1}$
for the Arches and ($\Delta$$\mu_{\alpha^*}$, $\Delta$$\mu_{\delta}$) = (-0.006 $\pm$ 0.030, -0.004 $\pm$ 0.02) mas yr$^{-1}$
for the Quintuplet.
This indicates that the HST-Gaia
and \emph{Gaia} EDR3 reference frames are consistent within
0.03 mas yr$^{-1}$ and 0.02 mas yr$^{-1}$ in the $\mu_{\alpha^*}$ and $\mu_{\delta}$ directions,
respectively.
As discussed in $\mathsection$\ref{sec:motion},
the uncertainty in the HST-Gaia reference frame is
the largest source of uncertainty in the absolute
proper motions of the clusters.

\subsection{Final Proper Motion Catalogs}
\label{sec:PMcat}

Proper motions are calculated from linear fits of the RA$\cdot \cos$(DEC) and DEC positions
(hereafter $\alpha^*$ and $\delta$)
as a function of time for all stars detected in at least 3 epochs:

\begin{equation}
\label{eq:pm_ra}
\alpha^* = \alpha^*_0 + \mu_{\alpha^*}(t - t_0)
\end{equation}
\begin{equation}
\label{eq:pm_dec}
\delta = \delta_0 + \mu_{\delta}(t - t_0)
\end{equation}

where ($\alpha^*$, $\delta$) is the observed position at time $t$,
($\alpha^*_0$, $\delta_0$) are the fitted ($\alpha^*$, $\delta$) position at $t_0$,
and ($\mu_{\alpha^*}$, $\mu_{\delta}$) are
the corresponding proper motions.
$t_0$ is the astrometric error-weighted average time
of the data points, and each data point is weighted by
1/$\sigma_{ast}^2$ in the fit.
The resulting proper motion catalogs contain 40,932 stars for the Arches
cluster field and 46,087 stars for the Quintuplet cluster field.
Proper motion uncertainties as a function of magnitude are shown in
Figure \ref{fig:ast_err}.
The best-measured stars have proper motion uncertainties
of $\sim$0.07 mas yr$^{-1}$, which is $\sim$2x higher than the errors achieved
by \citetalias{Hosek:2019kk} and \citetalias{2019ApJ...877...37R}.
As discussed in $\mathsection$\ref{sec:transforms},
this is due to the penalty in astrometric precision incurred when
transforming into an absolute reference frame.

We apply two quality cuts to the proper motion
catalogs for the final analysis.
First, we remove stars with proper motion errors larger than 1.4 mas yr$^{-1}$
in order to remove stars with large uncertainties.
Then, we eliminate stars with absolute proper motion values larger than 99.7\% of the rest of the sample,
as these sources are either bad measurements
or fast-moving foreground stars that don't represent the bulk field and cluster stellar
populations.
After these cuts, 34,600 stars remain in the Arches cluster sample ($\sim$85\% of original catalog) and 40,499
stars remain in the Quintuplet cluster sample ($\sim$88\% of original catalog).
Summary statistics for the proper motion catalogs are provided in Table \ref{tab:pm_cat},
and a sample of the Arches catalog is shown in Table \ref{tab:arches_cat}.
Full proper motion catalogs for both cluster fields are provided in
machine-readable form with this work.

\begin{deluxetable}{c c c c c c}
\tablecaption{HST Proper Motion Catalog: Summary Statistics}
\label{tab:pm_cat}
\tablehead{
\colhead{Cluster} & \colhead{N$_{stars}$} & \colhead{Depth} & \colhead{$\overline{\sigma_{ast}}$} & \colhead{$\overline{\sigma_{pm}}$} &   \colhead{$\overline{\sigma_{mag}}$}  \\
&   & F153M mag &  mas &  mas yr$^{-1}$ & mag
}
\startdata
Arches & 34,600 & 23.04 & 0.18 & 0.07 &  0.01 \\
Quintuplet & 40,499 & 22.60 & 0.25 & 0.09  & 0.01 \\
\enddata
\tablecomments{Description of columns: \emph{N$_{stars}$}: number of stars in catalog, \emph{Depth}: 95th percentile of F153M magnitudes,
\emph{$\overline{\sigma_{ast}}$}: median error in position for proper motion fits for stars with F153M $\leq$ 16 mag,
\emph{$\overline{\sigma_{pm}}$}:  median error in proper motion for proper motion fits for stars with F153M $\leq$ 16 mag,
\emph{$\overline{\sigma_{mag}}$}: median F153M photometric error for F153M $\leq$ 16 mag}
\end{deluxetable}

We perform several checks
of the proper motions in Appendix \ref{app:pm_cat},
examining the $\chi^2$ distribution of the
fits and the average astrometric residuals in each epoch.
In summary, we find that the $\chi^2$ distributions
are similar to their theoretical distribution,
indicating that the proper motion fit residuals are well
characterized by the astrometric errors.
We also find that the average astrometric
residuals for each epoch (which
we refer to as the residual distortion)
is less than half of the astrometric error for bright stars
and thus do not significantly impact our results.

\begin{rotatetable}
\begin{deluxetable*}{l r r r r r r r r r r r r r r r r r r r}
\tablewidth{0pt}
\tabletypesize{\tiny}
\tablecaption{Absolute Proper Motion Catalog of the Arches Cluster}
\tablehead{
\colhead{Name} & \colhead{F127M} & \colhead{$\sigma_{F127M}$} & \colhead{F139M} & \colhead{$\sigma_{F139M}$} & \colhead{F153M} & \colhead{$\sigma_{F153M}$} &
\colhead{$\alpha^*_0$\tablenotemark{a}} & \colhead{$\sigma_{\alpha^*_0}$} & \colhead{$\delta_0$\tablenotemark{a}} & \colhead{$\sigma_{\delta_0}$} & \colhead{$\mu_{\alpha^*}$} &  \colhead{$\sigma_{\mu_{\alpha^*}}$} &\colhead{$\mu_{\delta}$} &\colhead{$\sigma_{\mu_{\delta}}$} & \colhead{$t_0$} & \colhead{N$_{obs}$} & \colhead{$\chi^2_{\alpha^*}$} & \colhead{$\chi^2_{\delta}$} & \colhead{P$_{clust}$} \\
& mag & mag & mag & mag & mag & mag & $''$ & $''$ & $''$ & $''$ & mas yr$^{-1}$ & mas yr$^{-1}$ & mas yr$^{-1}$ & mas yr$^{-1}$ & year &  &   &   &
}
\startdata
A0000001 & 11.41 & 0.02 & 11.12 & 0.01 & 10.81 & 0.02 & -11.38042 & 0.00024 & -2.89786 & 0.00023 & 0.23 & 0.11 & -1.01 & 0.10 & 2013.0250 & 4 & 0.46 & 0.06 & 0.00 \\
A0000002 & 11.41 & 0.01 & 11.13 & 0.01 & 10.85 & 0.01 & 36.87441 & 0.00025 & 26.16057 & 0.00024 & 0.17 & 0.11 & -4.78 & 0.10 & 2012.9390 & 4 & 1.69 & 0.85 & 0.00 \\
A0000003 & 11.85 & 0.03 & 11.53 & 0.02 & 11.15 & 0.02 & -92.22162 & 0.00040 & -6.60737 & 0.00054 & 2.41 & 0.16 & -0.55 & 0.20 & 2013.4949 & 4 & 0.11 & 0.54 & 0.00 \\
A0000004 & 14.27 & 0.01 & 13.60 & 0.01 & 11.73 & 0.01 & -84.04126 & 0.00033 & -15.98480 & 0.00048 & -0.53 & 0.13 & -2.65 & 0.18 & 2013.5663 & 4 & 0.13 & 0.23 & 0.00 \\
A0000005 & 12.43 & 0.01 & 12.10 & 0.01 & 11.76 & 0.01 & 0.49685 & 0.00018 & -21.67100 & 0.00016 & -2.80 & 0.08 & -3.48 & 0.06 & 2013.0293 & 4 & 0.33 & 0.38 & 0.00 \\
\enddata
\tablecomments{Description of columns: \emph{Name}: star name, \emph{F127M, F139M, F153M}: mags in corresponding filters (Vega),
\emph{$\sigma_{F127M}$, $\sigma_{F139M}$, $\sigma_{F153M}$}: error in corresponding mags,
\emph{$\alpha^*_0$, $\delta_0$}: $\alpha^*$ and $\delta$ positions at $t_0$,
\emph{$\sigma_{\alpha^*_0}$, $\sigma_{\delta_0}$}: error in $\alpha^*_0$ and $\delta_0$,
\emph{$\mu_{\alpha^*}$, $\mu_{\delta_0}$}: proper motions in $\alpha^*$ and $\delta$,
\emph{$\sigma_{\mu_{\alpha^*}}$, $\sigma_{\mu_{\delta_0}}$}: error in $\mu_{\alpha^*}$ and $\mu_{\delta}$,
\emph{$t_0$}: reference time for proper motion fits (Eqns. \ref{eq:pm_ra}, \ref{eq:pm_dec}),
\emph{N$_{obs}$}: Number of epochs observed,
\emph{$\chi^2_{\alpha^*}$, $\chi^2_{\delta}$}: $\chi^2$ values for $\mu_{\alpha^*}$ and $\mu_{\delta}$ fits,
\emph{P$_{clust}$}: cluster membership probability calculated via the GMM
}
\tablecomments{This table, as well as a similar table for the Quintuplet cluster, is available in its entirety in the machine-readable format.}
\tablenotetext{a}{Positions are relative to ($\alpha$(J2000), $\delta$(J2000)) =
(17\textsuperscript{h}45\textsuperscript{m}50\textsuperscript{s}.65020, -28$^{\circ}$49$'$19$''$.51468)
for the Arches catalog and
($\alpha$(J2000), $\delta$(J2000)) = (17\textsuperscript{h}46\textsuperscript{m}14\textsuperscript{s}.68579, -28$^{\circ}$49$'$38$''$.99169)
for the Quintuplet catalog.}
\label{tab:arches_cat}
\end{deluxetable*}
\end{rotatetable}

\addtolength{\voffset}{5cm}

\clearpage

\addtolength{\voffset}{-5cm}

\section{The Absolute Proper Motion of the Arches and Quintuplet Clusters}
\label{sec:motion}
Star cluster members share a common proper motion on the sky
and thus form a distinct concentration in proper-motion space.
As a result, proper motions provide a reliable method for
identifying stars in the Arches and Quintuplet clusters \citep[e.g.][]{Stolte:2008qy, Clarkson:2012ty, Husmann:2012lr, Hosek:2015cs},
a difficult task to do from photometry alone
due to severe differential extinction across the field.
Following \citetalias{Hosek:2019kk} and  \citetalias{2019ApJ...877...37R}, we use a
Gaussian Mixture Model (GMM) to model the proper motion distribution of
the cluster and field star populations.
Because our proper motions are tied to the \emph{Gaia} EDR3 reference
frame, the center of the Gaussian distribution
that describes the cluster population corresponds to
the bulk proper motion of the cluster in ICRF.
In addition, a cluster membership probability for each star is
calculated using the GMM and is included in the proper
motion catalogs (Table \ref{tab:arches_cat}).
Details of the GMM analysis and the best-fit models
are provided in Appendix \ref{app:gmm}.

\begin{figure*}
\includegraphics[scale=0.35]{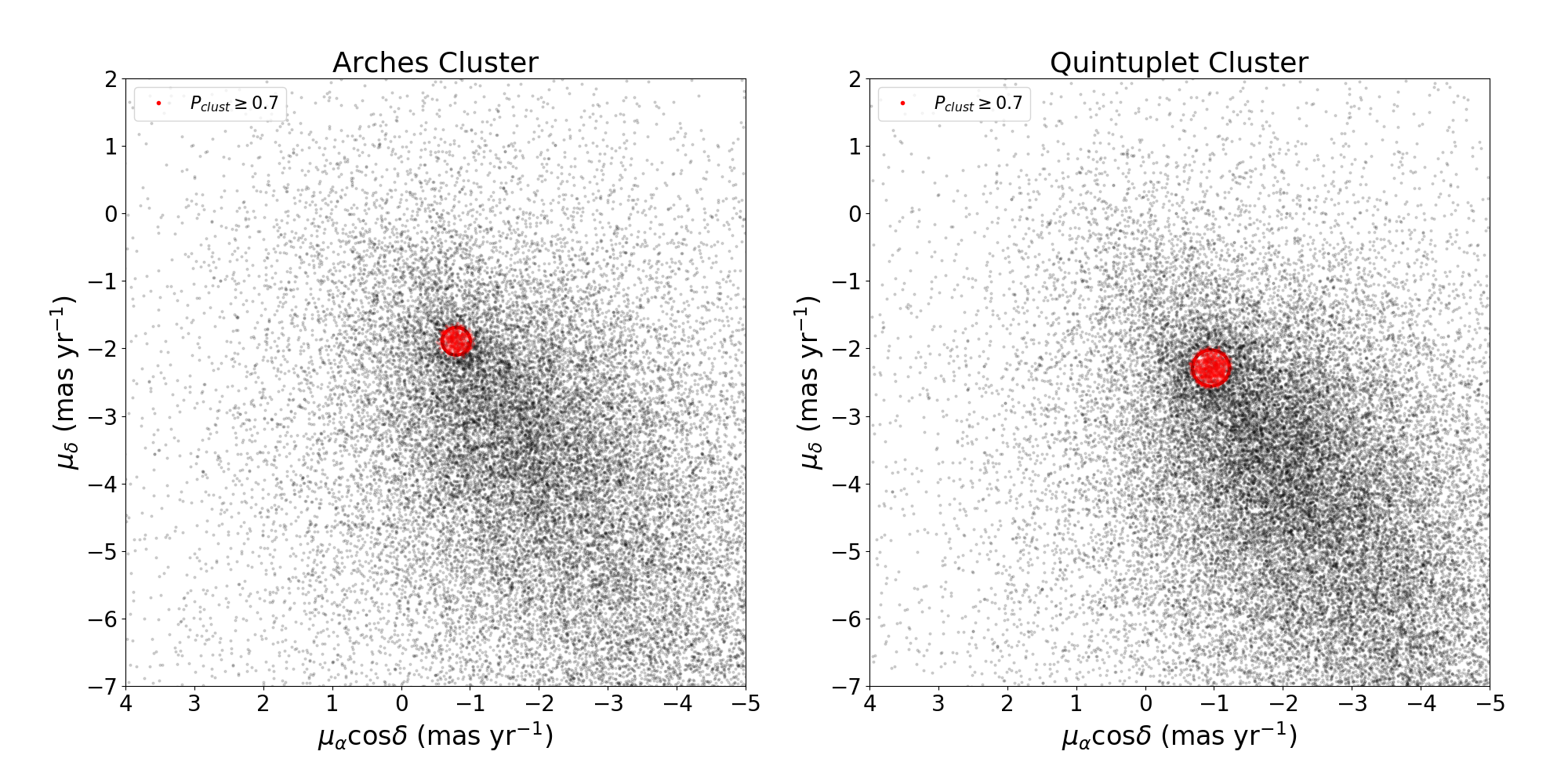}
\caption{The proper motions of stars in the Arches (left) and Quintuplet (right) fields. Cluster members form a concentrated distribution relative to the field stars. In each plot, the red circle represents the 2$\sigma$ probability contour of the cluster Gaussian in the GMM and stars with cluster membership probabilities greater than or equal to 0.7 are shown as red points. The velocity centroid of the cluster Gaussian represents the cluster's bulk proper motion in the HST-Gaia reference frame, which is tied to ICRF. }
\label{fig:gmm}
\end{figure*}

The distribution of cluster and field star proper motions in each field
is shown in Figure \ref{fig:gmm}.
The bulk proper motions of the clusters in ICRF are
$(\mu_{\alpha*}, \mu_{\delta})_{ICRF}$ = $(-0.80 \pm 0.032, -1.89 \pm 0.021) \textrm{ mas yr$^{-1}$}$
for the Arches and
$(\mu_{\alpha*}, \mu_{\delta})_{ICRF}$ = $(-0.96 \pm 0.032, -2.29 \pm 0.023) \textrm{ mas yr$^{-1}$}$
for the Quintuplet.
The uncertainty in this measurement is the quadratic sum of the
uncertainty in the centroid of the cluster Gaussian and
the uncertainty in the reference frame ($\mathsection$\ref{sec:transforms}).
The reference frame dominates the uncertainty in this measurement.

The cluster proper motion in ICRF
is a combination of the cluster's
motion in the rest frame of the galaxy
(i.e., its motion relative to SgrA*,
hereafter referred to as the ``SgrA*-at-Rest'' frame)
as well as the reflex motion from
the sun's orbit in the Galaxy.
To calculate the cluster motion in the SgrA*-at-Rest frame, we subtract
the observed proper motion of SgrA* in ICRF
\citep[($\mu_{\alpha*}, \mu_{\delta}$)$_{ICRF}$ =
(-3.156 $\pm$ 0.006, -5.585 +/- 0.010) mas yr$^{-1}$;][]{Reid:2020jo},
which we assume is induced entirely by the solar orbital motion.
The motions of the clusters in the SgrA*-at-Rest frame are thus
$(\mu_{\alpha*}, \mu_{\delta})_{int}$ = $(2.36 \pm 0.033, 3.70 \pm 0.024) \textrm{ mas yr$^{-1}$}$
for the Arches and
$(\mu_{\alpha*}, \mu_{\delta})_{int}$ = $(2.20 \pm 0.032, 3.30 \pm 0.025) \textrm{ mas yr$^{-1}$}$
for the Quintuplet.

A summary of the these proper motion measurements is
provided in Table \ref{tab:pm_measurements}.

\begin{deluxetable}{c c c c}
\tabletypesize{\small}
\tablecaption{Absolute Proper Motions of the Clusters}
\label{tab:pm_measurements}
\tablehead{
\colhead{Cluster} & \colhead{Ref Frame} & \colhead{$\mu_{\alpha^*}$} & \colhead{$\mu_{\delta}$}  \\
&  & mas yr$^{-1}$ & mas yr$^{-1}$
}
\startdata
Arches & ICRF & -0.80 $\pm$ 0.032 & -1.89 $\pm$ 0.021 \\
Arches & SgrA*-at-Rest & 2.36 $\pm$ 0.033 & 3.70 $\pm$ 0.024 \\
&  &  &  \\
Quintuplet & ICRF & -0.96 $\pm$ 0.032 & -2.29 $\pm$ 0.023 \\
Quintuplet & SgrA*-at-Rest & 2.20 $\pm$ 0.032 & 3.30 $\pm$ 0.025 \\
\enddata
\tablecomments{Description of columns: \emph{Ref frame}: reference frame of the measurement,
\emph{$\mu_{\alpha^*}$}: proper motion in $\alpha^*$, \emph{$\mu_{\delta}$}: proper motion in $\delta$}
\end{deluxetable}

\section{Modeling the Cluster Orbits}
\label{sec:orb_sims}
To constrain the orbital properties of the Arches and Quintuplet clusters,
we evaluate which orbits can
reproduce their observed \emph{present-day}
positions and motions within the uncertainties.
The probability of a given orbit model $\theta$ is calculated via Bayes equation:

\begin{equation}
\label{eq:bayes}
P(\theta | \vec{\mathbf{x}}) = \frac{P(\vec{\mathbf{x}}|\theta) P(\theta)}{P(\vec{\mathbf{x}})}
\end{equation}

where $\vec{\mathbf{x}}$ is a vector representing the observed
position and motion of the cluster in galactic longitude
coordinates
($\vec{\mathbf{x}}$ = \{l, b, $d_{los}$, $\mu_{l^*}$, $\mu_{b}$, $v_{los}$\}, where l, b are the
galactic longitude and latitude, $d_{los}$ is the line-of-sight distance,
$\mu_{l^*}$, $\mu_{b}$ are the proper motion in galactic longitude and latitude,
and $v_{los}$ is radial velocity).
$P(\vec{\mathbf{x}}|\theta)$ is the
likelihood of observing $\vec{\mathbf{x}}$ given $\theta$,
$P(\theta)$ is the prior probability on the free parameters in $\theta$,
and $P(\vec{\mathbf{x}})$ is the evidence.

The orbit model $\theta$ is described in $\mathsection$\ref{sec:orb_model}
while the observational constraints on the cluster positions
and motions are given in $\mathsection$\ref{sec:obs_constraints}.
The likelihood equation used to compare $\theta$
to the observations is defined in $\mathsection$\ref{sec:likelihood_eval},
while the adopted gravitational potential for the GC
is discussed in $\mathsection$\ref{sec:gc_gpot}.

\subsection{The Orbit Model}
\label{sec:orb_model}
We adopt an orbit model $\theta$ with 7 free parameters.
Six of the free parameters describe the birth position ($x_b$, $y_b$, $z_b$)
and birth velocity ($vx_b$, $vy_b$, $vz_b$) of the cluster
in Galactocentric coordinates\footnote{This is
a left-handed coordinate system centered on SgrA* where $x$ = $D\cdot \cos l \cdot \cos b$, $y$ = $D \cdot \sin l \cdot \cos b$, and $z$ = $D \cdot \sin b$,
where $D$ is the distance to the object.
$x$ is positive in the line-of-sight direction from the GC toward the Sun, $y$ is positive in the Galactic Plane
towards positive $l$, and $z$ is positive towards the North Galactic Pole \citep{Bovy_book}.}.
The final model parameter is the current cluster age $t_{clust}$.

The birth positions and velocities have uniform priors
with bounds based on the assumption that the clusters
formed within the CMZ.
The priors on $x_b$, $y_b$, and $z_b$ encompass the spatial
distribution of gas observed in the region.
$x_b$ and $y_b$, which define the location on the galactic plane,
conservatively extend $\pm$300 pc from SgrA* \citep[e.g.][]{Morris:1996db}.
$z_b$, which defines the location
perpendicular to the galactic plane, spans
$\pm$60 pc, corresponding to approximately twice the observed height
of the distribution of dense gas in the region \citep[e.g.][]{Molinari:2011dq}.
Similarly, the priors for $vx_b$ and $vy_b$
span $\pm$300 km s$^{-1}$,
enveloping the range
of observed line-of-sight velocities
of gas within $|l|$ $<$ 2$^{\circ}$ \citep[corresponding to a spatial scale of $|l|$ $\lesssim$ 280 pc;][]{Bitran:1997st, Dame:2001fe}.
The prior for $vz_b$ also covers $\pm$300 km s$^{-1}$,
allowing for significant vertical oscillations from the
Galactic Plane.
These priors are intentionally broad so
that the resulting constraints on the birth positions
and velocities (and thus orbits) of the clusters
are as agnostic as possible to proposed formation mechanisms,
and thus can be used to examine the viability of such
mechanisms ($\mathsection$\ref{sec:formation}).
Finally, the priors for t$_{clust}$ come from constraints in the literature,
which generally fall between 2.5 -- 3.5 Myr
for the Arches \citep[e.g.][]{Najarro:2004ij, Martins:2008hl, Clark:2018ij}
and 3 -- 5 Myr for the Quintuplet \citep[e.g.][]{Liermann:2012qq, Husmann:2012lr, Clark:2018qf}.
Thus, we adopt Gaussian priors for t$_{clust}$
with means and standard deviations of 3 $\pm$ 0.5 Myr and
4 $\pm$ 1 Myr for the Arches and Quintuplet, respectively.
The model parameters and priors are summarized
in Table \ref{tab:orb_priors}.

The parameter space is explored using
the multimodal nested sampling algorithm \texttt{Multinest} \citep{Feroz:2008yu, Feroz:2009lq},
called by the \texttt{python} wrapper code \texttt{PyMultinest} \citep{Buchner:2014wa}.
Values for the free parameters
are drawn from the priors
and the corresponding orbit is integrated in the GC
gravitational potential to t$_{clust}$ using \texttt{galpy} \citep{Bovy:2015bd}.
This makes a prediction for what the \emph{present-day} cluster position
and velocity should be
in Galactocentric coordinates (e.g., physical units relative to SgrA*).

\begin{deluxetable}{c c c c}
\tablecaption{Orbit Model: Parameters and Priors}
\label{tab:orb_priors}
\tablehead{
\colhead{Parameter} & \colhead{Arches Prior} & \colhead{Quintuplet Prior} & \colhead{Units}
}
\startdata
$x_b$ &  U(-300, 300) & U(-300, 300) & pc \\
$y_b$ &  U(-300, 300) & U(-300, 300) & pc \\
$z_b$ &  U(-60, 60) & U(-60, 60) & pc \\
$vx_b$ &  U(-300, 300) & U(-300, 300) & km s$^{-1}$ \\
$vy_b$ &  U(-300, 300) & U(-300, 300) & km s$^{-1}$  \\
$vz_b$ &  U(-300, 300) & U(-300, 300) & km s$^{-1}$  \\
$t_{clust}$ & G(6.48, 0.07) & G(6.60, 0.10) & log(years) \\
\enddata
\tablecomments{Uniform distributions: U(min, max), where min and max are bounds of the distribution; Gaussian distributions: G($\mu$, $\sigma$), where $\mu$ is the mean and $\sigma$ is the standard deviation}
\end{deluxetable}

\subsection{Observational Constraints}
\label{sec:obs_constraints}
To define $l$, $b$, $\mu_{l^*}$, and $\mu_b$
for $\vec{\mathbf{x}}$ in Equation \ref{eq:bayes},
we convert our results from $\mathsection$\ref{sec:motion}
into Galactic coordinates.
To calculate $\mu_{l^*}$ and $\mu_b$, the
($\mu_{\alpha^*}$, $\mu_{\delta}$)$_{ICRF}$ proper motions
are rotated clockwise by an angle of 90$^{\circ}$ - 31.40$^{\circ}$ = 58.60$^{\circ}$,
where 31.40$^{\circ}$ is the position angle of the galactic plane \citep[][]{Reid:2004xh}.
Following \citet{Libralato:2020ev}, we use a Monte Carlo approach to
propagate the proper motion uncertainties through the rotation\footnote{We take 5000 random samples from Gaussian distributions
describing the equatorial proper motions, with
the mean and standard deviation of each gaussian equal to the
measured proper motion and its uncertainty, respectively.
We then rotate each sample into Galactic coordinates as described
here, and then take the standard deviation of the rotated coordinates
to be the corresponding uncertainty in galactic coordinates.}.

The position of each cluster is
calculated from the median ($\alpha^*_0$, $\delta_0$)
position of all stars with cluster membership probabilities $\ge$ 0.7
in the proper motion catalogs.
This represents the cluster position at
2013.0750 for the Arches and 2013.3530 for the Quintuplet.\footnote{These times represent the median
$t_0$ for the proper motion fits for the stars used in this calculation.}
These positions are transformed into ($l$, $b$) using the
\texttt{astropy SkyCoord} package \citep{Astropy-Collaboration:2013kx, Astropy-Collaboration:2018ws}.
The uncertainty in $l$ and $b$ is conservatively estimated to
be the half-light radii for the clusters measured as
by \citet{Hosek:2015cs} and \citet{2019ApJ...877...37R}.

Constraints on $v_{los}$ and $d_{los}$ are taken from the literature.
The heliocentric $v_{los}$ of the clusters are measured via spectroscopic studies of the
brightest cluster members \citep{Figer:2002nr, Liermann:2009lr}.
Meanwhile, $d_{los}$ is allowed to range
$\pm$300 pc from SgrA*, making it the weakest
of the observational constraints.

This is justified by evidence that the clusters
currently reside within the CMZ, such as
the ionization of the nearby Arched
Filaments and Sickle structures \citep[e.g.][]{Lang:1997ah, Lang:2001qf, Simpson:2007tg, Wang:2010pr},
the orientation of surrounding ionized cloud edges and gas pillars \citep{Stolte:2014qf},
and the measured trends in gas and dust infrared luminosity and temperatures
in the region relative to the clusters \citep[e.g.][]{Cotera:2005gi, Hankins:2017sc}.
It has also been suggested that diffuse X-ray emission detected
near the Arches cluster may be a bow shock due to the cluster
colliding with a surrounding molecular cloud \citep[e.g.][]{Wang:2006wl}.
However, the localized position of the clusters within the CMZ
has not yet been determined.

A summary of the observational constraints used to define
$\vec{\mathbf{x}}$ is provided in Table \ref{tab:obs_vals}.

\begin{deluxetable*}{c c c c c c c}
\tablecaption{Observational Constraints on the Present-Day Cluster Positions and Motions}
\label{tab:obs_vals}
\tablehead{
&  & \multicolumn{2}{c}{\underline{Arches Cluster}} & & \multicolumn{2}{c}{\underline{Quintuplet Cluster}} \\
\colhead{Parameter} & \colhead{Units} & \colhead{Value} & \colhead{Reference} &  & \colhead{Value} & \colhead{Reference}
}
\startdata
$l$ & deg & 0.1230 $\pm$ 0.003 & This work &  & 0.1640 $\pm$ 0.005 & This work  \\
$b$ & deg & 0.0175 $\pm$ 0.003 & This work &  & -0.0602 $\pm$ 0.005 & This work  \\
$\mu_{l*}$ & mas yr$^{-1}$ & -2.03 $\pm$ 0.025 & This work &  & -2.45 $\pm$ 0.026 &This work \\
$\mu_b$ & mas yr$^{-1}$  & -0.30 $\pm$ 0.029 & This work &  & -0.37 $\pm$ 0.029 & This work \\
$d_{los}$ & pc & $\pm$300 & see $\mathsection$\ref{sec:obs_constraints} & & $\pm$300 & see $\mathsection$\ref{sec:obs_constraints} \\
$v_{los}$ & km s$^{-1}$  & 95 $\pm$ 8 & \citet{Figer:2002nr} &  & 102 $\pm$ 2 & \citet{Liermann:2009lr}  \\
\enddata
\tablecomments{Description of Parameters:
($l$, $b$)  = galactic longitude and latitude (ICRF coordinates; t = 2013.0750 for Arches, t = 2013.3530 for Quintuplet),
$d_{los}$ = the line-of-sight distance relative to SgrA*,
($\mu_{l^*}$, $\mu_{b}$) = proper motion in galactic longitude and latitude (ICRF coordinates),
and $v_{los}$ = heliocentric line-of-sight velocity}
\end{deluxetable*}

\subsection{Evaluating the Likelihood}
\label{sec:likelihood_eval}
The orbit model makes a prediction of the present-day position and motion of the cluster
in Galactocentric coordinates, which are
in physical units relative to SgrA* (pc, km s$^{-1}$).
These predictions must be converted into observable
units in order to compare
to the observed values in
Table \ref{tab:obs_vals} (deg, mas yr$^{-1}$ in Galactic coordinates).
To convert from ($y$, $z$) to ($l$, $b$),
we assume that the distance between Earth and SgrA* (R$_0$)
is 8090 $\pm$ 140 pc \citep[e.g. the average value and spread between different estimates of R$_0$ in the literature;][]{Do:2019gr, Gravity-Collaboration:2022mb}
and that the ICRF position of SgrA* is ($l$, $b$) = (-0.05576, -0.04617)$^{\circ}$
\citep{Reid:2004xh}\footnote{The positional uncertainty of $\sim$10 mas for SgrA* is negligible in this analysis and is ignored.}.
Similarly, to convert from ($vy$, $vz$) to ($\mu_{l*}$, $\mu_{b}$)
we use the same R$_0$ and assume an ICRF proper motion for SgrA* of
($\mu_{l*}$, $\mu_{b}$) = (-6.411, -0.219) mas yr$^{-1}$
\citep{Reid:2020jo}\footnote{The proper motion uncertainty of $\sim$0.008 mas yr$^{-1}$ for SgrA* is negligible in this analysis and is ignored.}.
Finally, we convert $vx$ into heliocentric v$_{los}$
by adopting a heliocentric radial velocity for SgrA* of -11.1 $\pm$ 1.2 km s$^{-1}$ \citep{Schonrich:2010ho},
where the uncertainty is quadratic sum of reported statistical and statistical uncertainties.

We define the likelihood of a given orbit model as:

\begin{equation}
\label{eq:likelihood}
P(\vec{\mathbf{x}}|\theta) = \prod_{i \neq d_{los}} \exp\left({-\frac{(x_{i,obs} - x_{i,\theta})^2}{2 (\sigma_{x_{i,obs}}^2 + \sigma_{x_{i,\theta}}^2)} }\right) \psi(d_{los,\theta})
\end{equation}

\begin{equation}
\label{eq:dlos}
\psi(d_{los, \theta}) =
\begin{cases}
\frac{1}{600} & |d_{los, \theta}| \leq 300 \\
-\infty & |d_{los,\theta}| > 300
\end{cases}
\end{equation}

The first term in Equation \ref{eq:likelihood} applies to all dimensions of $\vec{\mathbf{x}}$ except for $d_{los}$.
Within this term, $x_{i,obs}$ and $\sigma_{x_{i,obs}}$ are the observed value and uncertainty of
the $i$th dimension of $\vec{\mathbf{x}}$, respectively,
$x_{i,\theta}$ is the predicted value for the $i$th dimension from $\theta$,
and $\sigma_{x_{i,\theta}}$ is the uncertainty in $x_{i,\theta}$
incurred when converting from physical to observed units
due to the uncertainties in R$_{0}$ and the radial velocity of SgrA*.

The second term is a piecewise function that only depends on
the current line-of-sight distance of the cluster predicted by the orbit
model, $d_{los,\theta}$.
If $|d_{los,\theta}|$ $>$ 300 pc, then it forces the
likelihood to be $-\infty$ since this would violate
the constraint that the cluster is currently
within the CMZ.
If $|d_{los,\theta}|$ $\leq$ 300 pc,
then the second term is a constant value
of 1 / 600, corresponding to
a uniform probability distribution spanning $\pm$300 pc.
Thus, as long as $|d_{los,\theta}|$ $\leq$ 300 pc,
this parameter does not impact the relative likelihood
between orbit models.

\subsection{GC Gravitational Potential}
\label{sec:gc_gpot}
For the orbit integration, we adopt the same gravitational potential as
\citetalias{Kruijssen:2015fx},
namely an axisymmetric
potential based on the enclosed mass distribution from
\citet{Launhardt:2002hl} but flattened in the $z$ direction by
a factor $q_{\phi}$ = 0.63 (see $\mathsection$\ref{sec:methods_kdl15} for additional details).
However, the GC gravitational potential is uncertain,
and alternative potentials have been proposed \citep[e.g.][]{Sormani:2020my}.
We explore the impact of different gravitational
potentials on our results in $\mathsection$\ref{sec:other_gpot}.

We note that the adopted gravitational potential (as well as the alternative potentials examined in
$\mathsection$\ref{sec:other_gpot}) is axisymmetric, and thus ignores the non-axisymmetric
component of the potential due to the Galactic bar. However, for the range of galactic radii
considered in this analysis (r $\lesssim$ 300 pc) the potential is dominated by
the Nuclear Star Cluster and Nuclear Stellar Disk \cite[e.g.][]{Launhardt:2002hl}.
The observed properties of these structures
appear to be well reproduced by axisymmetric models \citep{Gerhard:2012he, Chatzopoulos:2015lq, Sormani:2022lz}.
Thus, the assumption of an axisymmetric potential is adequate for this analysis, especially for the
short timescales under consideration ($\lesssim$5 Myr).

\section{The Orbits of the Arches and Quintuplet Clusters}
\label{sec:orbit}

For both clusters, the posterior probability distributions for $\theta$
are bimodal and show significant degeneracies.
Orbits drawn from the two solution modes for the Arches and Quintuplet
are shown in Figures \ref{fig:orb_results_arch} and \ref{fig:orb_results_quint},
respectively.
The main feature distinguishing the modes is
the direction of the orbit.
As viewed from the North Galactic Pole
(e.g., the left panels in Figures \ref{fig:orb_results_arch} and \ref{fig:orb_results_quint}),
the first mode contains clockwise orbits
around SgrA* while the second mode contains
counter-clockwise orbits around SgrA*.
The clockwise orbits follow the general direction
of gas flow in the CMZ and are henceforth referred to
as prograde orbits.
The counter-clockwise orbits move in the
opposite direction and are thus referred to as
retrograde orbits.
A summary of each mode, including the
parameters for the maximum a posteriori (MAP) orbit,
is provided in Table \ref{tab:orb_results}.
The full posterior probability distributions for the free parameters
of $\theta$ are presented in Appendix \ref{app:posteriors}.

The range of allowed orbits is primarily driven by the uncertainty
in the \emph{present-day} line-of-sight distance to the cluster, $d_{los}$.
As found in previous work \citep[][]{Stolte:2008qy, Stolte:2014qf, Libralato:2020ev},
prograde orbits place the
clusters in front of SgrA* relative to Earth (negative $d_{los}$)
while retrograde orbits place the
clusters behind SgrA* (positive $d_{los}$).
Figure \ref{fig:orbit_comp} shows four orbit metrics
as a function of $d_{los}$:
the periapse distance,
the ratio of apoapse to periapse distance,
the radial period \citep[i.e., the time it takes for the cluster
to travel from apoapse to periapse and back;][]{Binney:2008bh},
and ratio of cluster age to radial period.
While the total range of values for these metrics is large,
they are reasonably well constrained for a given $d_{los}$.

It is important to note that although $d_{los}$ is not constrained by
our likelihood function (other than the requirement that $|d_{los}|$ $<$ 300 pc),
it does not mean that each value of $d_{los}$ is equally likely in the orbit posteriors.
The probability distributions of $d_{los}$ for the clusters
is shown in Figure \ref{fig:dlos_prob}.
While broad, the $d_{los}$ distributions are not uniform because the other kinematic parameters
of the clusters (three-dimensional velocity, two-dimensional sky position)
are constrained by observations.
Thus, the volume of parameter space
(initial position and velocity) that can produce suitable orbits with a similar value of $d_{los}$
is not necessarily equal. In addition, there are cases where the initial conditions
required to obtain an orbit with a given $d_{los}$ fall
outside the boundaries of our priors (i.e., they are not consistent with
our assumption that the clusters formed within the CMZ) and are not allowed by our model.

The ratio of the probability of a given mode to
the probability of the most likely mode can be
calculated from the Bayes factor:

\begin{equation}
\label{eq:bayes_factor}
\psi_{i} = \frac{P(\vec{\mathbf{x}})_i}{P(\vec{\mathbf{x}})_{max}}
\end{equation}

Where $P(\vec{\mathbf{x}})_i$ is the evidence of the $i$th mode
and $P(\vec{\mathbf{x}})_{max}$ is the evidence of the most likely mode
(i.e., the mode with the largest evidence).
The mode probabilities are calculated by renomalizing
these ratios across all modes to sum to 1:

\begin{equation}
Prob_{i} = \frac{\psi_{i}}{\sum_j^m \psi_{j}}
\end{equation}

where $m$ is the total number of modes for the cluster.
The probabilities for the modes are reported in the plot titles of
Figures \ref{fig:orb_results_arch} and \ref{fig:orb_results_quint}
as well as in Table \ref{tab:orb_results}.
For the Arches cluster, the prograde and retrograde solution
modes have equal probability, while for the Quintuplet cluster
the prograde mode is slightly favored.
The asymmetry in the Quintuplet mode probabilities is largely
because more retrograde orbits are disallowed
by the prior boundaries compared to the prograde orbits.

While both prograde and retrograde orbit solutions
are allowed by this analysis, it is unclear how the clusters
could have formed on retrograde orbits.
This would imply that their natal molecular clouds,
which must have been quite massive in order to create the clusters,
must also have been on retrograde orbits.
Such high-mass retrograde clouds have
not been found in observations or simulations.
\emph{As a result, we restrict the remaining analysis in the paper
to the prograde solution mode only.}
When statistical quantities are calculated,
such as 3$\sigma$ limits or probability contours,
the probability of retrograde mode is
set to zero and only the prograde mode
is considered.

\subsection{Orbit Properties}
\label{sec:orb_prop}
We integrate the orbits from the model
posteriors forward 10 Myr in order to place
statistical constraints on several properties of interest.
The probability distributions for the closest approach
of each cluster to SgrA* is shown in left panel of Figure \ref{fig:orb_closest}.
The closest approach distance is directly related to
$d_{los}$; the smaller the value for $d_{los}$,
the closer the cluster approaches SgrA*.
We place 3$\sigma$ lower limits of 24.7 pc and 29.8 pc
for the closest approach of the Arches and Quintuplet, respectively.
This raises the question of whether dynamical friction,
which is ignored in these orbit calculations,
could cause either cluster to spiral inward and
merge with either the Nuclear Star Cluster (NSC) or Young Nuclear Cluster (YNC)
before being tidally disrupted.
This is discussed further in $\mathsection$\ref{sec:future}.

The ratio of apoapse to periapse distance
($r_{apo}$ / $r_{peri}$) provides a measure of
the eccentricity of the orbits.
The clusters share a similar probability distribution
for this ratio, with a 50th percentile
of $\sim$1.9 (equivalent to an
eccentricity of $\sim$0.31) and a tail that
extends to larger values (Figure \ref{fig:orb_closest}, right panel).
The distributions have 3$\sigma$ lower limits
of  $r_{apo}$ / $r_{peri}$ $\sim$1.4,
indicating that neither cluster
can be on a circular orbit,
in agreement with past studies \citep[][]{Stolte:2008qy, Stolte:2014qf, Libralato:2020ev}.
The largest values of $r_{apo}$ / $r_{peri}$ occur as $d_{los}$
decreases.\footnote{Figure \ref{fig:orbit_comp} also reveals
a sharp increase in $r_{apo}$ / $r_{peri}$ for small positive
values of $d_{los}$. This is caused by highly eccentric retrograde orbits that
become possible if the clusters formed nearly 300 pc from SgrA*
(the edge of the birth location prior).
While such orbits could bring the clusters as close as $\sim$15 pc
to SgrA*, it would not be close enough to cause a
merger with the NSC ($\mathsection$\ref{sec:future}).
As discussed in the text,
we do not investigate the retrograde orbits in further detail.}

The radial period of the clusters
also depends on $d_{los}$, ranging from $\sim$2 Myr
for small $d_{los}$ to $\sim$6 Myr for large $d_{los}$ (Figure \ref{fig:orbit_comp}, lower left panel).
The Arches cluster has completed at least 1 complete radial period if
$d_{los}$ $\lesssim$100 pc
while the Quintuplet has completed at least 1 radial period if
$d_{los}$ $\lesssim$ 140 pc (Figure \ref{fig:orbit_comp}, lower right panel).

\subsection{The Clusters Do Not Share A Common Orbit}
\label{sec:same_orbit}
\begin{figure*}
\begin{center}
\includegraphics[scale=0.25]{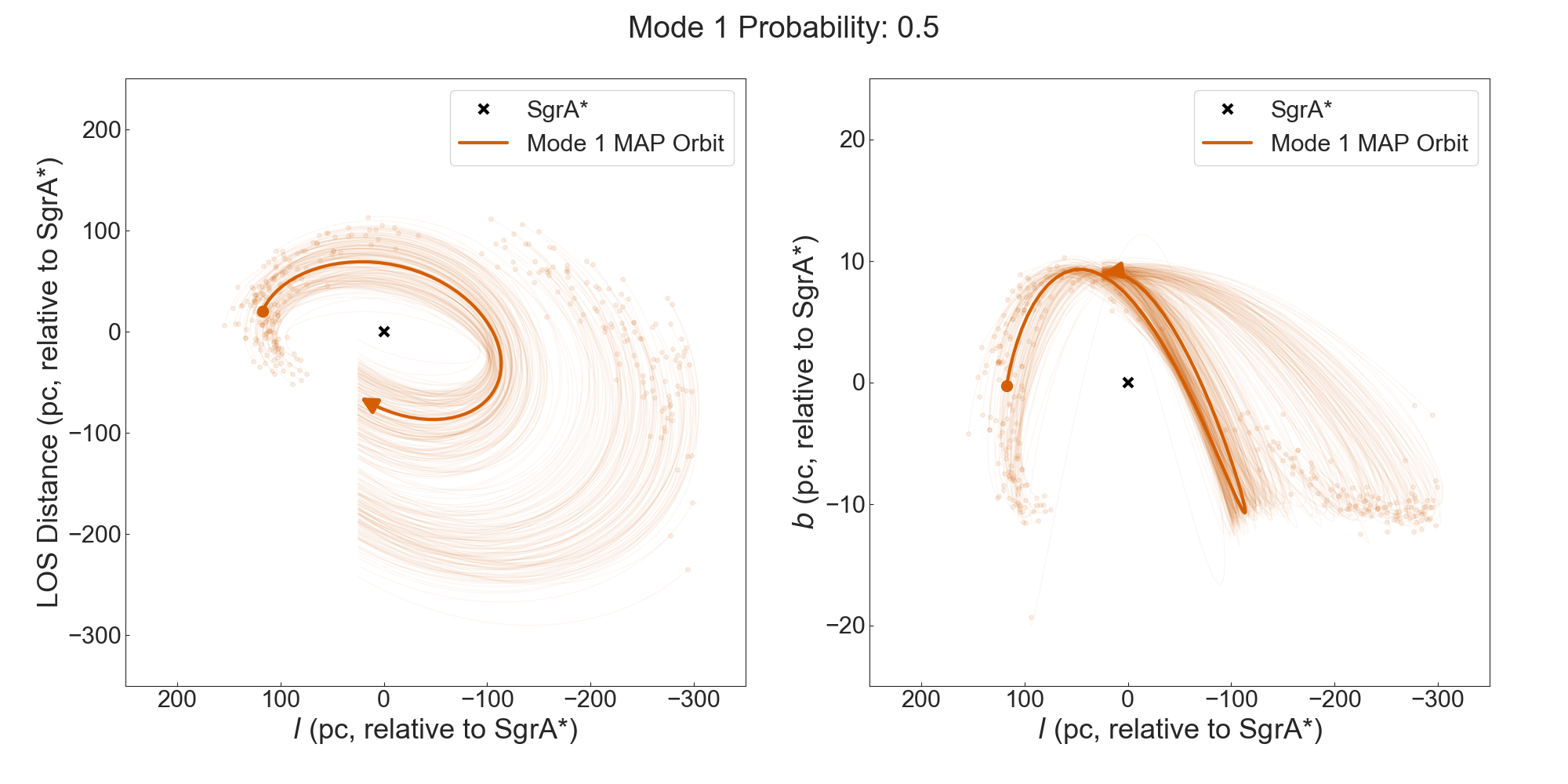}
\includegraphics[scale=0.25]{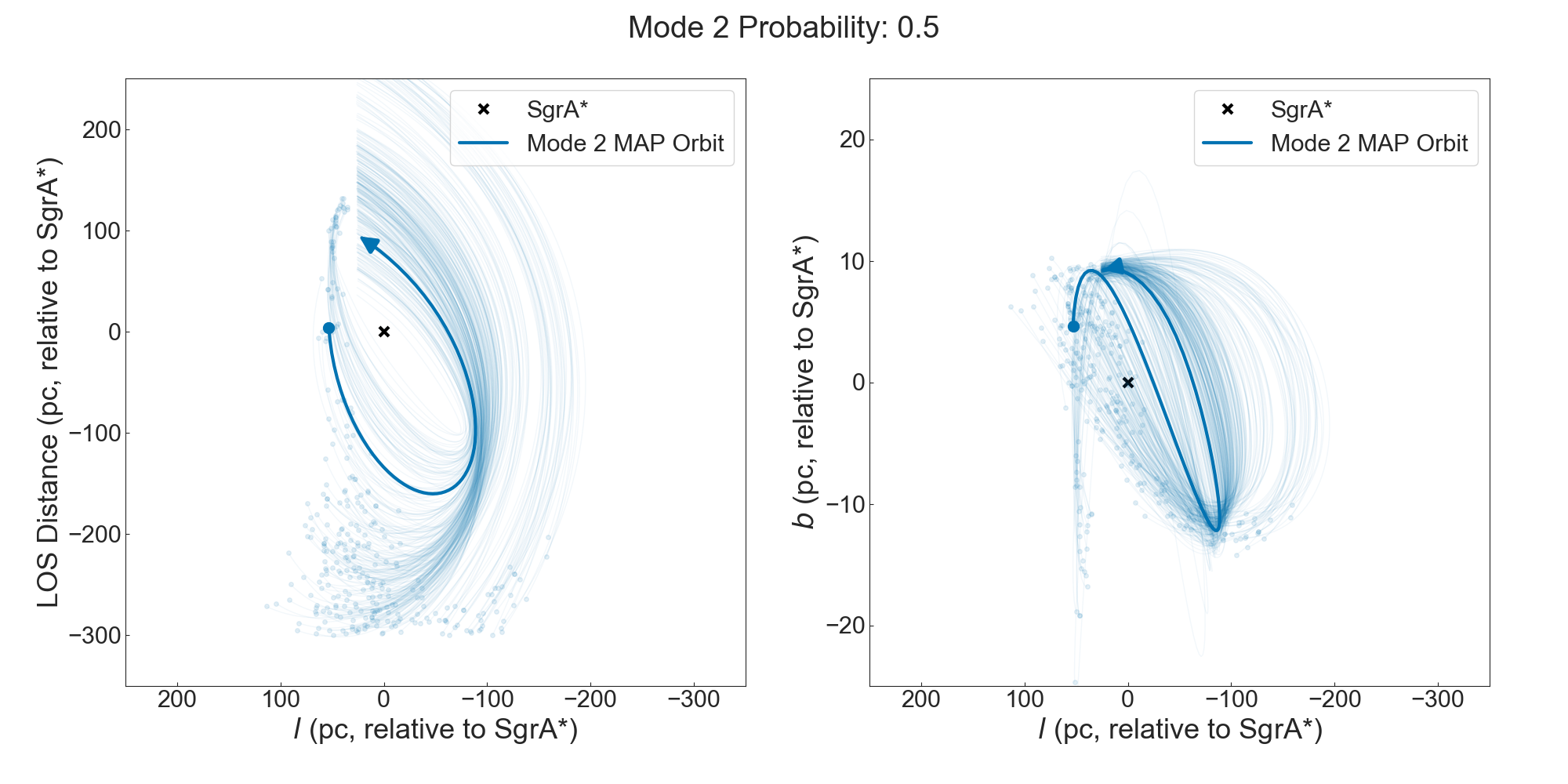}
\caption{Orbits drawn from the two solution modes for the Arches cluster. Each row represents a different mode, with the left panel showing a ``top-down'' view of the orbits
from the North Galactic Pole (with the Earth at negative LOS distance) and the right panel showing an ``edge-on'' view of the orbits from the Galactic Plane.
The maximum a posteriori (MAP) orbit for each mode is represented by the thick line, with a distribution
of orbits drawn from the posterior shown by the thin lines.
On the MAP orbit, the arrowhead shows the current position of the cluster while the circle shows the birth location of the cluster.
On the sample of orbits drawn from the posterior, the birth locations of the individual orbits are represented by the smaller and fainter circles.
The overall probability of each mode is given in the title of each row.}
\label{fig:orb_results_arch}
\end{center}
\end{figure*}

\begin{figure*}
\begin{center}
\includegraphics[scale=0.25]{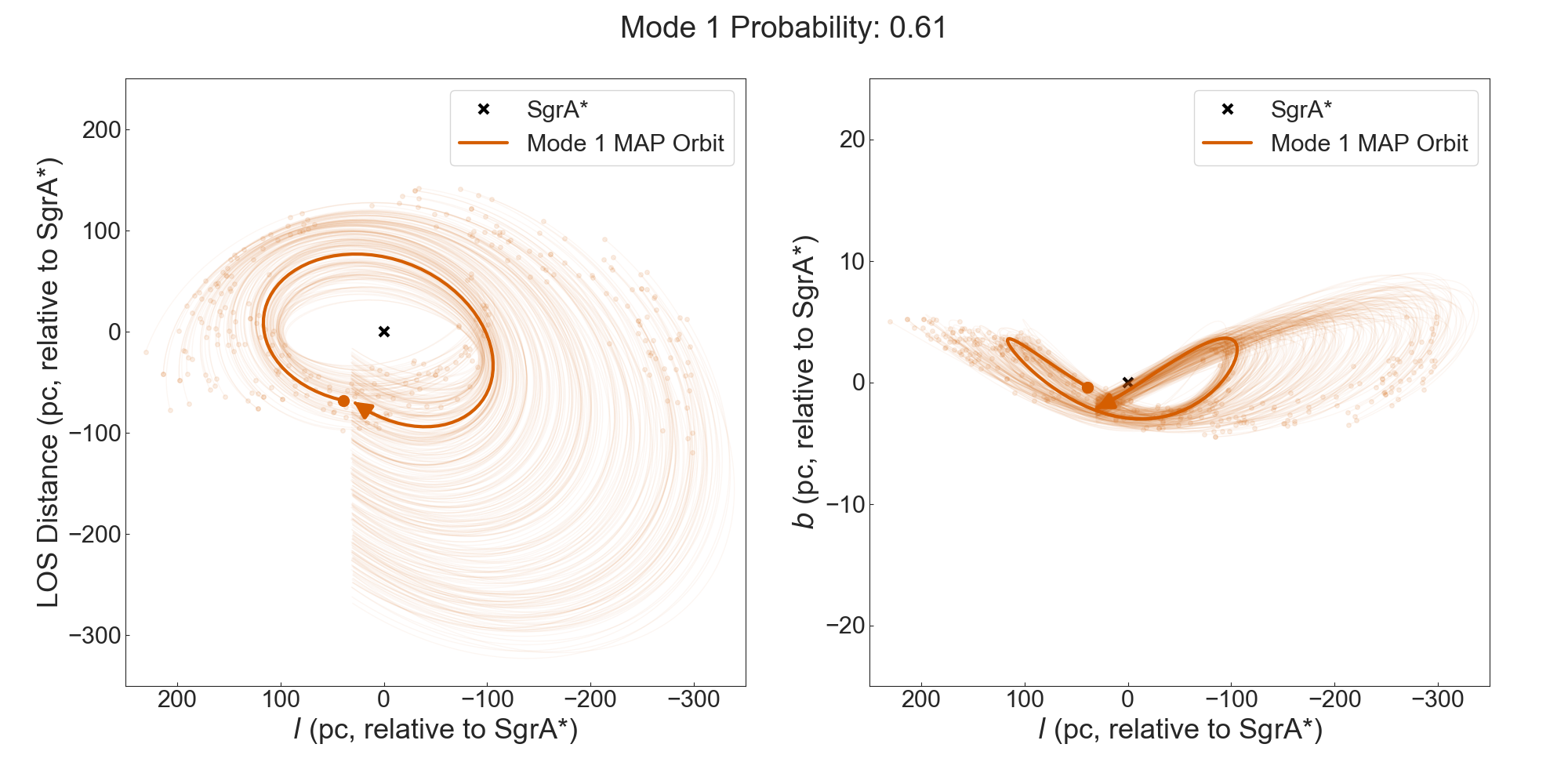}
\includegraphics[scale=0.25]{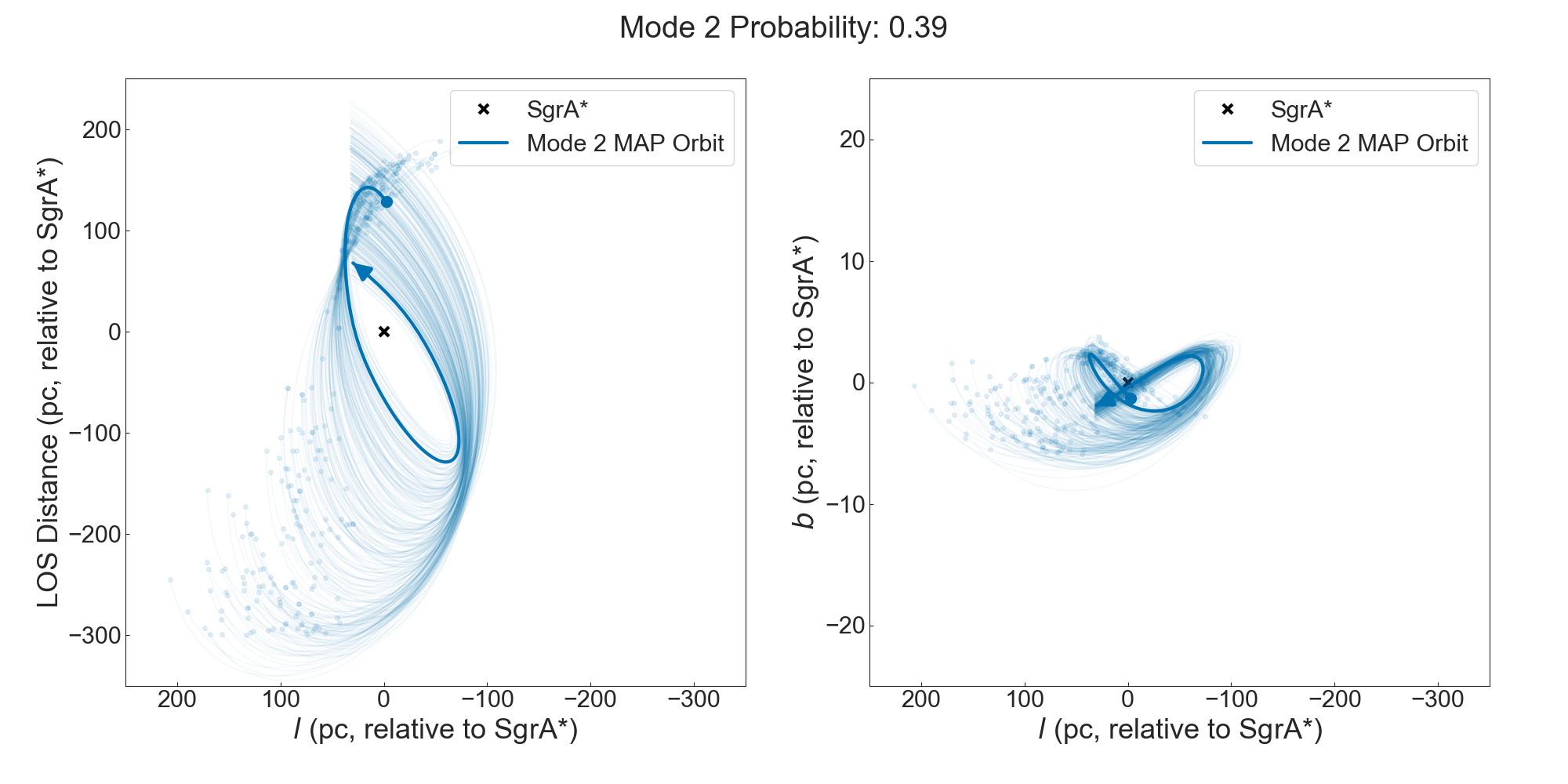}
\caption{Orbits drawn from the two solution modes for the Quintuplet cluster, plotted in same manner as Figure \ref{fig:orb_results_arch}.}
\label{fig:orb_results_quint}
\end{center}
\end{figure*}

\begin{figure*}
\begin{center}
\includegraphics[scale=0.3]{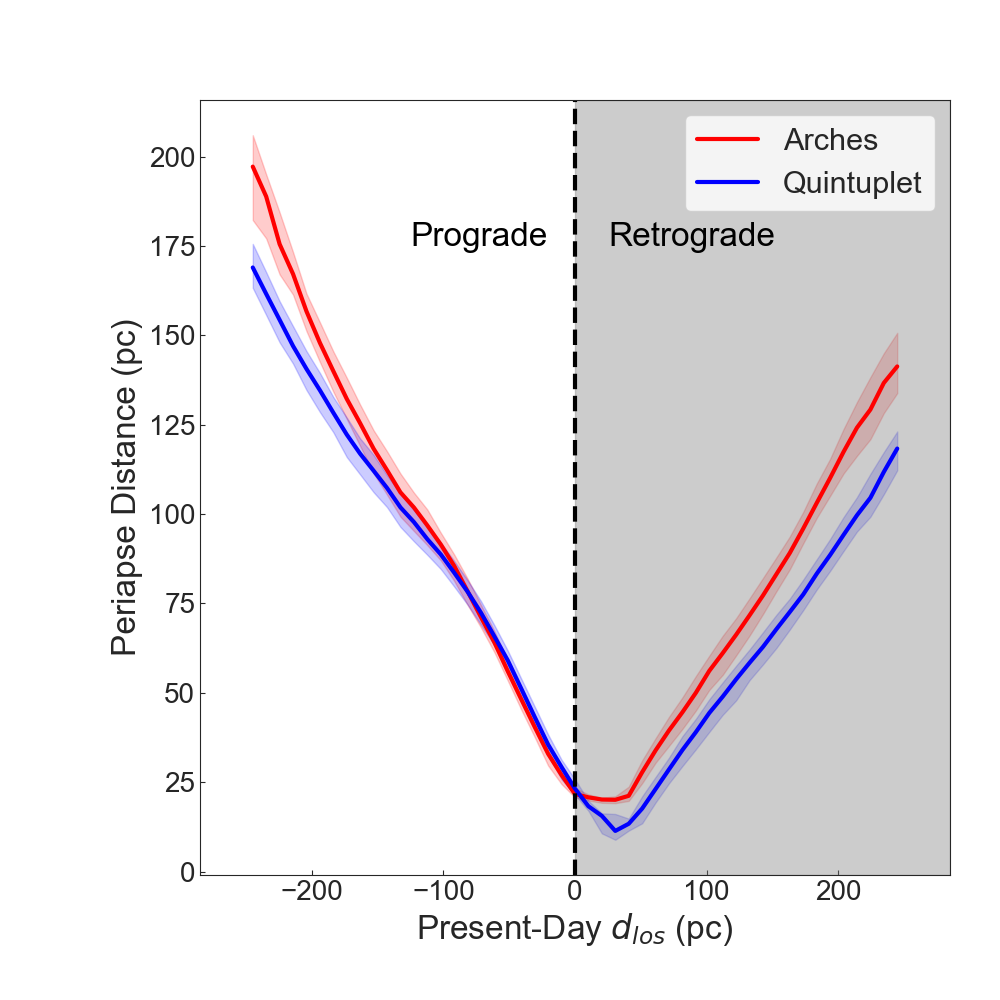}
\includegraphics[scale=0.3]{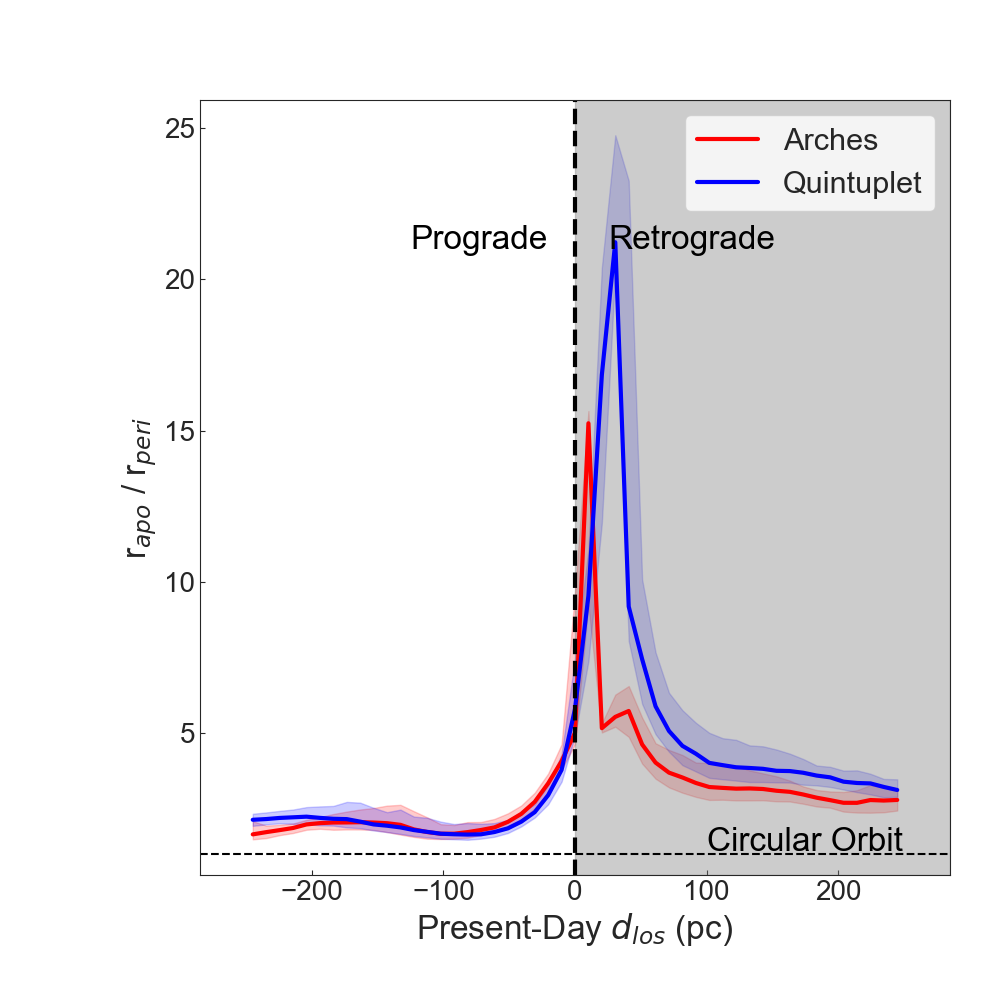}
\includegraphics[scale=0.3]{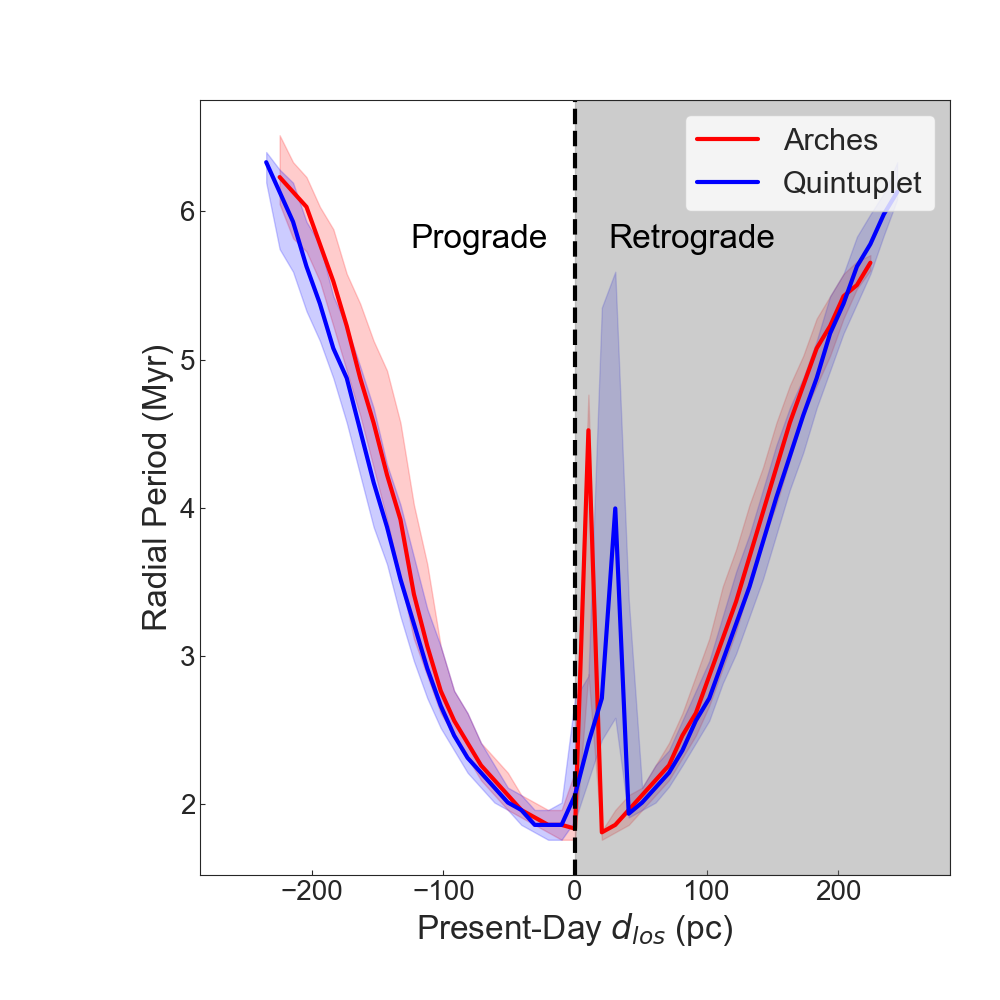}
\includegraphics[scale=0.3]{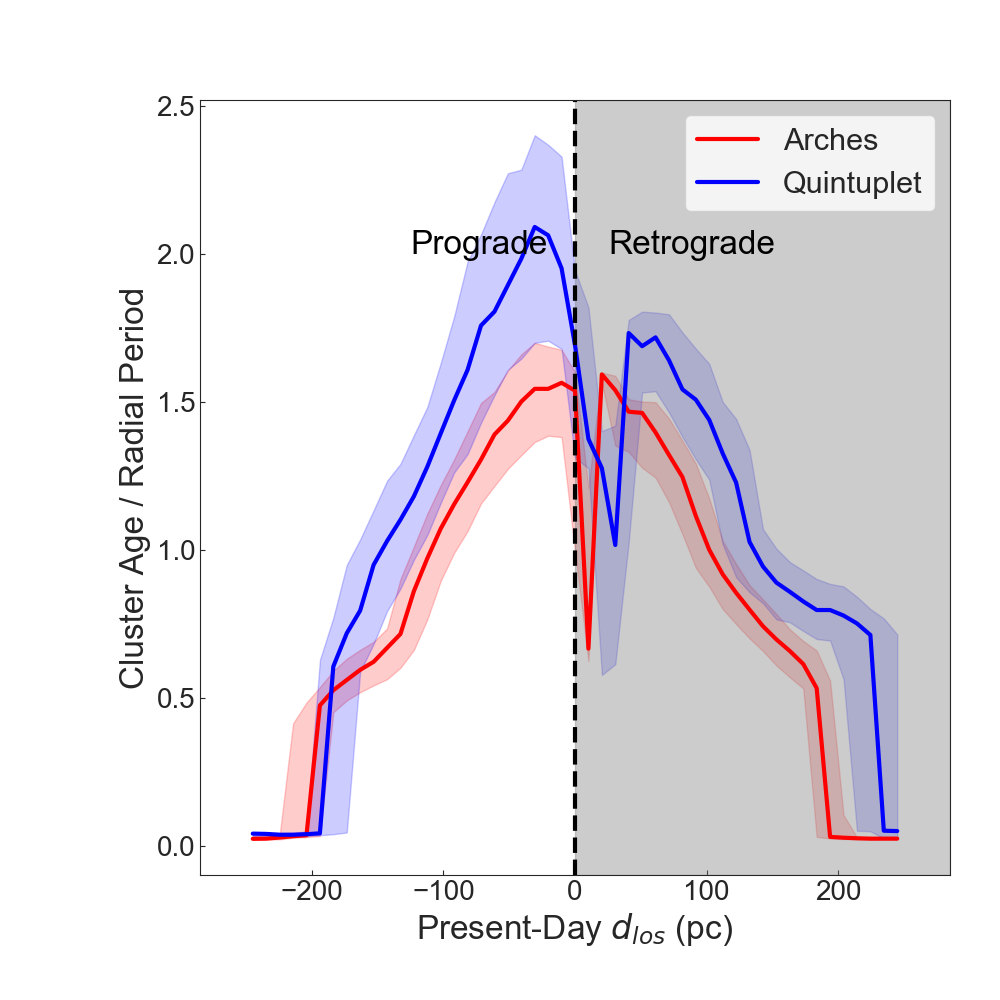}
\caption{The periapse distance (top left), ratio of apoapse to periapse (top right), radial period (bottom left), and ratio of cluster
age to radial period (bottom right) as a function of $d_{los}$
for the Arches (red) and Quintuplet (blue) clusters. For each cluster, the solid line represents the 50\% percentile in each radius bin,
while the shaded region represents the range between the 15.9\% and 84.1\% percentiles (1$\sigma$) in that bin.
The black dotted line separates the prograde orbits from the retrograde orbits.
For the remainder of the paper, only the prograde orbits are considered.}
\label{fig:orbit_comp}
\end{center}
\end{figure*}

Whether the Arches and Quintuplet clusters share a common orbit has
significant implications for their formation mechanism.
Our results reveal that the cluster orbits
appear to be similar in the
the ``top-down'' view
(left panels of Figures \ref{fig:orb_results_arch} and \ref{fig:orb_results_quint}),
as has been noted in past work \citep[e.g.][]{Stolte:2014qf, Libralato:2020ev}.
This suggests that the clusters likely share a similar formation mechanism.
However, the ``edge-on'' view of the orbits reveals that
the Arches tends to exhibit larger vertical oscillations in the Galactic
Plane compared to the Quintuplet
(right panels of Figures \ref{fig:orb_results_arch}
and \ref{fig:orb_results_quint}).
This difference is more clearly seen
in a comparison of the probability distributions
for the maximum deviation of the orbit from the
Galactic Plane, which we define as
$b_{max}$ (Figure \ref{fig:same_orbit}, left panel).

The probability that $b_{max}$ is consistent between
the Arches and Quintuplet, $P(b_{max, arch=quint})$,
can be calculated as:
\begin{equation}
\label{eq:zdist}
P(b_{max, arch=quint}) = \int_{-\infty}^{\infty} P_{arch}(b_{max}) P_{quint}(b_{max})  db_{max}
\end{equation}
where $P_{arch}(b_{max})$ and $P_{quint}(b_{max})$ are the probability distributions
of $b_{max}$ for the Arches and Quintuplet, respectively.
We find that $P(b_{max, arch=quint})$ = 0.01\%,
which corresponds to a difference of $\sim$3.9$\sigma$.
We therefore conclude that the $b_{max}$ distributions of the
clusters are inconsistent.

The difference between the cluster
orbits is also reflected in their birth properties,
in particular the joint probability distribution of
birth $v_b$ vs. birth $b$
(Figure \ref{fig:same_orbit}, right panel).
While the one-dimensional distributions
of these properties have significant overlap
(e.g., if one were to examine either birth $b$
or birth $v_b$ alone and marginalize over the other dimension),
they distinctly separate in 2-dimensional space.
For a given value of $b$, the distribution of
birth $v_b$ for the Arches has significantly higher
absolute values than the same distribution for the Quintuplet.

Thus, we conclude that the clusters
cannot share a common orbit.
The difference in their orbits
can be traced to the difference between
their present-day $b$ positions ($\Delta$b = 0.0789 $\pm$ 0.008$^{\circ}$),
which corresponds to a difference of $\sim$11 pc at the distance of the GC.
We will discuss the implications of this result
in the context of possible cluster
formation scenarios in $\mathsection$\ref{sec:formation}.

\begin{figure}
\includegraphics[scale=0.3]{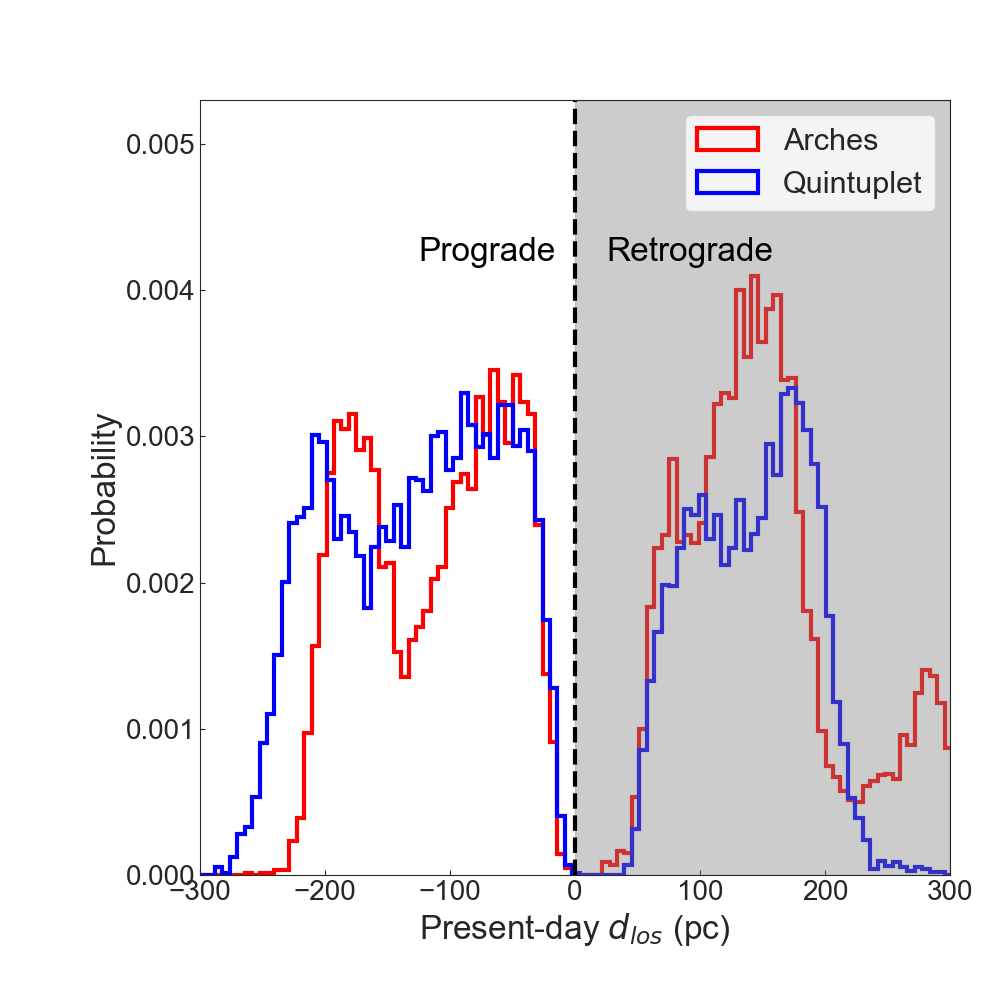}
\caption{The probability distributions for the present-day $d_{los}$ of the Arches (red) and Quintuplet (blue) clusters
calculated from the orbit model posteriors. Although $d_{los}$ is not constrained by Eqns. \ref{eq:likelihood} and \ref{eq:dlos},
these distributions are not uniform because the other kinematic parameters of the clusters (three-dimensional velocity,
two-dimensional sky position) are constrained by observations. In addition, the boundaries on the model priors
(set by the assumption that the clusters formed within the CMZ, see $\mathsection$\ref{sec:orb_model})
disfavor certain values of $d_{los}$.}
\label{fig:dlos_prob}
\end{figure}

\begin{figure*}
\begin{center}
\includegraphics[scale=0.3]{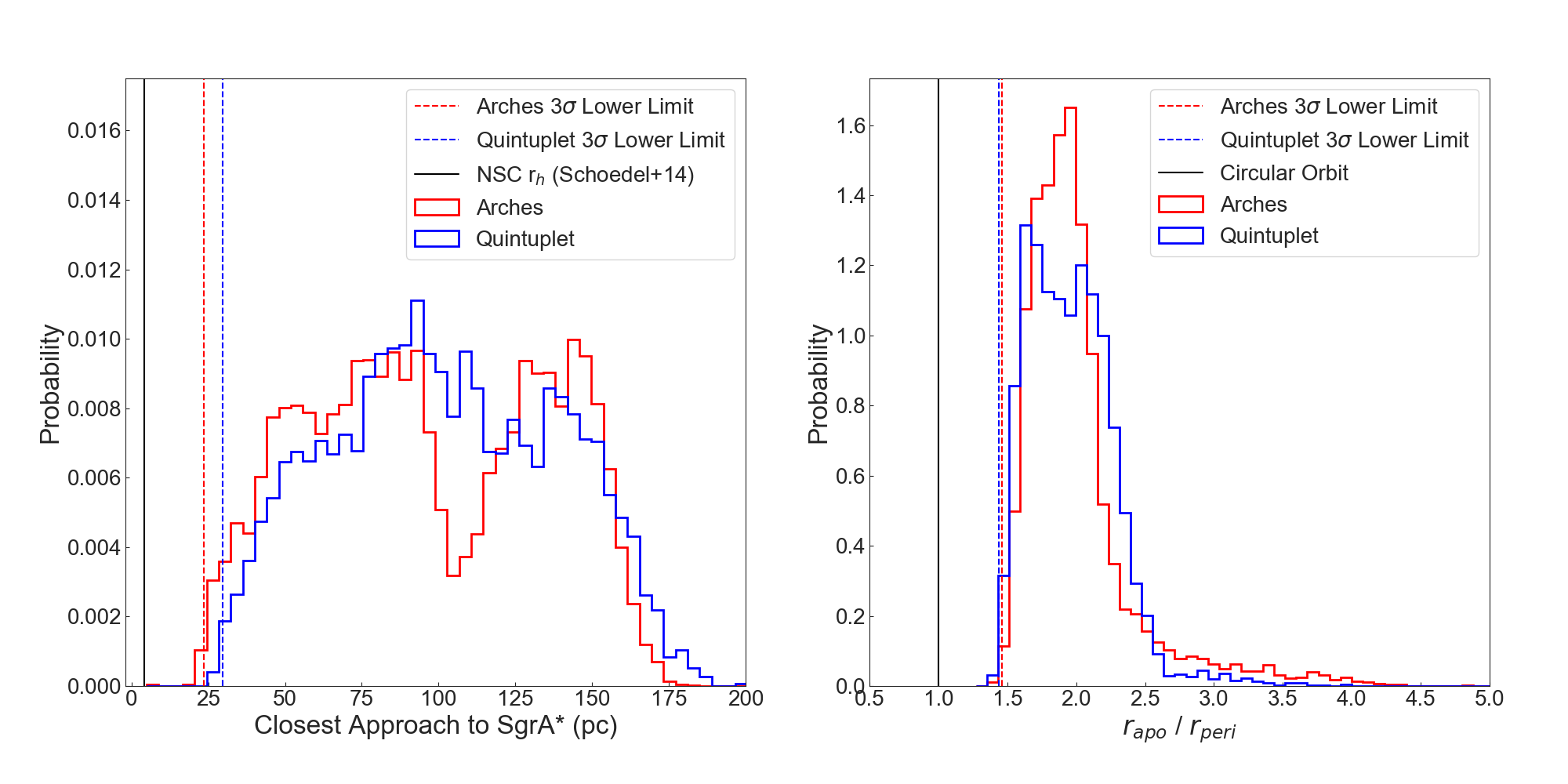}
\caption{\emph{Left:} The probability distributions for the closest approach to SgrA* for the Arches (red histogram) and Quintuplet (blue histogram) orbits,
compared to the half-light radius of the NSC \citep[black line, 4.2 pc;][]{Schodel:2014zl}. We obtain 3$\sigma$ limits of 24.7 pc for the
closest approach of the Arches (red dotted line) and 29.8 pc for the closest approach of the Quintuplet (blue dotted line).
\emph{Right:} The probability distributions for $r_{apo}$ / $r_{peri}$, which gives a measure of the orbit eccentricity.
Neither cluster is consistent with a circular orbit ($r_{apo}$ / $r_{peri}$ = 1.0, black line); the 3$\sigma$ lower limit is
$\sim$1.4 for both clusters (blue and red dotted lines). The 50th percentile value for $r_{apo}$ / $r_{peri}$ $\sim$ 1.9 for both clusters, which
is approximately equal to an eccentricity of $\sim$0.31. Note that only the prograde orbit modes are considered in these plots.}
\label{fig:orb_closest}
\end{center}
\end{figure*}

\begin{figure*}
\begin{center}
\includegraphics[scale=0.3]{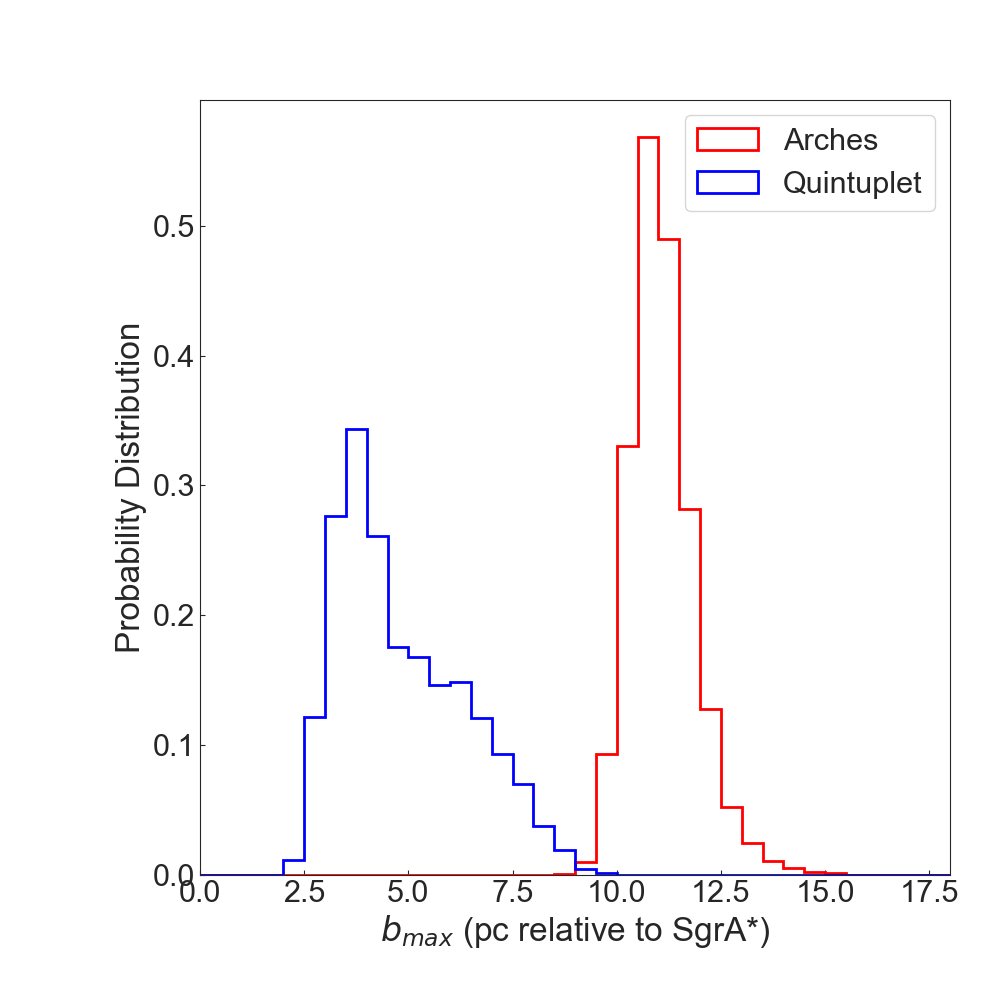}
\includegraphics[scale=0.3]{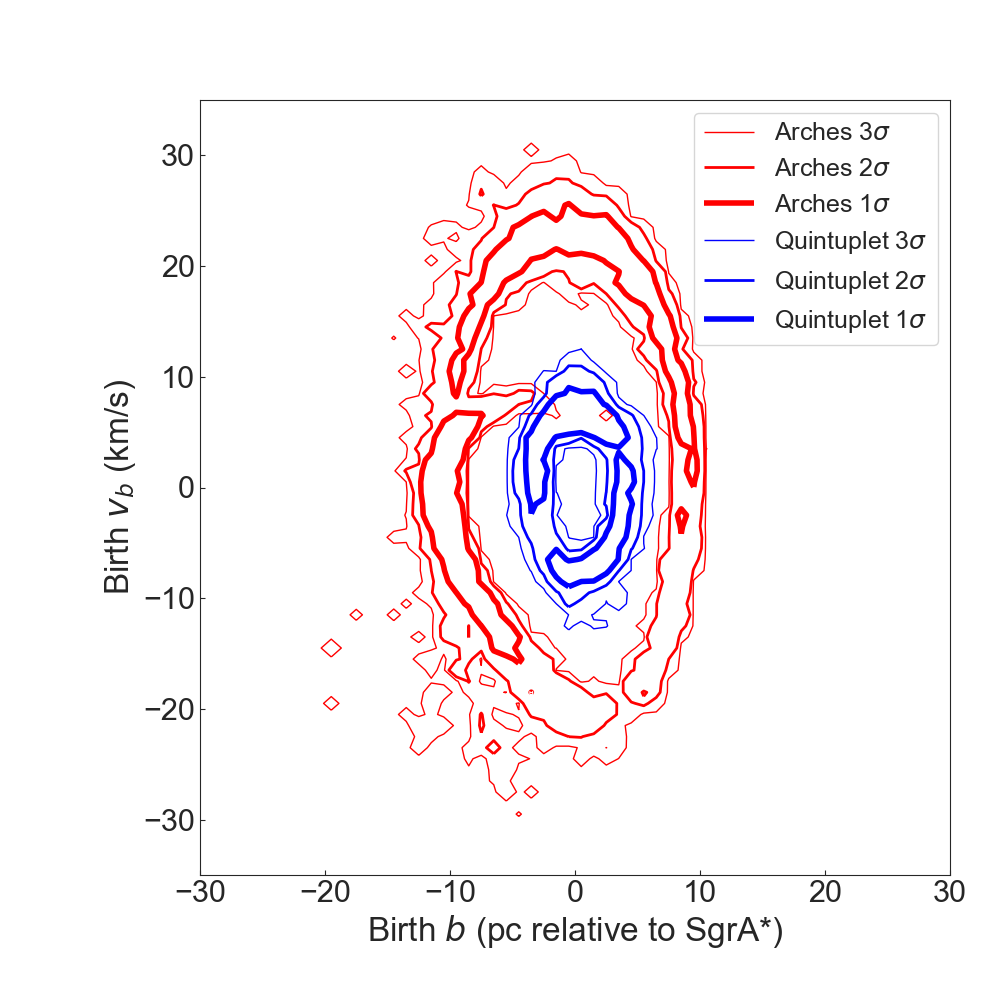}
\caption{A comparison of the orbital properties of the Arches (red) and Quintuplet (blue)
clusters that are associated with the vertical
oscillations of their orbits in the Galactic Plane.
\emph{Left:} The probability distributions for the maximum vertical
deviation of the orbits from the Galactic Plane ($b_{max}$).
The distributions are discrepant by $\sim$3.9$\sigma$
and are thus inconsistent.
\emph{Right:} The joint probability distributions of the birth $v_b$ vs. birth $b$
for the clusters. For a given $b$, the Arches formed
with a significantly larger birth $v_b$ (in terms of absolute value) than the Quintuplet.
Note that only the prograde orbit modes are considered in these plots.
}
\label{fig:same_orbit}
\end{center}
\end{figure*}

\begin{deluxetable*}{c c c c c r r r r r r r}
\tablecaption{Summary of Cluster Orbit Solution Modes}
\label{tab:orb_results}
\tablehead{
&  &  &  &  & \multicolumn{7}{c}{\underline{Maximum a Posteriori Parameters}} \\
\colhead{Cluster} & \colhead{Mode} & \colhead{Direction} & \colhead{log-evidence} & \colhead{Prob} &
\colhead{$x_b$} & \colhead{$y_b$} & \colhead{$z_b$} & \colhead{$vx_b$} & \colhead{$vy_b$} & \colhead{$vz_b$} & $t_{clust}$ \\
&  &  &  &  & \colhead{pc} & \colhead{pc} & \colhead{pc} & \colhead{km s$^{-1}$} & \colhead{km s$^{-1}$} & \colhead{km s$^{-1}$} & \colhead{log(years)}
}
\startdata
Arches & 1 & Prograde & -64.10 & 0.50 & -20.3 & 117.4 & -0.3 & -106.8 & -46.7 & 23.7 & 6.47  \\
Arches & 2 & Retrograde & -64.10   & 0.50 & -4.1 & 53.2 & 4.6 & 249.3 & -15.3 & 26.2 & 6.49  \\
& & & & &  &  &  &  &  & \\
Quintuplet & 1 & Prograde & -63.72 & 0.61 & 68.0 & 39.3 & -0.4 & -41.2 & 181.8 & 9.2 & 6.59 \\
Quintuplet & 2 & Retrograde & -64.15  & 0.39 & -128.1 & -2.3 & -1.3 & -82.2 & 50.8 & 4.1 & 6.60  \\
\enddata
\tablecomments{Description of Columns: \emph{Mode}: number of mode, \emph{Direction}: the direction of the orbits relative to the general gas flow observed in the CMZ, \emph{log-evidence}: ln($P(\vec{\mathbf{d}})$), as defined in Equation \ref{eq:bayes}, \emph{Prob}: the probability of the solution mode,
$x_b$, $y_b$, $z_b$, $vx_b$, $vy_b$, $vz_b$, $t_{clust}$: parameters of MAP orbit within the mode, in the same coordinate system as Table \ref{tab:orb_priors}. }
\end{deluxetable*}

\section{Discussion}
\label{sec:discussion}

In this section we compare our measurements of the absolute proper motion
of the Arches and Quintuplet clusters to past work ($\mathsection$\ref{sec:lit_comp})
and place our constraints on the cluster orbits
in the context of proposed formation scenarios ($\mathsection$\ref{sec:formation}).
We also discuss the whether the clusters are likely to inspiral
into the Nuclear Star Cluster within their lifetimes ($\mathsection$\ref{sec:future})
and explore the impact that our assumptions about the GC gravitational
potential have on the orbit results ($\mathsection$\ref{sec:other_gpot}).

\subsection{Arches and Quintuplet Absolute Motion: Comparison to Literature}
\label{sec:lit_comp}

\begin{figure*}
\includegraphics[scale=0.35]{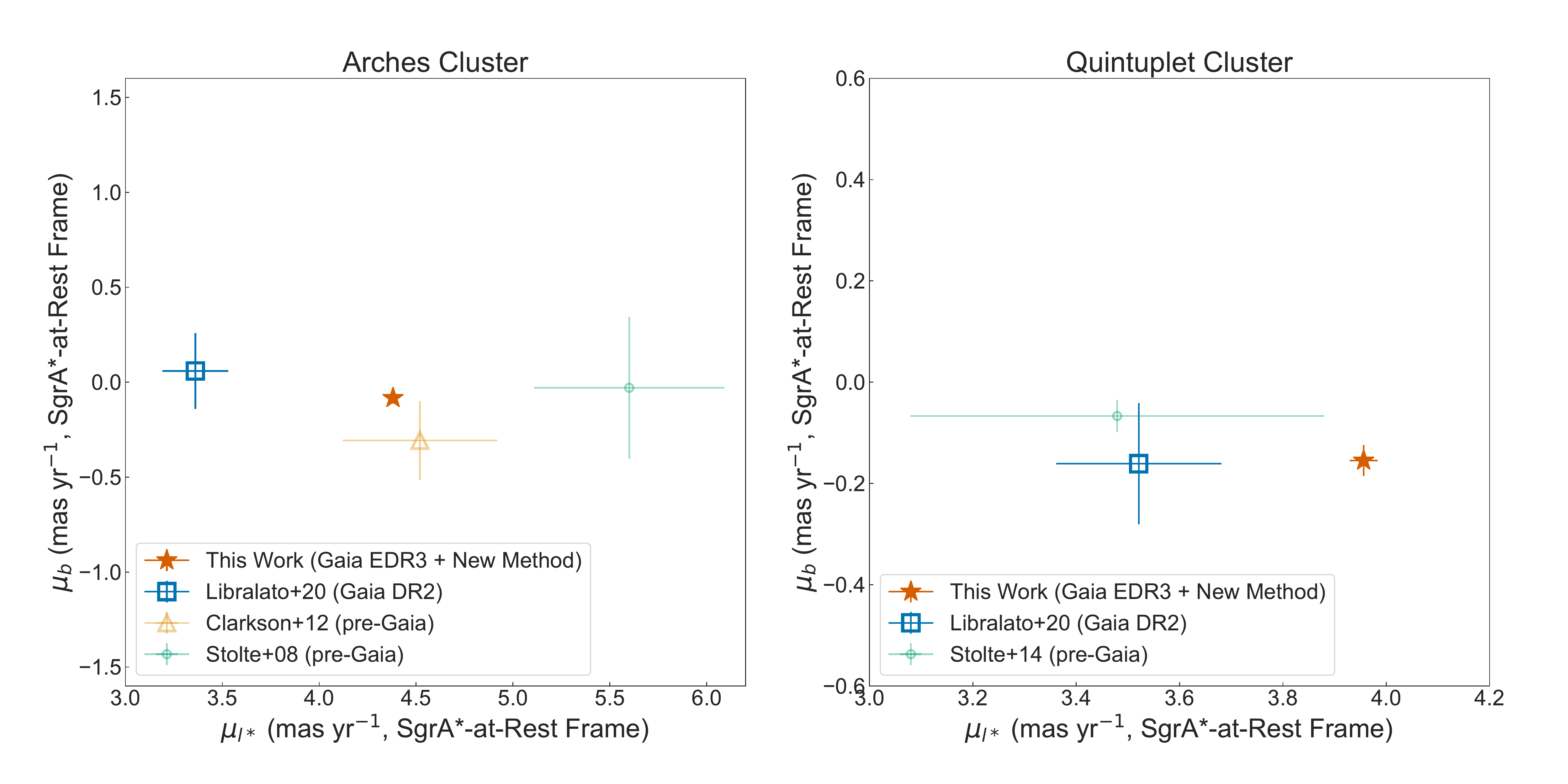}
\caption{Proper motion measurements of the Arches (left) and Quintuplet (right) clusters in Galactic coordinates in the SgrA*-at-Rest reference frame.
The measurements from this work (red filled stars) and \citet[][open blue squares]{Libralato:2020ev} are derived using \emph{Gaia}.
Pre-\emph{Gaia} measurements by \citet{Stolte:2008qy}, \citet{Clarkson:2012ty}, and \citet{Stolte:2014qf} (open circles and triangles) are derived from
the relative proper motion between the clusters and field stars. We make the most precise measurement of the absolute proper motions of the clusters to date.
}
\label{fig:pm_comp}
\end{figure*}

We present the most precise measurements of the absolute proper motions
of the Arches and Quintuplet clusters made to date.
A comparison between our measurements and those in the literature
is shown in Figure \ref{fig:pm_comp}.
Note that we convert
all proper motion measurements into Galactic coordinates ($\mu_{l^*}$, $\mu_b$)
relative to SgrA* (e.g., the SgrA*-at-Rest reference frame)
for this comparison.

\citet{Stolte:2008qy}, \citet{Clarkson:2012ty}, and \citet{Stolte:2014qf}
measure the \emph{relative} proper motion of the clusters with respect to
the field star population, and assume that
this is equivalent to the motion of the clusters
in the SgrA*-at-Rest reference frame.
To convert to Galactic coordinates, we rotate their measurements
and corresponding uncertainties
to the position angle of the Galactic Plane (31.40$^\circ$).
Although our measurements are somewhat different than
these past studies (for example, our value for $\mu_{l*}$
is generally smaller for the Arches and generally larger for the Quintuplet),
they are consistent to within 3$\sigma$ of the combined uncertainty.
However, our measurements are significantly more precise, with $\gtrsim$10x
smaller uncertainties then these past measurements.
In addition, it is unclear whether the
assumption that the relative
proper motion of the clusters is equivalent to
their motion in the SgrA*-at-Rest reference frame is valid.
This requires that the average field star motion
is at rest relative to SgrA*; in other words,
that the observations
are deep enough such that the streaming motion
of the field stars in front of SgrA* cancels
out with the streaming motion of the field
stars behind SgrA*.
Further, the field star population has been shown to exhibit
multiple kinematic substructures,
making it challenging to interpret
the average field star proper motion \citep{Hosek:2015cs, 2019ApJ...877...37R}.

Similar to this work, \citet{Libralato:2020ev} measure the absolute proper motion
of the clusters in the \emph{Gaia} reference frame.
In the same manner as our measurements,
we subtract the absolute motion of SgrA* \citep{Reid:2020jo}
to get the cluster motions in the SgrA*-at-Rest
reference frame.
For the Arches, our value for $\mu_{l*}$
is significantly larger
(4.38 $\pm$ 0.026 mas yr$^{-1}$ vs.
3.36 $\pm$ 0.17 mas yr$^{-1}$),
which represents a 6$\sigma$ difference in terms of the
combined uncertainty.
All other measurements agree within 3$\sigma$.

Our proper motion uncertainties are
$\sim$5x smaller than those of \citet{Libralato:2020ev},
which can be explained by differences in our data and
methodology.
Briefly, \citet{Libralato:2020ev} identify a set of stars in the \emph{Gaia} DR2
catalog that are also found in the published proper motion catalogs of the
clusters from \citet{Stolte:2015rr}\footnote{\citet{Stolte:2015rr} derive proper motions
using ground-based adaptive optics observations of the clusters using the
NAOS-CONICA system on the VLT \citep[][]{Rousset:2003tz, Lenzen:2003fj}
made over a 3 -- 5 year baseline.}.
They find 4 suitable matches in the Arches field and 12 suitable matches
in the Quintuplet field.
They then calculate the error-weighted difference between the \emph{Gaia}
proper motions and the \citet{Stolte:2015rr} proper motions for these stars.
Since the \citet{Stolte:2015rr} catalog is constructed in a reference
frame where the cluster is at rest, the average difference between the
\emph{Gaia} and \citet{Stolte:2015rr} proper motions is
interpreted to be the difference between the cluster-at-rest
and \emph{Gaia} reference frames, and thus
represents the absolute proper motions of the clusters.

Our data and methodology offer several improvements to this approach.
First, the \emph{HST} WFC3-IR data provides a larger field-of-view
of the clusters, and thus contains significantly more
\emph{Gaia} stars (26 reference stars
for the Arches field and 28
reference stars for the Quintuplet field; $\mathsection$\ref{sec:gaia_ref}).
As a result, we transform the \emph{HST} astrometry into
the \emph{Gaia} reference frame with higher precision and accuracy.
The fact that \citet{Libralato:2020ev} was limited to only 4 \emph{Gaia}
reference stars for the Arches cluster likely explains the
sizable discrepancy with our proper motion measurements.
Second, we convert the \emph{HST}
astrometry into the \emph{Gaia}
reference frame on an epoch-by-epoch basis,
using spatially-dependent transformations to
correct for optical distortions and other biases.
This allows us to measure proper motions
in the \emph{Gaia} frame directly,
explore the quality of the proper motion fits
and corresponding uncertainties,
and search for remaining distortions
or systematics in the astrometry ($\mathsection$\ref{sec:PMcat} and Appendix \ref{app:pm_cat}).
Third, the \emph{Gaia} EDR3 catalog
offers improved position and proper motion
measurements compared to DR2.
These improvements include
the use of more data (34 months vs. 22 months of observations) and
the introduction of color-dependent correction terms to
reduce astrometric biases.
As a result, the position and proper motion uncertainties in
the EDR3 catalog are generally improved by
a factor of $\sim$0.8 and $\sim$0.5 compared to DR2, respectively \citep{Lindegren:2021ae}.

\subsection{Comparing Possible Formation Scenarios for the Clusters}
\label{sec:formation}

We evaluate our results in the context of two proposed formation scenarios
for the Arches and Quintuplet clusters:
the $x_1$ - $x_2$ collision scenario,
in which the clusters formed in collisions between
gas on $x_1$ and $x_2$ orbits,
and the open stream scenario,
where the clusters formed from the collapse of molecular
clouds on the proposed \citetalias{Kruijssen:2015fx} orbit.
Both scenarios would place the clusters on
prograde orbits, and so only the prograde
solution modes are considered in this analysis.
We also consider whether the clusters could have formed
within the dust lanes before the $x_1$ -- $x_2$ collision region,
another location where star formation might occur \citep[e.g.][]{Sormani:2020sa}.

\subsubsection{The $x_1$ - $x_2$ Collision Scenario}
\label{sec:x1x2_discussion}
The regions where gas from $x_1$ orbits
intersect with gas on $x_2$ orbits have
long been identified as locations where
gas will shock and compress \citep[e.g.][]{Binney:1991cr, Morris:1996db}.
It is often suggested that
these collisions trigger local enhancements in star
formation activity, which has been invoked
to explain ring-like structures of star formation commonly
found in the nuclei of spiral galaxies \citep[e.g.][]{Boker:2008uq}.
Recent hydrodynamic simulations of the CMZ
predict that a ring of dense gas forms along the $x_2$ orbits with
a radius of $\sim$100 -- 200 pc, and that collisions
with infalling $x_1$ gas occur near the apoapse of this ring \citep{Tress:2020zj}.
These collisions produce dense molecular clouds that
grow progressively more massive and compact due to self-gravity.
Star formation can occur throughout
the gas ring, but peaks in the area between the apoapse and
following periapse of the ring \citep{Sormani:2020sa}.

\citet{Stolte:2008qy, Stolte:2014qf} argue that this is a viable
formation scenario for the Arches and Quintuplet clusters because
(1) the birth locations they derive for the clusters are consistent
with the apoapse of the $x_2$ orbits where these collisions occur, and
(2) their measurements of the \emph{present-day} three-dimensional velocities of the clusters
(172 $\pm$ 15 km s$^{-1}$ and 167 $\pm$ 15 km s$^{-1}$ relative to SgrA* for the Arches and Quintuplet, respectively)
are significantly higher
than the maximum expected velocity of gas on $x_2$ orbits \citep[120 km s$^{-1}$;][]{Englmaier:1999wd}.
They suggest that this could be the result of a
gas collision where the clusters gained momentum
from the higher-velocity infalling $x_1$ cloud.
We re-examine these arguments in light of our updated
constraints on the birth locations and velocities of
the clusters.

\begin{figure*}
\begin{center}
\includegraphics[scale=0.3]{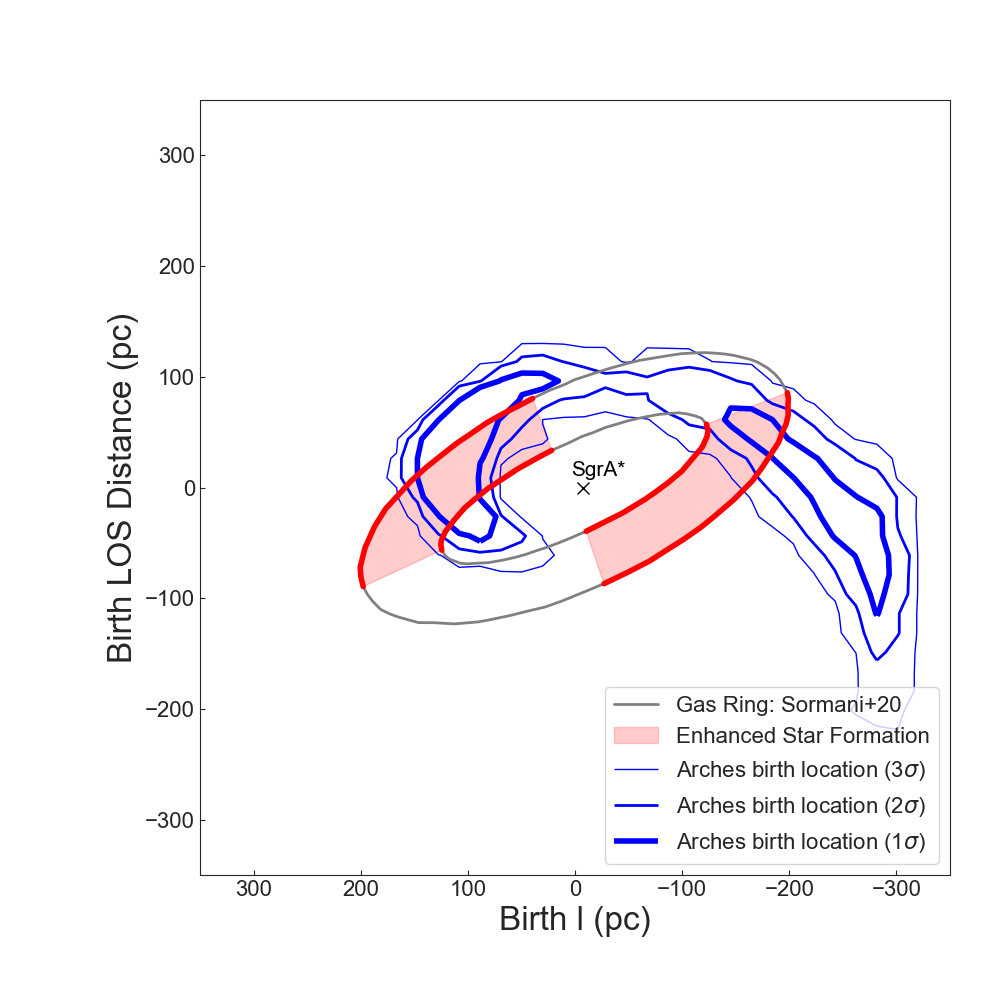}
\includegraphics[scale=0.3]{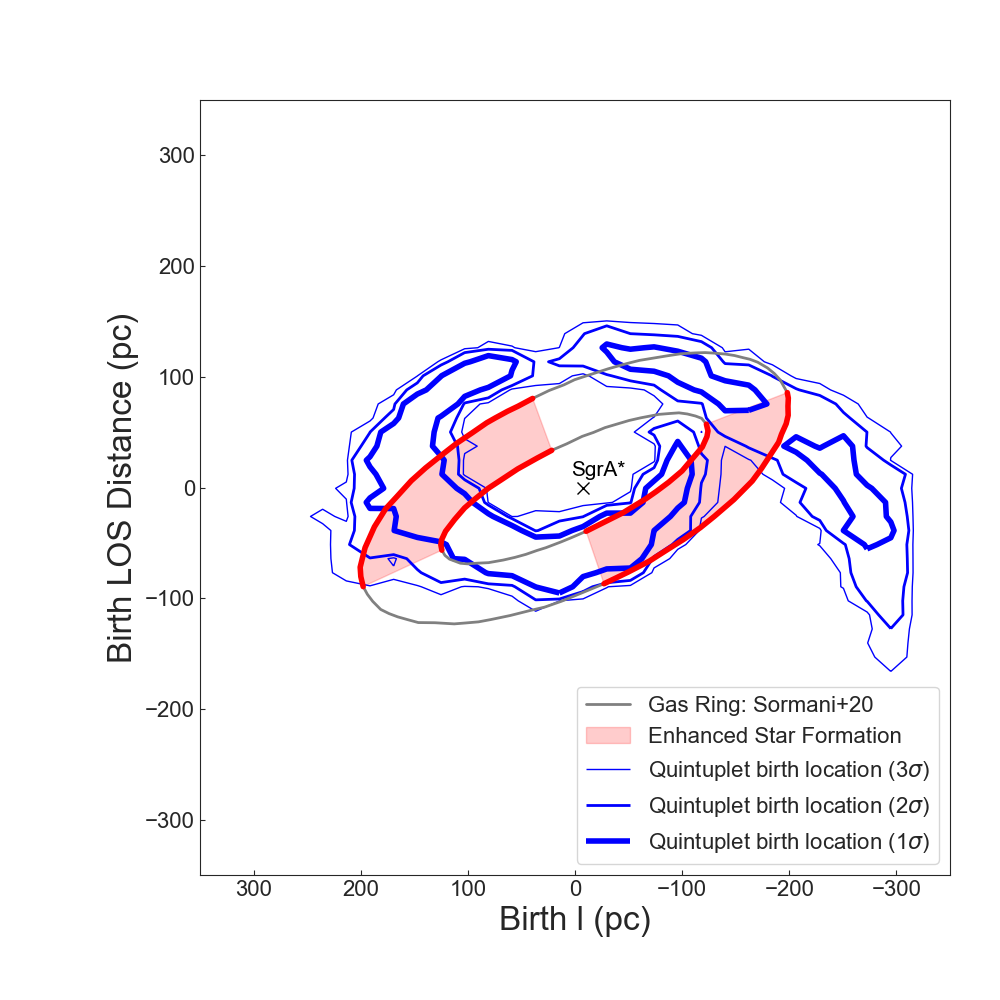}
\caption{Constraints on the cluster birth location (blue contours, representing 1$\sigma$, 2$\sigma$, and 3$\sigma$ contours with lines of decreasing thickness)
for the Arches (left) and Quintuplet (right) clusters, compared to the dense CMZ gas ring predicted by simulations \citep[grey ring;][]{Sormani:2020sa}.
In the $x_1$ -- $x_2$ gas collision scenario, enhanced star formation activity is expected between the apoapse and following periapse of the
gas ring (red shaded regions).
We find that both clusters are consistent with forming in these areas of enhanced formation,
although the uncertainties are large.
The cluster constraints are for the prograde solution mode only.
}
\label{fig:x2_pos}
\end{center}
\end{figure*}

In Figure \ref{fig:x2_pos}, we compare
our constraints on
the birth location of the Arches and
Quintuplet clusters to the CMZ gas ring
predicted by \citet{Sormani:2020sa}\footnote{The shape of the gas ring is taken from Figure 12 of \citet{Sormani:2020sa}.
Note that in their figure, the x-axis is oriented along the major axis of the bar such that the sun
is located in the positive x direction with an Earth-GC-bar angle of 20$^{\circ}$ \citep{Sormani:2018pk}.
In our Figure \ref{fig:x2_pos}, the axes are oriented such
that the Earth is located at (0, 8.09) kpc. Thus, we rotate the \citet{Sormani:2020sa} gas ring clockwise by 90 + 20 = 110$^{\circ}$.
While this structure will gradually rotate with time due to the pattern speed of the
bar, we note that
it would have only rotated by $\sim$12$^{\circ}$ -- 18$^{\circ}$ at most since the time the cluster formed,
assuming a cluster age of $\sim$5 Myr and a bar pattern speed of 40 -- 63 km s$^{-1}$ kpc$^{-1}$ \citep[][]{Bland-Hawthorn:2016cu}.
This additional rotation is insignificant compared to the uncertainties in the cluster birth location constraints and thus is ignored.
}.
We find that the highest probability contours
for the birth location of the Arches
generally fall between apoapse
and the following periapse of the ring,
consistent with the expected region of enhanced star formation
in the $x_1$ - $x_2$ collision scenario.
The birth location of the Quintuplet
is also generally consistent with the $x_1$ - $x_2$ collision
scenario, although the constraints are weaker.
Unlike the Arches, there is significant probability
that the Quintuplet formed at nearly any point throughout the gas ring.
This is likely because the Quintuplet
is older and has a larger uncertainty on its age (Table \ref{tab:orb_priors}),
and so its birth location is less constrained.

\begin{figure*}
\begin{center}
\includegraphics[scale=0.3]{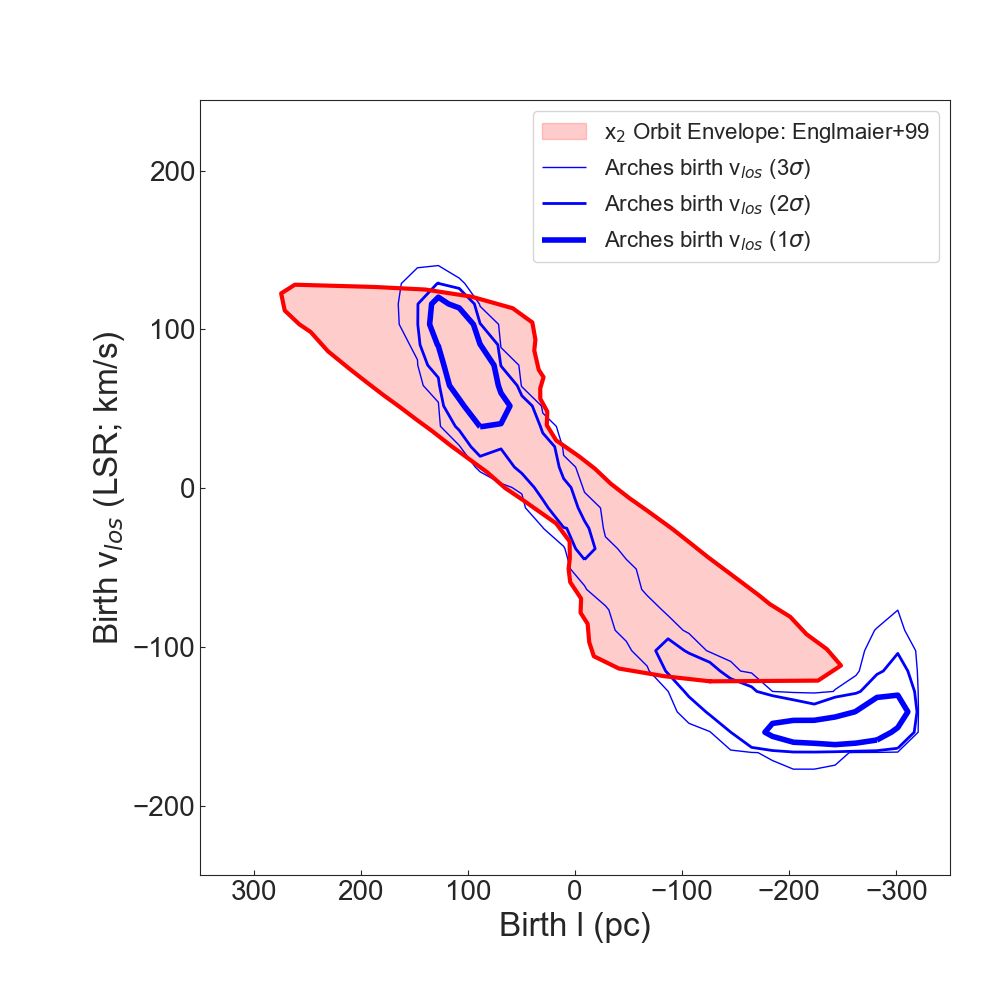}
\includegraphics[scale=0.3]{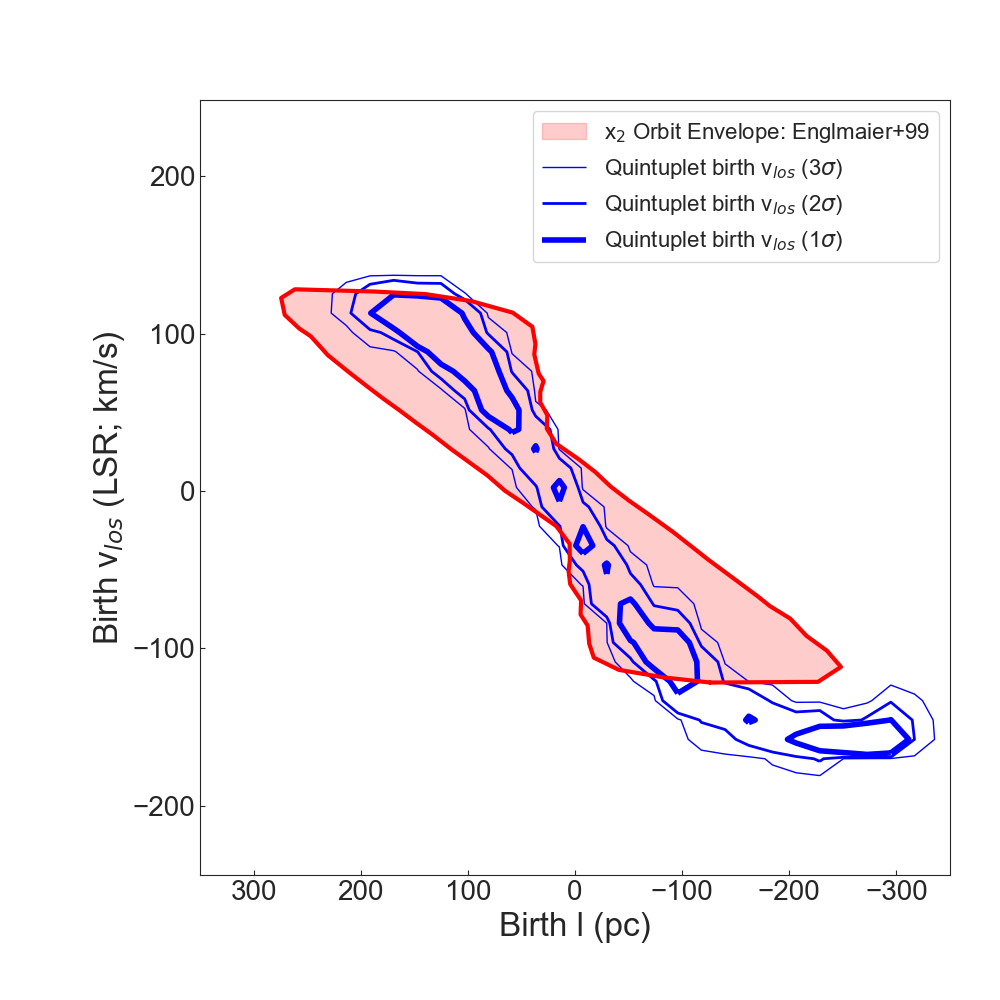}
\caption{Constraints on the birth $v_{los}$ versus birth $l$ (blue contours, representing 1$\sigma$, 2$\sigma$, and 3$\sigma$ contours with lines of decreasing thickness)
for the Arches (left) and Quintuplet (right) clusters, compared to the envelope of expected $v_{los}$ values for gas on $x_2$ orbits \cite[red shaded region;][]{Englmaier:1999wd}.
For both clusters, the birth $v_{los}$ values are consistent with the $x_2$ envelope
where the birth locations of the clusters are consistent with forming in the dense gas ring in the first place ($|l|$ $\lesssim$ 200 pc).
The cluster constraints are for the prograde solution mode only.
}
\label{fig:x2_vlos}
\end{center}
\end{figure*}

In Figure \ref{fig:x2_vlos}, we compare our constraints on
the birth $v_{los}$ of the
clusters to the predicted envelope of
$v_{los}$ values for gas on
$x_2$ orbits from \citet{Englmaier:1999wd}.
We obtain a similar result for both clusters:
for birth locations that are consistent with
the gas ring ($|l|$ $\lesssim$ 200 pc),
the birth $v_{los}$ for the clusters
appear to be fully consistent with the
$x_2$ orbits, with no evidence of velocity enhancement.
Only when the cluster birth locations
are outside the gas ring ($l$ $\lesssim$ -200 pc)
is there evidence of an enhanced
birth $v_{los}$ at 3$\sigma$ significance.
However, the clusters would no longer be
consistent with the $x_1$ - $x_2$ collisions at these radii,
and so we do not interpret this
as evidence in favor of this scenario.

We conclude that our results offer mild support for the
$x_1$ -- $x_2$ collision scenario.
The birth locations of the clusters appear
consistent with the expected regions of enhanced star
formation,
although the uncertainty in the constraints are admittedly large.
We do not find evidence for an enhancement
in the birth $v_{los}$ of the clusters relative
to the $x_2$ orbits at birth locations
where the clusters are consistent with
forming within the dense gas ring.
However, it is not yet clear how large
of a velocity enhancement might be expected due
to a gas collision and whether
or not it could be observed.
\citet[][]{Stolte:2008qy} point out that
the degree to which the cluster's birth
velocity is enhanced would
depend on many factors including the geometry
of the collision and the relative
densities of the clouds involved;
additional theoretical work is needed to explore
these effects.
Finally, we do not consider the fact that the
Arches and Quintuplet clusters have different
orbits to be a challenge to the $x_1$ - $x_2$ collision scenario.
It would be no surprise that two clusters that formed via gas collisions
at different times in
the tumultuous CMZ would have sufficiently different
initial conditions in $b$ and $v_b$
to account for the difference in the vertical oscillations
seen in their orbits.

\subsubsection{The Open Stream Scenario}
\label{sec:kdl_discussion}

\begin{figure}
\includegraphics[scale=0.32]{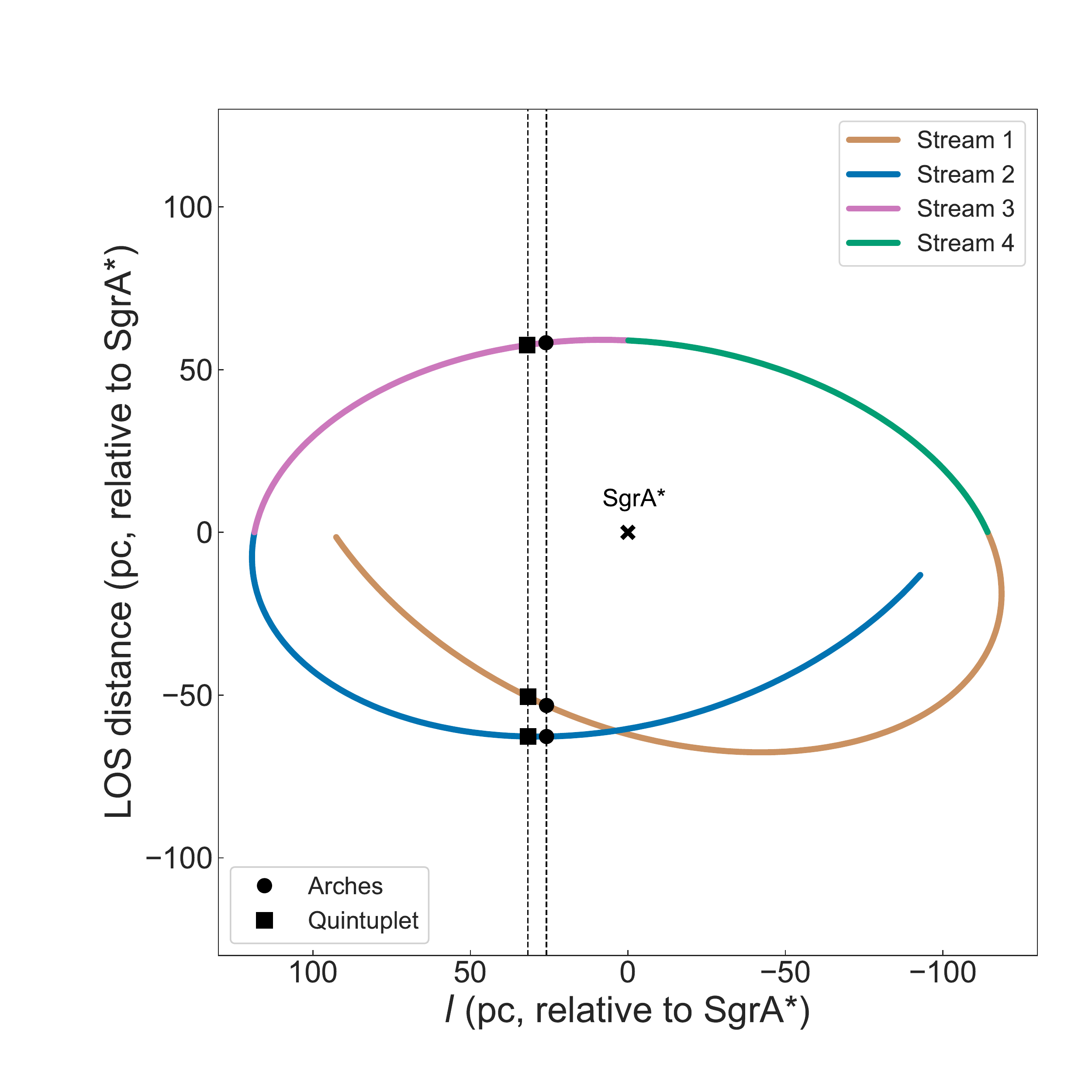}
\caption{The proposed \citetalias{Kruijssen:2015fx} orbit, plotted as the
line-of-sight distance from SgrA* vs. galactic longitude $l$. The points at which
the Arches and Quintuplet clusters intersect the orbit are marked by circles and squares,
respectively. Using the nomenclature from \citetalias{Kruijssen:2015fx}, these intersections
occur on streams 1, 2, and 3 (different color line segments).}
\label{fig:kdl_orb}
\end{figure}

If the clusters formed via the open stream scenario,
then we would expect their \emph{present-day} positions and
motions to be consistent with the \citetalias{Kruijssen:2015fx} orbit,
as it is unlikely that the clusters could have significantly
deviated from their natal orbit by their current age ($\lesssim$5 Myr).
Due to the uncertainty in $d_{los}$,
there are three possible locations for the Arches and Quintuplet clusters
on the \citetalias{Kruijssen:2015fx} orbit (Figure \ref{fig:kdl_orb}).
At each of these intersection points,
we calculate the probability of obtaining the observed properties
$\overrightarrow{\mathbf{x_{int}}}$ = \{b, $\mu_{l*}$, $\mu_{b}$, $v_{los}$\}
of the clusters under the assumption that they are on the proposed orbit.
Note that $\overrightarrow{\mathbf{x_{int}}}$ doesn't include the $l$
or $d_{los}$;
this is because the intersection points
are defined such that $l$ and $d_{los}$ match
the values on the orbit.
We define $\overrightarrow{\mathbf{x_{int}}}$ using
the values in Table \ref{tab:obs_vals}.

For each dimension of $\overrightarrow{\mathbf{x_{int}}}$, the probability of obtaining the
observed value ($x_{int,i}$) given the predicted value on the \citetalias{Kruijssen:2015fx} orbit
($x_{kdl,i}$) is:

\begin{equation}
P(x_{int, i}) = \frac{1}{\sigma_{tot}\sqrt{2\pi}} e^{-\frac{1}{2} \left( \frac{x_{int,i} - x_{kdl,i}}{\sigma_{tot}} \right)^2}
\end{equation}

where $\sigma_{tot}^2$ = $\sigma_{x_{int,i}}^2$ + $\sigma_{x_{kdl,i}}^2$, and
$\sigma_{x_{int,i}}$ and $\sigma_{x_{kdl,i}}$ are the uncertainties in the observed data
and \citetalias{Kruijssen:2015fx} orbit value \footnote{We calculate $\sigma_{x_{kdl,i}}$
via a Monte-Carlo simulation over the best-fit orbit
parameters from \citetalias{Kruijssen:2015fx}
as described in Appendix \ref{sec:methods_kdl15}.}.
The probability over all dimensions is then:

\begin{equation}
\label{eq:prob}
P(\overrightarrow{\mathbf{x_{int}}}) = \prod_i P(x_{int,i})
\end{equation}

\begin{figure*}
\begin{center}
\includegraphics[scale=0.32]{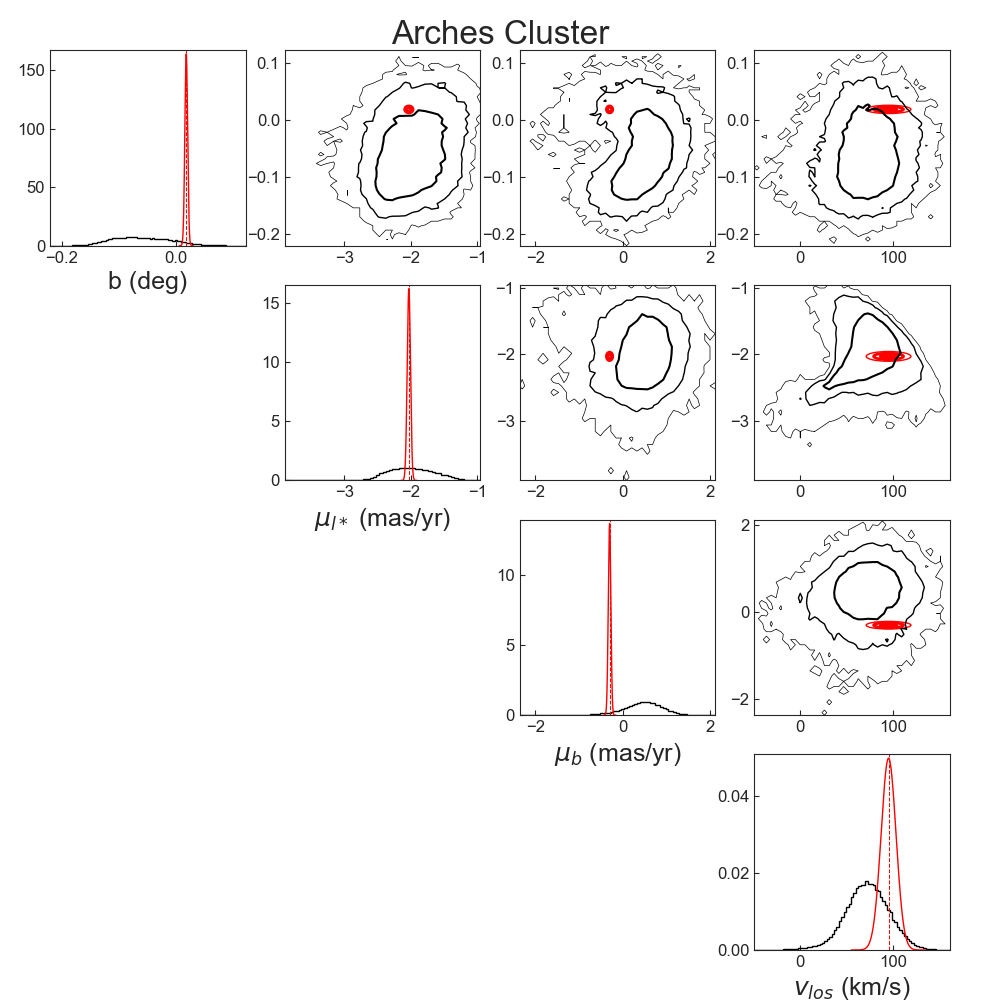}
\includegraphics[scale=0.32]{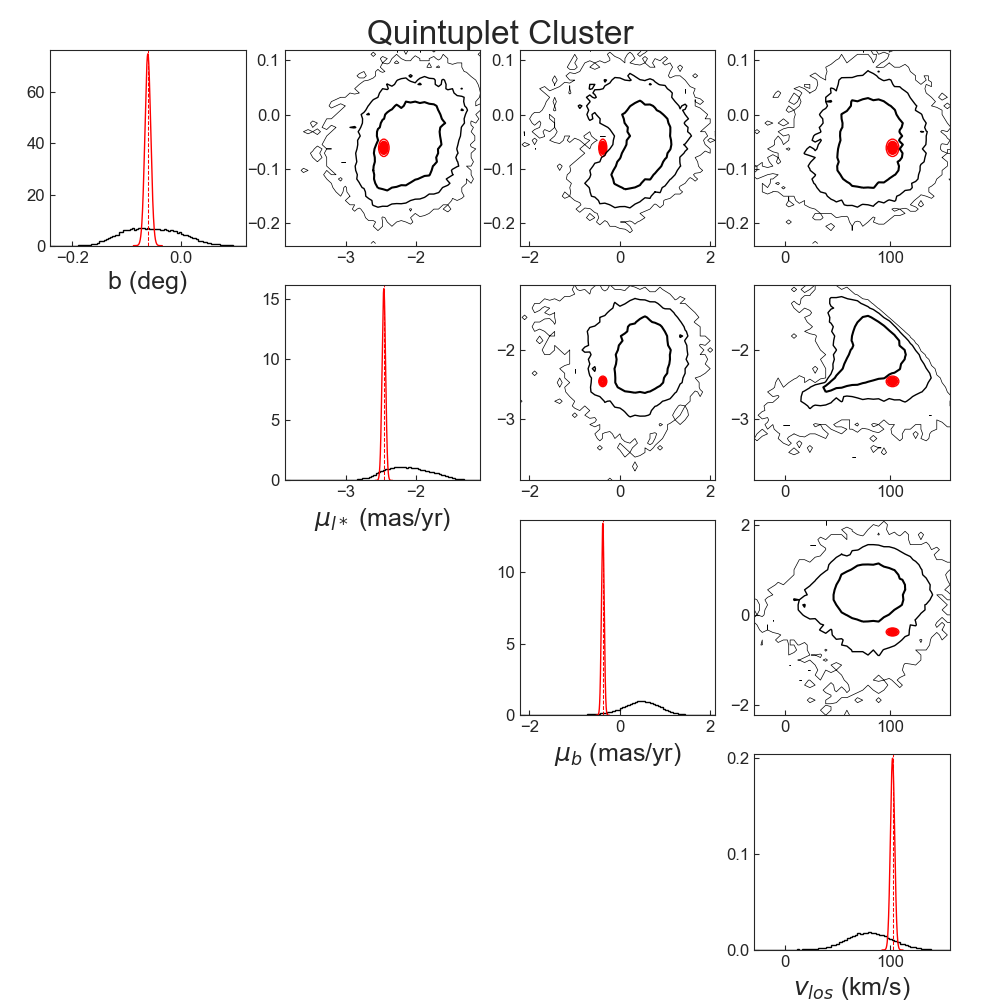}
\caption{
The comparison between the observed values of $\overrightarrow{\mathbf{x_{int}}}$ for the Arches (left) and Quintuplet (right)
clusters and their predicted values at the stream 1 intersection points on the \citetalias{Kruijssen:2015fx} orbit.
In all plots, the
cluster measurements are in red and the predicted orbit values are in black. In the 2D contour
plots, the thick, moderate, and thin lines correspond to the 1$\sigma$, 2$\sigma$, and 3$\sigma$ probability contours, respectively.
We find the clusters to be marginally consistent with stream 1, with the observations matching the orbit predictions
within 2.71$\sigma$ and 2.57$\sigma$ for the Arches and Quintuplet, respectively.}
\label{fig:kdl_comp}
\end{center}
\end{figure*}

Table \ref{tab:kdl_comp} shows the values of $P(\overrightarrow{\mathbf{x_{int}}})$
at the intersection points for both clusters.
Both clusters are inconsistent with the
stream 2 and stream 3 intersection points,
which are discrepant with observations by
4$\sigma$ -- 5$\sigma$ and $>$10$\sigma$, respectively (Appendix \ref{app:other_kdl15}).
However, we find that the clusters are marginally consistent with the stream 1 intersection point
at the 2.71$\sigma$ and 2.57$\sigma$ level
for the Arches and Quintuplet, respectively (Figure \ref{fig:kdl_comp}).
Note that the uncertainty in this comparison is
almost entirely due to the
\citetalias{Kruijssen:2015fx} orbit model
rather than the cluster measurements themselves.

Although the clusters are individually consistent with the
\citetalias{Kruijssen:2015fx} orbit within the considerable uncertainties,
a challenge for the open stream scenario is that the clusters
do not share a common orbit ($\mathsection$\ref{sec:same_orbit}).
Thus, the difference between the vertical oscillations
of the cluster orbits would
need to be explained by intrinsic latitudinal differences between
individual molecular clouds on the open stream.
However, it is unclear if this is possible
while maintaining the overall structure of a
coherent orbital stream.
Therefore, we conclude that
while it is possible that \emph{either} the Arches
or Quintuplet could have formed on the open stream proposed by
\citetalias{Kruijssen:2015fx},
it is unlikely that \emph{both} clusters
could have formed this way due to the
difference in their orbits.

\begin{deluxetable}{c c c c c}
\tablecaption{Probability of Obtaining the Observed Measurements at the KDL15 Orbit Intersection Points}
\label{tab:kdl_comp}
\tablehead{
& \multicolumn{2}{c}{\underline{Arches Cluster}} & \multicolumn{2}{c}{\underline{Quintuplet Cluster}} \\
\colhead{Stream} & \colhead{log(Prob)} & \colhead{Sigma} & \colhead{log(Prob)} & \colhead{Sigma}
}
\startdata
1 & -5.00 & 2.71 & -4.58 & 2.57 \\
2 & -10.74 & 4.25 & -14.07 & 4.94 \\
3 & -95.63 & $>$10 & -65.23 & $>$10 \\
\enddata
\tablecomments{Description of Columns: Stream = \citetalias{Kruijssen:2015fx} orbit stream, log(Prob) = Natural log of the probability described in Equation \ref{eq:prob}, Sigma = The probability from Equation \ref{eq:prob} converted into $\sigma$, assuming Gaussian statistics.}
\end{deluxetable}

\subsubsection{Star Formation Along the Dust Lanes}
\label{sec:dust_lanes}

\citet{Sormani:2020sa} find that star formation can also occur along the
dust lanes before the $x_1$ - $x_2$ collisions take place.
This has been invoked to explain the
formation of the Sgr E complex,
which comprises HII regions located at projected distances of $\sim$170 pc from the GC
that exhibit -220 km s$^{-1}$ $\lesssim$ $v_{los}$ $\lesssim$ -190 km s$^{-1}$ \citep{Anderson:2020mr}.
We do not find evidence that this formation mechanism could produce either the
Arches or Quintuplet cluster, as it would require
highly eccentric prograde orbits with larger values of $v_{los}$ at their current location than is
observed \citep[$|v_{los}|$ $\gtrsim$ 150 km s$^{-1}$, based on Figure 13 of][]{Anderson:2020mr}.
While highly eccentric orbits are possible for a small fraction of retrograde orbits
(see Figures \ref{fig:orbit_comp} and \ref{fig:dlos_prob}), it is unclear how a retrograde orbit could be produced by
this mechanism.

\subsection{The Clusters Are Unlikely to Merge With the NSC or YNC}
\label{sec:future}
Given the strong tidal field at the GC, the Arches and
Quintuplet clusters are not expected to have long
lifetimes before being tidally disrupted.
N-body simulations predict that massive clusters
with galactocentric radii between 30~--~150 pc
should tidally disrupt within $\sim$20 Myr \citep{Kim:1999fk, Kim:2000wd, Portegies-Zwart:2002hc}.
However, dynamical friction can cause clusters to inspiral
closer to SgrA*, potentially driving mergers
with the NSC \citep[half-light radius $\sim$ 4.2 pc;][]{Schodel:2014zl}
or even the YNC \citep[r $\lesssim$ 0.5 pc;][]{Stostad:2015ht}.
Indeed, it has been proposed that the inward migration of massive clusters
could contribute significant mass to the NSC \citep[e.g.][]{Antonini:2012vy, Arca-Sedda:2020fc}
or be the source of the young stars found
in the YNC  \citep[e.g.][]{Gerhard:2001lr}.
Here we examine whether the Arches or Quintuplet clusters
could undergo such mergers within their lifetimes.

Our results (which ignore dynamical friction and tidal disruption)
show that the Arches and Quintuplet
are allowed to be on orbits that bring
them as close as $\sim$25 -- 30 pc from SgrA* (Figure \ref{fig:orb_closest}).
\citet{Kim:2003bl} simulate massive clusters
at similar radii and evaluate
how close they migrate towards SgrA* before disrupting.
They find that only 10$^{6}$ M$_{\odot}$ clusters with
central densities of $\rho_c$ $\gtrsim$ 10$^{6}$ $M_{\odot}$ pc$^{-3}$,
or 10$^{5}$ M$_{\odot}$ clusters with initial radii of 10 pc and similar or
higher central densities,
can bring significant numbers of stars into the YNC or NSC before dissolving.
At $\sim$10$^4$ M$_{\odot}$, the Arches and Quintuplet
are 1 -- 2 orders of magnitude less massive than these
models, and have central densities at least an order
of magnitude lower as well \citep[Arches $\rho_c$ = 2.0 $\pm$ 0.4 x 10$^5$ M$_{\odot}$ pc$^{-3}$;][]{Espinoza:2009bs}.
Therefore, it appears unlikely that either the Arches or Quintuplet
will merge with either the NSC or YNC before
being tidally disrupted, even if they are at the innermost
orbits allowed by our analysis.

\subsection{Orbit Analysis Using Different Gravitational Potentials}
\label{sec:other_gpot}
A significant assumption in our analysis of the cluster
orbits is the gravitational potential of the GC.
As discussed in $\mathsection$\ref{sec:gc_gpot},
we adopt the potential from \citet{Kruijssen:2015fx}:
an axisymmetric potential with the enclosed mass
distribution from \citet{Launhardt:2002hl} that is flattened
in the $z$-direction.
We refer to this as the L02\_flat potential.
However, \citet[][hereafter S20]{Sormani:2020my} present 3 alternative potentials
for the inner $\sim$300 pc of the galaxy
based on axisymmetric Jeans modeling of the Nuclear Stellar Disk.
To examine the sensitivity of our results to the gravitational
potential, we repeat our analysis with these
alternative potentials in Appendix \ref{app:other_gpot}
and summarize the results here.

While the choice of gravitational potential does impact the orbital properties
of the clusters, the effect is relatively small.
For the Arches cluster, the S20 potentials
produce orbits that don't extend as close to SgrA*
and are slightly more eccentric.
The 3$\sigma$ lower limits on the closest approach distance to SgrA*
are between $\sim$25 -- 50 pc,
compared to 24.7 pc for the L02\_flat potential.
The ratio $r_{apo}$ / $r_{peri}$
has average (50\% percentile) values between $\sim$2.0 -- 2.5,
compared to 1.9 for the L02\_flat potential.
For the Quintuplet cluster, the effect of the S20 potentials
on the closest approach to SgrA* is opposite
compared to the Arches; the 3$\sigma$ lower limits
are closer ($\sim$13 -- 25 pc) than for the L02\_flat
potential (29.8 pc).
However, the effect on $r_{apo}$ / $r_{peri}$ is similar,
with the Quintuplet orbits also having average values
between $\sim$2.0 -- 2.5.
Overall, we find that the choice of gravitational
potential does not alter our results that
neither cluster will pass close enough to SgrA* to
merge into the NSC within their lifetimes,
and that the clusters are inconsistent with a circular orbit.

To determine the robustness of the result
that the clusters do not share a common orbit,
we repeat the calculation in Equation \ref{eq:zdist}
for the different gravitational potentials.
The Arches always exhibits larger vertical oscillations in the Galactic Plane,
and $P(b_{max, arch=quint})$ $\leq$ 0.2\% for all potentials.
Thus, the clusters do not share a common
orbit for all choices of gravitational potentials
examined here.

In addition, the choice of potential does not impact
our conclusions regarding the viability of the
cluster formation scenarios discussed in $\mathsection$\ref{sec:formation}.
The constraints on the cluster birth locations do not significantly
change across the different potentials,
and so there is a significant probability that the clusters formed
in the regions of enhanced star formation predicted by
the $x_1$ -- $x_2$ gas collision scenario
in all cases.\footnote{This statement assumes that the
locations of $x_1$ -- $x_2$ collision regions don't
significantly change for the different potentials examined here, as well.
Exploring the impact of the potential on the locations of $x_1$ -- $x_2$ collision
regions is beyond the scope of this paper.}
The choice of potential has a larger influence on the
birth $v_{los}$ of the clusters,
especially at negative galactic longitudes.
However, there is little evidence that the clusters exhibit
an enhanced birth $v_{los}$ compared to the $x_2$ gas orbits,
with the possible exception of the Quintuplet
cluster in the S20\_2 potential,
which may show a possible enhancement
near -200 pc $<$ $l$ $<$ -150 pc.
Thus, our conclusion that our results provide
mild support for the $x_1$ -- $x_2$ formation
scenario remains unchanged.

For the open stream formation scenario,
we note that the \citetalias{Kruijssen:2015fx}
orbit was derived assuming the \citet{Kruijssen:2015fx} potential.
Thus, changing the potential would naturally
change the orbit model.
However, recalculating the \citetalias{Kruijssen:2015fx} orbit
for different potentials is beyond the scope of
this paper.
That said, the Arches and Quintuplet
clusters do not share a common orbit regardless of the potential used,
and so this remains as a challenge for the open stream formation scenario.

\section{Conclusions}
\label{sec:conclusions}
We use multi-epoch \emph{HST} WFC3-IR observations
and the \emph{Gaia} EDR3 catalog to measure the absolute
proper motion of the Arches and Quintuplet clusters
and constrain their orbital properties.
Using 26 -- 28 \emph{Gaia} stars in each field,
we transform the \emph{HST} astrometry into the
\emph{Gaia} reference frame (which is tied to ICRF)
and calculate absolute proper motions for $\sim$35,000
stars in the Arches field and $\sim$40,000 stars in the Quintuplet
field, achieving a depth of F153M $\sim$ 23 mag in both.
Using these catalogs,
we measure bulk proper motions of
($\mu_{\alpha*}$, $\mu_{\delta}$)$_{ICRF}$ = (-0.80 $\pm$ 0.032, -1.89 $\pm$ 0.021) mas~yr$^{-1}$
for the Arches cluster and ($\mu_{\alpha*}$, $\mu_{\delta}$)$_{ICRF}$ = (-0.96 $\pm$ 0.032, -2.29 $\pm$ 0.023) mas~yr$^{-1}$
for the Quintuplet cluster.
The uncertainties on these measurements are $\gtrsim$5x smaller than
previous work and are dominated by the
construction of the reference frame.

For each cluster, we forward-model the set of orbits within the CMZ
that can replicate its observed position and motion while
taking into account the uncertainties in the cluster's present-day
line-of-sight distance ($d_{los}$), age, and
observational measurements.
The posterior probability distributions for the
the birth positions and birth velocities of the clusters
are highly degenerate,
primarily due to the uncertainty in $d_{los}$.
Two solution modes are found for each cluster,
one representing prograde orbits and the other
representing retrograde orbits
relative to the general gas flow in the CMZ.
The retrograde orbits are deemed unlikely
due to lack of evidence (in either observations or simulations)
for gas on retrograde orbits in the CMZ
that might be capable of forming a cluster
similar to the Arches or Quintuplet.
Therefore, we restrict our analysis to the prograde
solution mode only.

From the distribution of possible prograde orbits for the clusters,
we find that:

\begin{itemize}
\item The Arches and Quintuplet will not approach closer than
24.7 pc and 29.8 pc to SgrA*, respectively.
These values represent the 3$\sigma$ lower limits on the
distributions of closest approach distances.
While this calculation ignores dynamical friction,
previous simulations suggest that neither cluster will merge
with the NSC or YNC before they are tidally disrupted \citep{Kim:2003bl}.
\item Both clusters are inconsistent with a circular orbit,
with a 3$\sigma$ lower limit on $r_{apo}$ / $r_{peri}$
of $\sim$1.4.
The orbits have a typical (i.e., 50th percentile) value of $r_{apo}$ / $r_{peri}$ $\sim$1.9,
which is equivalent to an orbital eccentricity of $\sim$0.31.
\item The clusters do not share a common orbit,
as the Arches exhibits significantly larger vertical oscillations in the Galactic Plane
then the Quintuplet.
\end{itemize}

These results are not significantly
altered by the choice of gravitational potential.

The distribution of cluster orbits are examined
in the context of two proposed formation scenarios for the clusters:
the $x_1$ - $x_2$ collision scenario, in which the clusters
formed when infalling gas from $x_1$ orbits collided with gas
on $x_2$ orbits in the CMZ \citep[e.g.][]{Stolte:2008qy, Stolte:2014qf},
and the open stream scenario, where the clusters formed
from the collapse of molecular clouds along the proposed \citetalias{Kruijssen:2015fx}
orbit \citep[e.g.][]{Kruijssen:2015fx, Kruijssen:2019kx}.

We conclude that our constraints on the birth position
and location of the clusters are in mild agreement
with the $x_1$ -- $x_2$ scenario.
The birth locations of the clusters are consistent
with the expected region of enhanced star formation
due to gas collisions, although the uncertainties are large.
We do not find evidence that either cluster
formed with an enhanced line-of-sight
velocity ($v_{los}$) compared to
typical $x_2$ gas orbits, as might
occur due to momentum transfer
from the higher-velocity $x_1$ cloud during a collision.
However, this does not yet discount
the $x_1$ -- $x_2$ scenario, as
the significance of this effect
likely depends on several factors
(e.g., the collision geometry and relative densities
of the clouds).
Thus, additional work is needed to determine
if a $v_{los}$ enhancement would be observable.

On the other hand, our results present a challenge for the
open stream formation scenario.
While the present-day positions
and motions of the clusters are marginally
consistent with their predicted values on
``stream 1'' of the \citetalias{Kruijssen:2015fx} orbit individually,
it is unclear how \emph{both} clusters could have formed
this way given the difference between their orbits.
This would require the individual clouds on the
\citetalias{Kruijssen:2015fx} stream to span an
intrinsic range of orbits that encompasses
the difference in vertical oscillation between the clusters,
and yet still maintain the appearance of an
orbital stream.

Future progress on constraining the formation
mechanism(s) for the Arches and Quintuplet clusters
can be made by reducing the uncertainties
in several key areas.
For the clusters themselves,
improved constraints on $d_{los}$ would
reduce the range of allowed orbits
and thus provide more stringent constraints on
their orbital properties, birth positions, and birth velocities.
Further, the evaluation of
whether the present-day position and
motion of the clusters are consistent
with the \citetalias{Kruijssen:2015fx} orbit
is dominated by the uncertainty \citetalias{Kruijssen:2015fx}
orbit itself.
Reducing the uncertainty of the orbit model would
provide a stronger test of whether the individual clusters
are indeed consistent with forming via the open stream
scenario.
Finally, the GC gravitational potential itself remains
a source of uncertainty which impacts both the
orbits of the clusters as well as the predictions of
the different star formation scenarios.

\acknowledgements
The authors thank Jesus Salas, Mattia Sormani, and Perry Hatchfield for helpful discussions regarding the gravitational potential and possible star formation mechanisms at the Galactic center as well as the anonymous referee whose feedback improved this paper. M.W.H. is supported by the Brinson Prize Fellowship. This work was supported by \emph{HST} AR-16121 and is based on observations made with the NASA/ESA Hubble Space Telescope, obtained at the Space Telescope Science Institute, which is operated by the Association of Universities for Research in Astronomy, Inc., under NASA contract NAS 5-26555. The observations are associated with programs GO-11671, GO-12318, GO-12667, and GO-14613. This work used computational and storage services associated with the Hoffman2 Shared Cluster provided by UCLA Institute for Digital Research and Education's Research Technology Group, and made extensive use of the NASA Astrophysical Data System.

\facilities{HST (WFC3-IR), Gaia}

\software{AstroPy \citep{Astropy-Collaboration:2013kx, Astropy-Collaboration:2018ws}, numPy \citep{oliphant2006guide}, Matplotlib \citep{Hunter:2007}, SciPy \citep{ScipyGuide}, \texttt{KS2} \citep{Anderson:2008qy}, \texttt{galpy} \citep{Bovy:2015bd}, \texttt{Multinest} \citep{Feroz:2008yu, Feroz:2009lq}, \texttt{PyMultinest} \citep{Buchner:2014wa}}
\clearpage

\bibliography{ms.bib}

\clearpage

\appendix

\section{Testing The Proper Motion Catalogs}
\label{app:pm_cat}

To check the proper motions in the Arches and Quintuplet field catalogs,
we examine the distribution of $\chi^2$ values and astrometric
residuals for the proper motion fits.
The $\chi^2$ statistic is a measure of how large the proper motion fit residuals are relative
to the astrometric errors.
For each star, the $\chi^2$ statistic is calculated as:

\begin{equation}
\chi^2 = \sum_{i=0}^{i=N} \frac{(x_{obs_i} - x_{pred_i})^2}{\sigma_{ast_{x_i}}^2}
\end{equation}

where $x_{obs_i}$ and $x_{pred_i}$ are the observed and predicted position of the star in the $i$th epoch,
$\sigma_{ast_{x_i}}$ is the astrometric error in the $i$th epoch, and $N$ is the total number of epochs.

\begin{figure*}
\begin{center}
\includegraphics[scale=0.3]{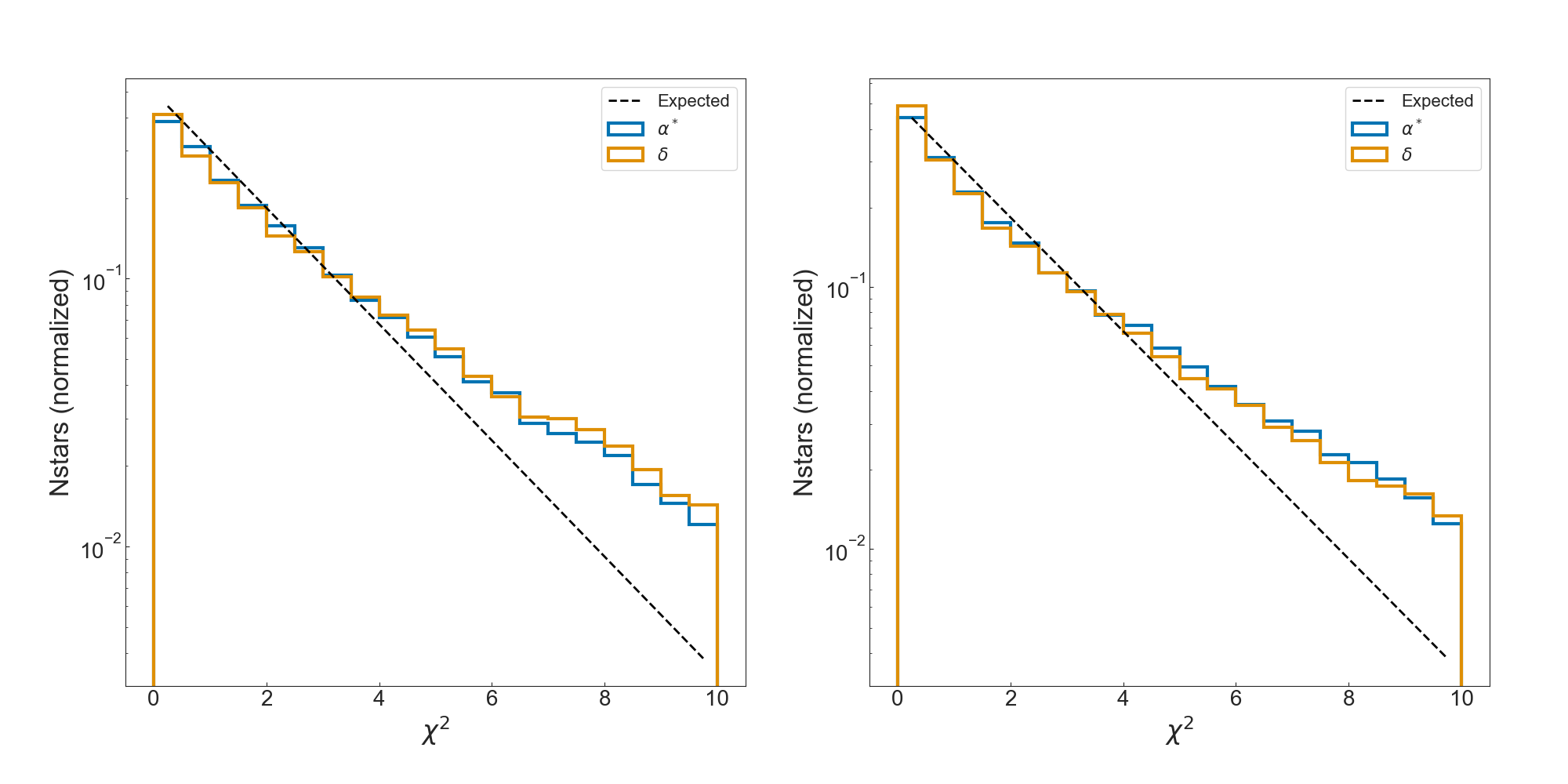}
\caption{$\chi^2$ distributions for the proper motion fits in the Arches (left) and Quintuplet (right) catalogs. The $\chi^2$ distribution
in the $\alpha^*$ and $\delta$ directions (blue and orange lines, respectively) are similar to the expected $\chi^2$ distribution with 2 degrees of
freedom (black dotted line).}
\label{fig:chi2}
\end{center}
\end{figure*}

The distribution of $\chi^2$ values for both cluster fields is shown in Figure \ref{fig:chi2}.
The distributions are found to be similar to the expected distribution in both the $\alpha^*$ and $\delta$
directions, indicating that the
astrometric errors are a good representation of the fit residuals.
There is a slight overabundance of stars with high $\chi^2$ values for both clusters,
which are dominated by fainter sources.
This is likely due to systematic errors in the astrometry caused by stellar crowding,
which impacts fainter sources more strongly than brighter sources.

\begin{figure*}
\begin{center}
\includegraphics[scale=0.3]{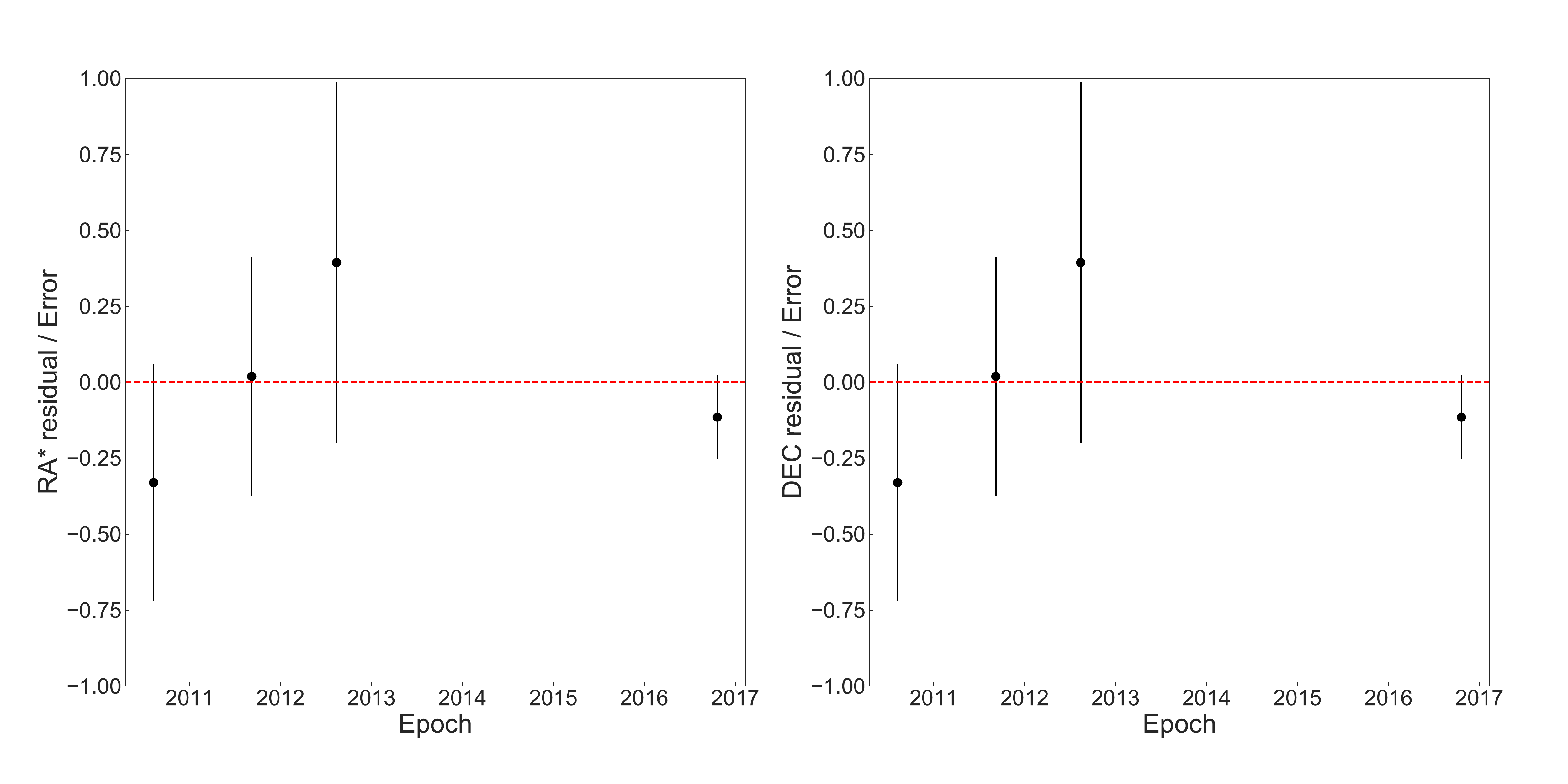}
\includegraphics[scale=0.3]{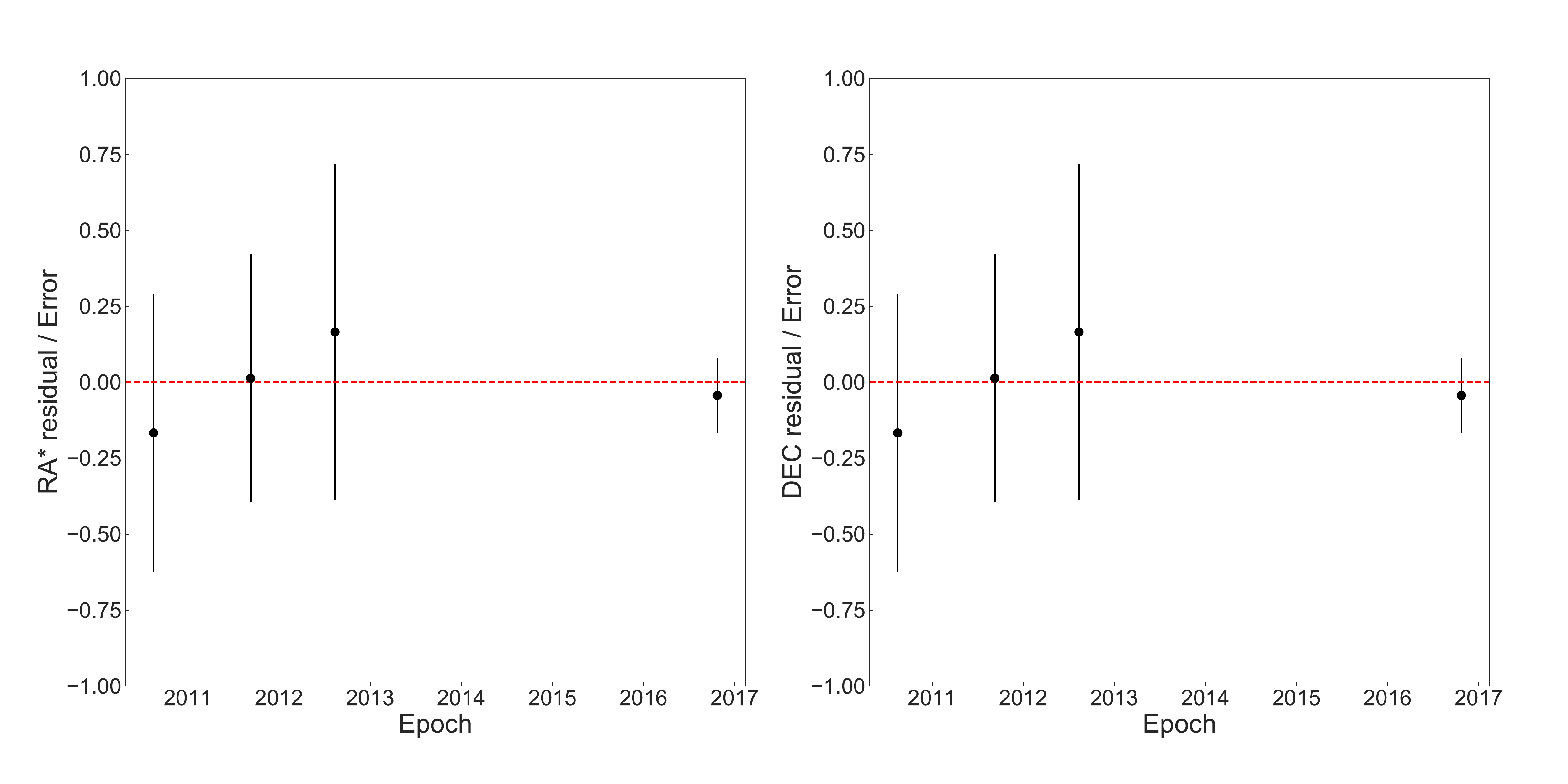}
\caption{Mean and standard deviation of the ratio of the proper motion fit residuals to the astrometric error for bright stars (F153M $<$ 17.5 mag) as a function
of epoch for the Arches (upper panels) and Quintuplet (lower panels) catalogs. The average residual is less than half of the astrometric error for all epochs, and so we conclude that residual distortion
does not significantly impact our measurements.}
\label{fig:distortion}
\end{center}
\end{figure*}

Within a given epoch, the distribution of proper motion fit residuals reveals whether there
are systematic offsets in the astrometry that are not captured by the transformations.
If significant, these offsets, which we refer to as residual distortion, could indicate that more complex
transformations are required to correct the distortions in the field.
Figure \ref{fig:distortion} shows the average ratio of proper motion fit residuals
to the astrometric error for bright stars (F153M $<$ 17.5 mag) in the Arches
and Quintuplet fields for each epoch.
Bright stars are used for this analysis because their astrometric errors are the
smallest, thus making them the most sensitive to residual distortion.
We find that the average residual distortion is less than half of the bright-star astrometric errors
in all epochs, and conclude that residual distortion does not significantly impact our
measurements.

\section{Modeling Cluster and Field Star Proper Motions With Gaussian Mixture Models}
\label{app:gmm}
Here we describe the methodology for
modeling the cluster and field star proper motions using
Gaussian Mixture Models, report the parameters of the best-fit
models for the Arches and Quintuplet clusters,
and discuss how cluster membership
probabilities are calculated for individual stars.

\subsection{Gaussian Mixture Models: Methodology}
We follow the methodology of \citetalias{2019ApJ...877...37R} to find the GMM
that best matches the data.
Briefly, we use multiple Gaussians to describe the field star populations and a single
Gaussian to describe the cluster members.
Each Gaussian used to describe the field population has 6 free parameters:
the fraction of the total sample described by that Gaussian ($\pi$),
the proper motion centroid of the Gaussian ($\mu_{\alpha^*}$, $\mu_{\delta}$
for the $\alpha^*$ and $\delta$ directions, respectively),
the standard deviation along the semi-major axis ($\sigma_{a}$),
the ratio between the semi-minor and semi-major axis ($f$),
and the angle between the semi-major axis and the $\alpha^*$ axis ($\theta$).
In contrast, the Gaussian used to described the star cluster has only 4 free parameters
($\pi$, $\mu_{\alpha^*}$, $\mu_{\delta}$, $\sigma_{a}$),
as we require it to be circular (e.g., $f$ = 1 and $\theta$ = 0 deg).
We adopt the same likelihood equation
as \citetalias{2019ApJ...877...37R}
and use \texttt{Multinest}/\texttt{PyMultinest}
to search the parameter space and calculate posterior
probability distributions for the free parameters in the GMM.

The priors used for the GMM parameters are given in Tables \ref{tab:gmm_arch}
and \ref{tab:gmm_quint}.
For most parameters, we adopt uniform priors with the
same minimum and maximum ranges as \citetalias{2019ApJ...877...37R}.
The exception to this is the proper motion centroid of the
cluster Gaussian, for which we adopt a Gaussian prior
with a mean equal to the median HST-\emph{Gaia} proper motion of high probability cluster members
established via previous analyses
(e.g., stars previously identified as having P$_{clust}$ $>$ 0.7 in \citetalias{Hosek:2019kk} and \citetalias{2019ApJ...877...37R}).
We adopt a conservative standard deviation of 2 mas for the Gaussian priors,
over 10x larger than the actual standard deviation of the proper motion values for these stars,
in order to minimize the influence of this prior on the final result.

The final choice to be made is the number of Gaussian components
to use in the GMM.
To do this, we repeat the analysis using 3, 4, and 5 total Gaussians,
and then use the Bayesian Information Criterion \citep[BIC;][]{Schwarz+78}
to identify which model represents the best fit of the data.
The BIC strongly prefers the GMM with
4 components for both clusters.

\subsection{Best-Fit GMM Model Parameters}
The best-fit parameters for the 4-component GMMs
for the proper motion distribution in the Arches and Quintuplet fields are
given in Tables \ref{tab:gmm_arch}
and \ref{tab:gmm_quint}, respectively.
Note that these GMMs are different than the
ones presented in \citetalias{Hosek:2019kk} and \citetalias{2019ApJ...877...37R}
because (1) the proper motions in this paper are in an absolute
reference frame (ICRF) rather than a relative one, and (2) the proper motion
uncertainties are larger as discussed in $\mathsection$\ref{sec:PMcat}.
However, the cluster membership probabilities
we calculate in Appendix \ref{app:membership} are generally consistent
with those in \citetalias{Hosek:2019kk} and \citetalias{2019ApJ...877...37R}
within 0.05.

\begin{deluxetable}{c |c c |c r | c r | c r }
\tabletypesize{\scriptsize}
\tablecaption{Arches Gaussian Mixture Model: Free Parameters, Priors, and Results}
\tablehead{
& \multicolumn{2}{c}{Cluster Gaussian} & \multicolumn{2}{c}{Field Gaussian 1} & \multicolumn{2}{c}{Field Gaussian 2} & \multicolumn{2}{c}{Field Gaussian 3}\\
\colhead{Parameter} & \colhead{Prior} & \colhead{Result} & \colhead{Prior} & \colhead{Result} & \colhead{Prior} & \colhead{Result} & \colhead{Prior} & \colhead{Result}
}
\startdata
$\pi_{k}$ & U(0, 1) & 0.039 $\pm$ 0.003 & U(0, 1) & 0.46 $\pm$ 0.02 & U(0, 1) & 0.34 $\pm$ 0.02 & U(0, 1) & 0.16 $\pm$ 0.01 \\
$\mu_{\alpha^*, k}$ (mas yr$^{-1}$) & G(-0.8, 2) & -0.80 $\pm$ 0.01 & U(-6, 6) & -2.43 $\pm$ 0.05 & U(-6, 6) & -2.62 $\pm$ 0.06 & U(-6, 6) & -1.41 $\pm$ 0.05 \\
$\mu_{\delta, k}$ (mas yr$^{-1}$) & G(-1.88, 2) & -1.89 $\pm$ 0.01 & U(-6, 6) & -4.49 $\pm$ 0.08 & U(-6, 6) & -4.84 $\pm$ 0.07 & U(-6, 6) & -2.83 $\pm$ 0.05 \\
$\sigma_{a, k}$ (mas yr$^{-1}$) & U(0, 3) & 0.08 $\pm$ 0.02 & U(0, 8) & 2.71 $\pm$ 0.05 & U(0, 8) & 3.26 $\pm$ 0.05 & U(0, 8) & 1.16 $\pm$ 0.06 \\
$f$ & --- & 1.0 & U(0, 1) & 0.48 $\pm$ 0.02 & U(0, 1)  & 0.93 $\pm$ 0.02 & U(0, 1) & 0.50 $\pm$ 0.04 \\
$\theta_{k}$ (rad) & --- & 0 & U(0, $\pi$) & 0.98 $\pm$ 0.02 & U(0, $\pi$) & 1.06 $\pm$ 0.13 & U(0, $\pi$) & 0.92 $\pm$ 0.04 \\
\enddata
\label{tab:gmm_arch}
\tablecomments{Description of parameters: $\pi_k$ = fraction of stars in Gaussian; $\mu_{\alpha^*, k}$ = $\alpha^*$ velocity centroid of Gaussian; $\mu_{\delta,k}$ = $\delta$ velocity centroid of Gaussian; $\sigma_{a,k}$ = semi-major axis of Gaussian; $f$ = ratio of semi-minor to semi-major axis; $\theta_{k}$ = angle between $\sigma_{a,k}$ and the $\alpha^*$ axis}
\tablecomments{Description of priors: Uniform distributions: U(min, max), where min and max are bounds of the distribution; Gaussian distributions: G($\mu$, $\sigma$), where $\mu$ is the mean and $\sigma$ is the standard deviation}
\end{deluxetable}

\begin{deluxetable}{c |c c |c r | c r | c r }
\tabletypesize{\scriptsize}
\tablecaption{Quintuplet Gaussian Mixture Model: Free Parameters, Priors, and Results}
\tablehead{
& \multicolumn{2}{c}{Cluster Gaussian} & \multicolumn{2}{c}{Field Gaussian 1} & \multicolumn{2}{c}{Field Gaussian 2} & \multicolumn{2}{c}{Field Gaussian 3}\\
\colhead{Parameter} & \colhead{Prior} & \colhead{Result} & \colhead{Prior} & \colhead{Result} & \colhead{Prior} & \colhead{Result} & \colhead{Prior} & \colhead{Result}
}
\startdata
$\pi_{k}$ & U(0, 1) & 0.047 $\pm$ 0.003 & U(0, 1) & 0.53 $\pm$ 0.02 & U(0, 1) & 0.29 $\pm$ 0.02 & U(0, 1) & 0.13 $\pm$ 0.01 \\
$\mu_{\alpha^*, k}$ (mas yr$^{-1}$) & G(-0.96, 2) & -0.96 $\pm$ 0.01 & U(-6, 6) & -2.58 $\pm$ 0.04 & U(-6, 6) & -2.70 $\pm$ 0.06 & U(-6, 6) & -1.61 $\pm$ 0.06 \\
$\mu_{\delta, k}$ (mas yr$^{-1}$) & G(-2.26, 2) & -2.29 $\pm$ 0.01 & U(-6, 6) & -4.79 $\pm$ 0.06 & U(-6, 6) & -4.97 $\pm$ 0.07 & U(-6, 6) & -3.12 $\pm$ 0.07 \\
$\sigma_{a, k}$ (mas yr$^{-1}$) & U(0, 3) & 0.11 $\pm$ 0.012 & U(0, 8) & 2.44 $\pm$ 0.04 & U(0, 8) & 3.42 $\pm$ 0.06 & U(0, 8) & 1.20 $\pm$ 0.08 \\
$f$ & --- & 1.0 & U(0, 1) & 0.49 $\pm$ 0.02 & U(0, 1)  & 0.89 $\pm$ 0.02 & U(0, 1) & 0.42 $\pm$ 0.04 \\
$\theta_{k}$ (rad) & --- & 0 & U(0, $\pi$) & 0.99 $\pm$ 0.01 & U(0, $\pi$) & 0.95 $\pm$ 0.07 & U(0, $\pi$) & 0.92 $\pm$ 0.03 \\
\enddata
\label{tab:gmm_quint}
\tablecomments{Description of parameters: $\pi_k$ = fraction of stars in Gaussian; $\mu_{\alpha^*, k}$ = $\alpha^*$ velocity centroid of Gaussian; $\mu_{\delta,k}$ = $\delta$ velocity centroid of Gaussian; $\sigma_{a,k}$ = semi-major axis of Gaussian; $f$ = ratio of semi-minor to semi-major axis; $\theta_{k}$ = angle between $\sigma_{a,k}$ and the $\alpha^*$ axis}
\tablecomments{Description of priors: Uniform distributions: U(min, max), where min and max are bounds of the distribution; Gaussian distributions: G($\mu$, $\sigma$), where $\mu$ is the mean and $\sigma$ is the standard deviation}
\end{deluxetable}

\subsection{Stellar Cluster Membership Probabilities}
\label{app:membership}
For each star, a cluster membership probability P$_{clust}$ is calculated based on the best-fit GMM in the same manner as \citetalias{Hosek:2019kk} and \citetalias{2019ApJ...877...37R}:

\begin{equation}
P_{clust} = \frac{\pi_c P_{c}}{\pi_c P_{c} + \sum_k^K{\pi_k P_{k}}}
\end{equation}

\noindent where $\pi_c$ and $\pi_k$ are the fraction of total stars in the cluster and $k$th field Gaussian, respectively, and P$_{c}$ and P$_{k}$ are the probability of the given star being part of the cluster and $k$th field Gaussian, respectively, based on its observed proper motion. If we consider stars with $P_{clust} \geq 0.7$ to be high-probability cluster members, then we find that our sample includes 577 high-probability members for the Arches cluster and 977 high-probability members for the Quintuplet cluster. If we sum the cluster membership probabilities, we obtain $\sum P_{clust}$ = 1503.9 for the Arches sample and $\sum P_{clust}$ = 2238.5 for the Quintuplet sample. The cluster membership probabilities are included in the proper motion catalogs (Table \ref{tab:arches_cat}).

\section{Full Posteriors for Cluster Orbit Models}
\label{app:posteriors}
\begin{figure*}
\begin{center}
\includegraphics[scale=0.32]{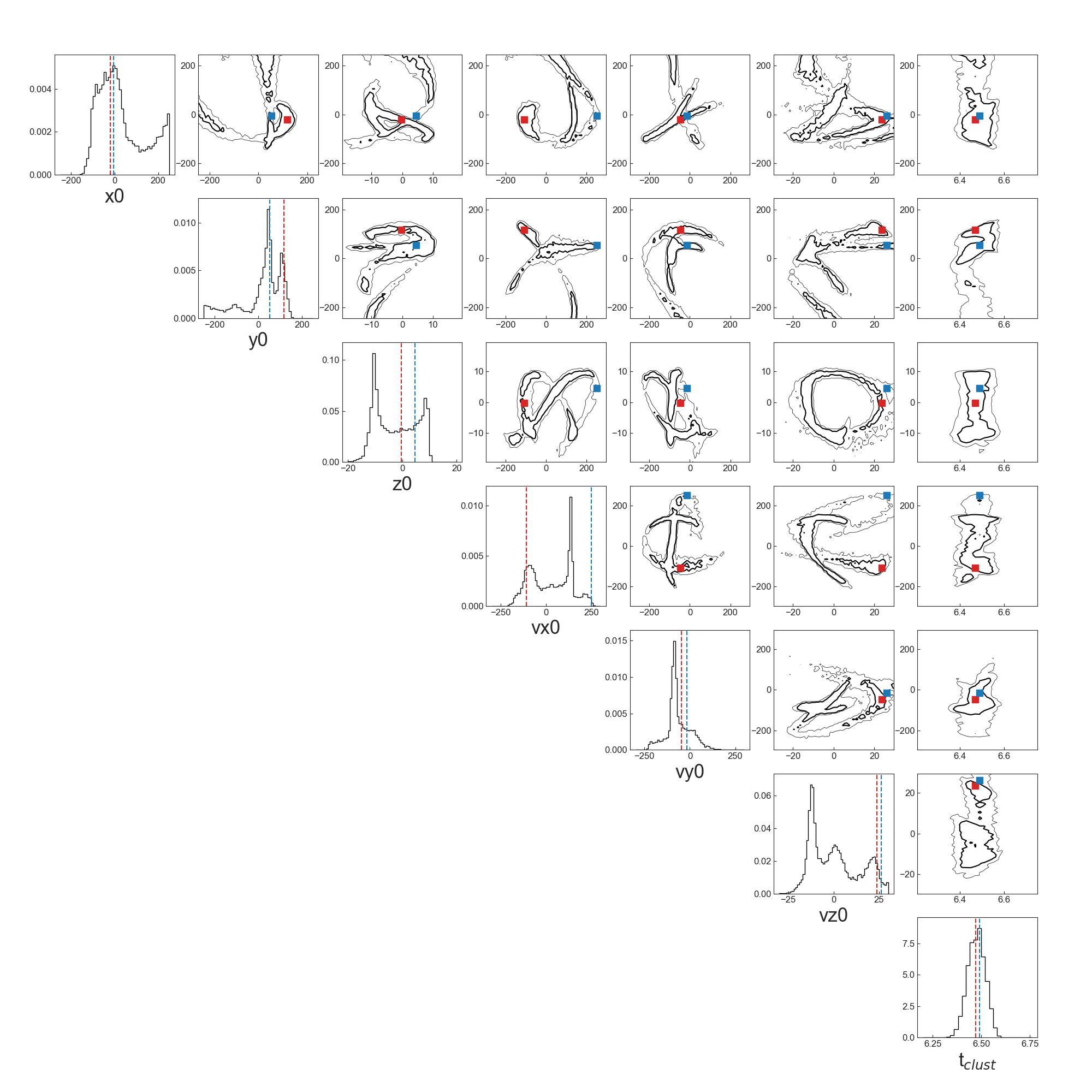}
\caption{Posterior probability distributions for the free parameters in the orbit model for the Arches cluster (priors and units as defined in Table \ref{tab:orb_priors}). The black histograms show the
marginalized one-dimensional posterior for a given parameter, while the black contours show the 1$\sigma$ (thick) and 2$\sigma$ (thin) contours for the joint posteriors. The dotted lines and squares represent the values of the maximum likelihood orbit in each solution mode for the one-dimensional and joint posterior plots, respectively, with the red color corresponding to mode 1 and the blue color corresponding to mode 2.}
\label{fig:triangle_results_arch}
\end{center}
\end{figure*}

\begin{figure*}
\begin{center}
\includegraphics[scale=0.32]{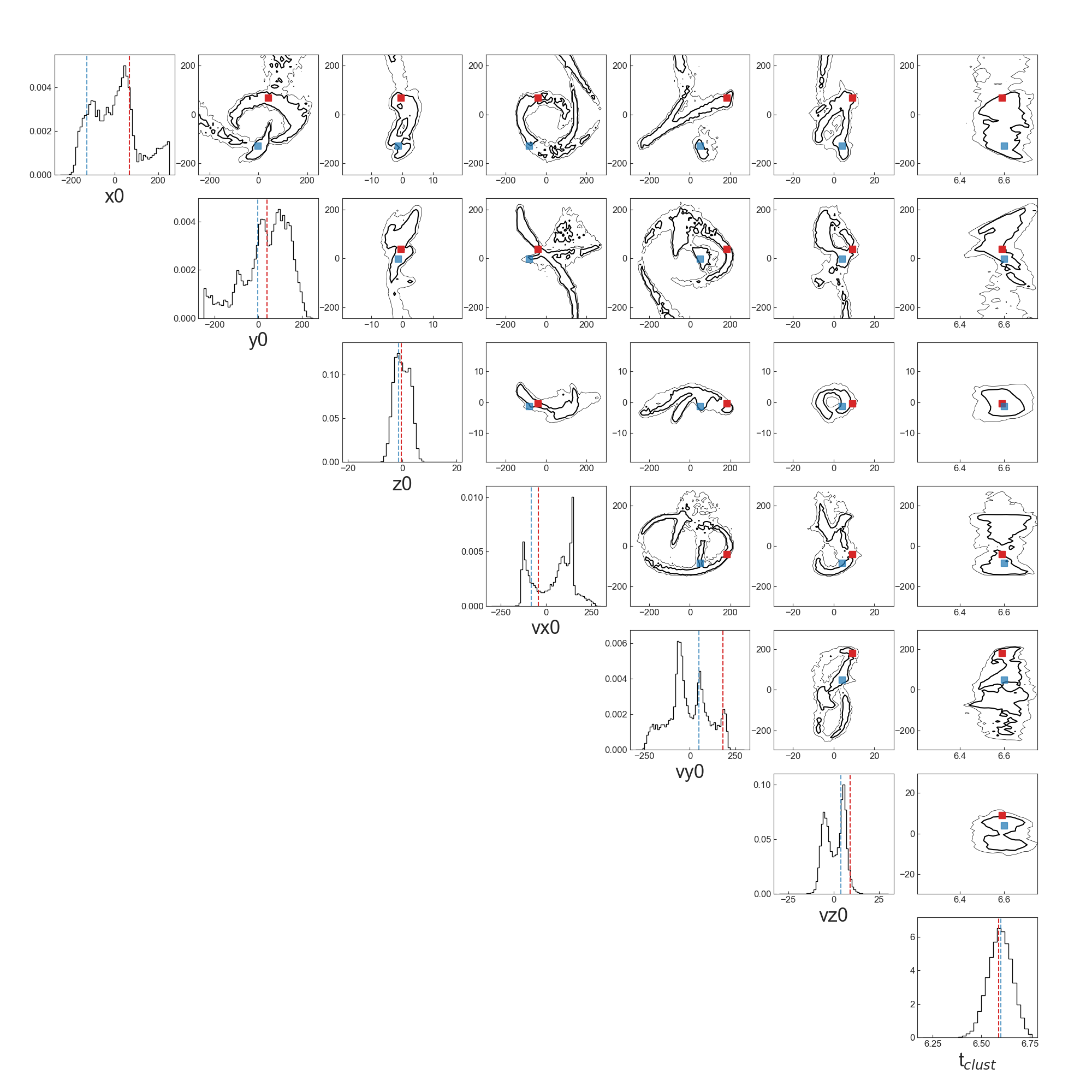}
\caption{Posterior probability distributions for the free parameters in the orbit model for the Quintuplet cluster, constructed in the same manner as Figure \ref{fig:triangle_results_arch}.}
\label{fig:triangle_results_quint}
\end{center}
\end{figure*}

The posterior probability distributions for the free parameters in the
Arches and Quintuplet orbit models are shown in
Figures \ref{fig:triangle_results_arch} and \ref{fig:triangle_results_quint}, respectively.

\section{KDL15 Analysis}

\subsection{Calculating the Uncertainty in the KDL15 Orbit Model}
\label{sec:methods_kdl15}
Because of the uncertainty in $d_{los}$,
there are three possible locations of the Arches and Quintuplet clusters
on the \citetalias{Kruijssen:2015fx} orbit
based on their galactic longitudes (Figure \ref{fig:kdl_orb}).
In $\mathsection$\ref{sec:kdl_discussion},
we calculate the probability that the clusters are at
these locations by comparing the present-day cluster
galactic latitude and motion ($b$, $\mu_{l*}$, $\mu_b$, and $v_{los}$)
to their predicted values
on the \citetalias{Kruijssen:2015fx} orbit.
We use orbit simulations to determine the uncertainty
in the \citetalias{Kruijssen:2015fx} orbit model,
which is a key component of this analysis.

First, we construct the same gravitational potential for the GC
that is used by \citetalias{Kruijssen:2015fx}.
This potential is based on the enclosed mass distribution for
the inner 300 pc of the Milky Way from \citet{Launhardt:2002hl},
but is flattened in the vertical direction
by a factor $q_{\phi}$:

\begin{equation}
\Phi(x, y, z) = \Phi_{S}(r(q_{\phi}))
\end{equation}

where $\Phi_{S}(r(q_{\phi}))$ is a spherical potential calculated
at the modified radius:

\begin{equation}
r(q_{\phi})^2 = x^2 + y^2 + \frac{z^2}{q_{\phi}^2}
\end{equation}

\citetalias{Kruijssen:2015fx} require a flattened potential in order to produce
the vertical oscillations in their orbit model. They fit $q_{\phi}$ as a free parameter
in their analysis, obtaining $q_{\phi}$ = 0.63$^{+0.07}_{-0.06}$.
While $q_{\phi}$ is not well constrained by direct
measurements, models of the structure of the Galactic bulge
are significantly flattened in the $z$
direction \citep[e.g.][]{Rodriguez-Fernandez:2008od, Wegg:2013pd},
suggesting that a flattened potential is not an unreasonable assumption.

The \citetalias{Kruijssen:2015fx} orbit is parameterized by six parameters:
the apoapse and periapse (R$_a$ and R$_p$),
the height above the galactic plane (z$_p$),
the velocity angle at pericenter ($\theta_{kdl}$), the projection
angle between the origin-observer and origin-pericenter vectors ($\phi$),
and the ratio of the vertical-to-planar axes of the gravitational potential ($q_{\phi}$).
We draw 50,000 sets of these parameters
from Gaussian distributions, each with a mean and standard deviation
equal to the corresponding best-fit values and uncertainties reported
in Table 1 of \citetalias{Kruijssen:2015fx}\footnote{Some of the reported parameters have mildly asymmetric
error bars. For these, we adopt a symmetric uncertainty that is equal to the average of the positive and negative
error values.}.
From these parameters
we calculate the three-dimensional position and velocity of a particle at
periapse for each of the orbits.
We then integrate the orbits for $\pm$2.5 Myr
from periapse (the same time range as \citetalias{Kruijssen:2015fx})
with a timestep of 0.025 Myr using \texttt{galpy},
each with a gravitational potential flattened by
a value for $q_{\phi}$ that is also drawn from a Gaussian
distribution with a mean and standard deviation corresponding
to the constrain reported in \citetalias{Kruijssen:2015fx}.

We use this set of orbits to calculate the uncertainty in the
predicted position and motion of the \citetalias{Kruijssen:2015fx} orbit at each of the
intersection points in Figure \ref{fig:kdl_orb}.
First, we convert the orbit positions and velocities at each timestep from
\texttt{galpy} Galactocentric coordinates into observed
quantities ($l_{kdl}$, $b_{kdl}$, $\mu_{l*,kdl}$, $\mu_{b,kdl}$, $v_{los,kdl}$)
using the distance, location, proper motion,
and radial velocity of SgrA* as described in $\mathsection$\ref{sec:orb_sims}.
Next, we linearly interpolate
$b_{kdl}$, $\mu_{l*,kdl}$, $\mu_{b,kdl}$, and $v_{los,kdl}$
as a function of $l_{kdl}$ in order to
get their values at at each intersection point.
The mean and standard deviation of a given quantity across all 50,000
orbits thus represents its predicted value and uncertainty
in the \citetalias{Kruijssen:2015fx} orbit model.

\subsection{Cluster Comparison to KDL15 Streams 2 and 3}
\label{app:other_kdl15}

\begin{figure*}
\begin{center}
\includegraphics[scale=0.32]{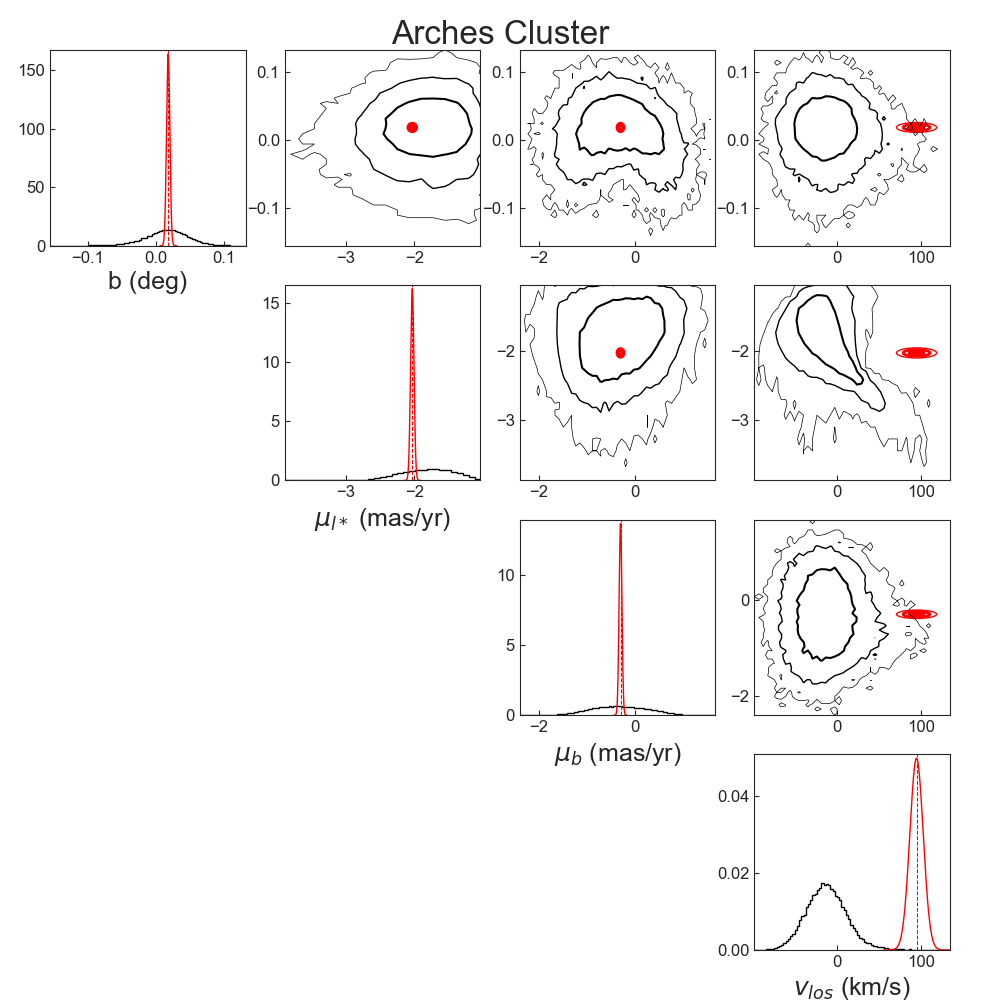}
\includegraphics[scale=0.32]{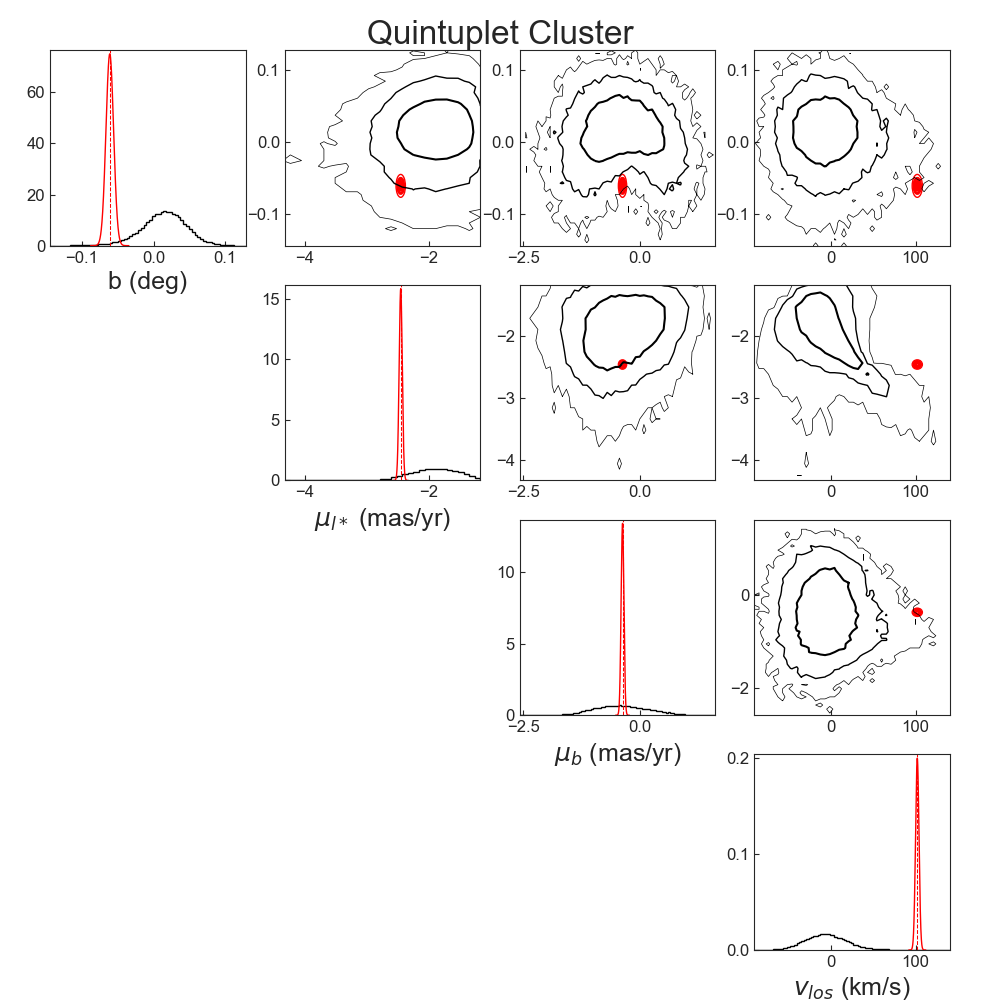}
\caption{The comparison between the observed values of $\overrightarrow{\mathbf{x_{int}}}$ for the
Arches (left) and Quintuplet (right) clusters compared to
the predicted values for stream 2 of the \citetalias{Kruijssen:2015fx} orbit. The plots are
constructed in the same manner as Figure \ref{fig:kdl_comp}.
The Arches and Quintuplet are discrepant with Stream 2 by 4.25$\sigma$ and 4.94$\sigma$,
respectively.}
\label{fig:kdl_comp_s2}
\end{center}
\end{figure*}

\begin{figure*}
\begin{center}
\includegraphics[scale=0.32]{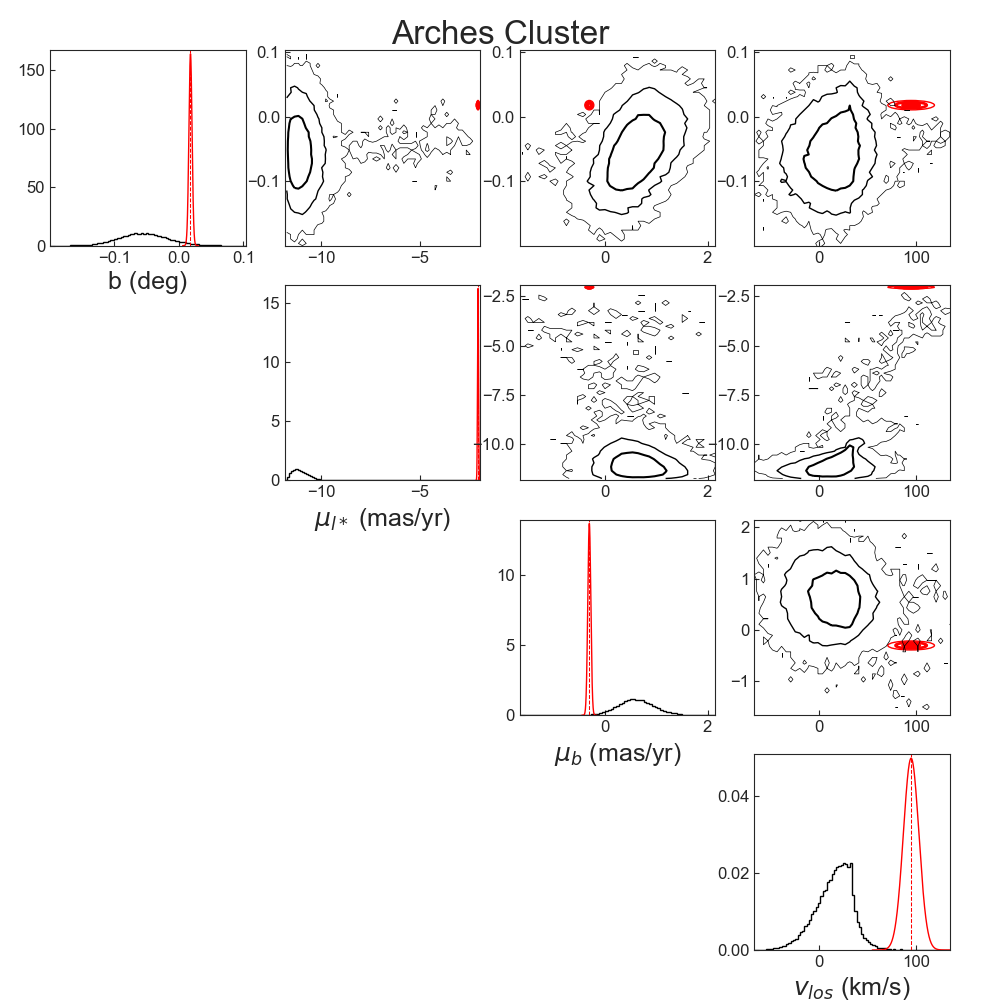}
\includegraphics[scale=0.32]{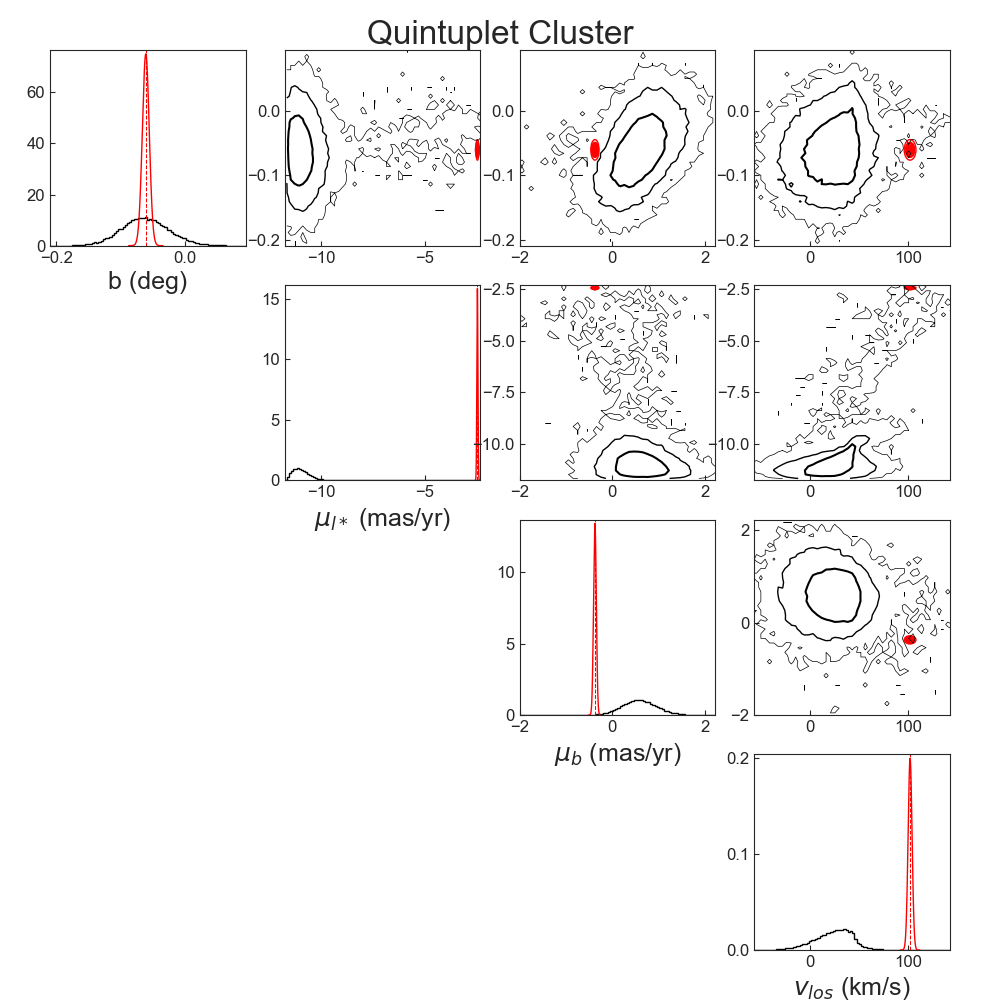}
\caption{The comparison between the observed values of $\overrightarrow{\mathbf{x_{int}}}$ for the
Arches (left) and Quintuplet (right) clusters compared to
the predicted values for stream 3 of the \citetalias{Kruijssen:2015fx} orbit. The plots are
constructed in the same manner as Figure \ref{fig:kdl_comp}.
The Arches and Quintuplet are discrepant with Stream 3 by $>$10$\sigma$.}
\label{fig:kdl_comp_s3}
\end{center}
\end{figure*}

Figures \ref{fig:kdl_comp_s2} and \ref{fig:kdl_comp_s3} show a
comparison between the observed values of $\overrightarrow{\mathbf{x_{int}}}$
for the clusters and their corresponding predicted values
at the Stream 2 and 3 intersection points, respectively.
The \citetalias{Kruijssen:2015fx} orbit predictions for both streams are significantly different
than the observations: stream 2 is
discrepant by 4.35$\sigma$ and 5.03$\sigma$ for the Arches and Quintuplet,
respectively, and stream 3 is discrepant by $>$10$\sigma$ for both clusters.
For Stream 2, the largest discrepancy is found in the $v_{los}$ dimension;
the observed values for the clusters are significantly higher than what
the \citetalias{Kruijssen:2015fx} orbit calls for.
For Stream 3, the observed values for $\mu_{l*}$ are in the opposite
direction then what is predicted by the \citetalias{Kruijssen:2015fx} orbit.

\section{Orbit Constraints Using Different GC Gravitational Potentials}
\label{app:other_gpot}

We repeat the orbit analysis described in $\mathsection$\ref{sec:orb_sims} using the three alternative
gravitational potentials for the GC proposed by \citet[][hereafter S20]{Sormani:2020my}.
These potentials have two components, one for the
Nuclear Star Cluster (NSC) and the other for the
Nuclear Stellar Disk (NSD).
The NSC component is the same for each potential,
generated from the axisymmetric cluster model from \citet{Chatzopoulos:2015lq}.
Each NSD component has a different
functional form for the mass distribution:
the best-fit model from \citet{Launhardt:2002hl} (Equation 24 from \citetalias{Sormani:2020my}),
the best-fit model from \citet{Chatzopoulos:2015lq} (Equations 25 and 26 from \citetalias{Sormani:2020my}),
and a deprojection of the stellar density profile from \citet{Gallego-Cano:2020uo} (Equation 27 from \citetalias{Sormani:2020my}).
We refer to these potentials as S20\_1, S20\_2, and S20\_3, respectively.
\citetalias{Sormani:2020my} fit a scale factor to the NSD component of
each potential based on Jeans modeling of the NSD.
We refer to the potential used in the main body of paper as L02\_flat,
as it is a flattened version as the potential from the
\citet{Launhardt:2002hl} mass distribution.

\begin{figure*}
\begin{center}
\includegraphics[scale=0.3]{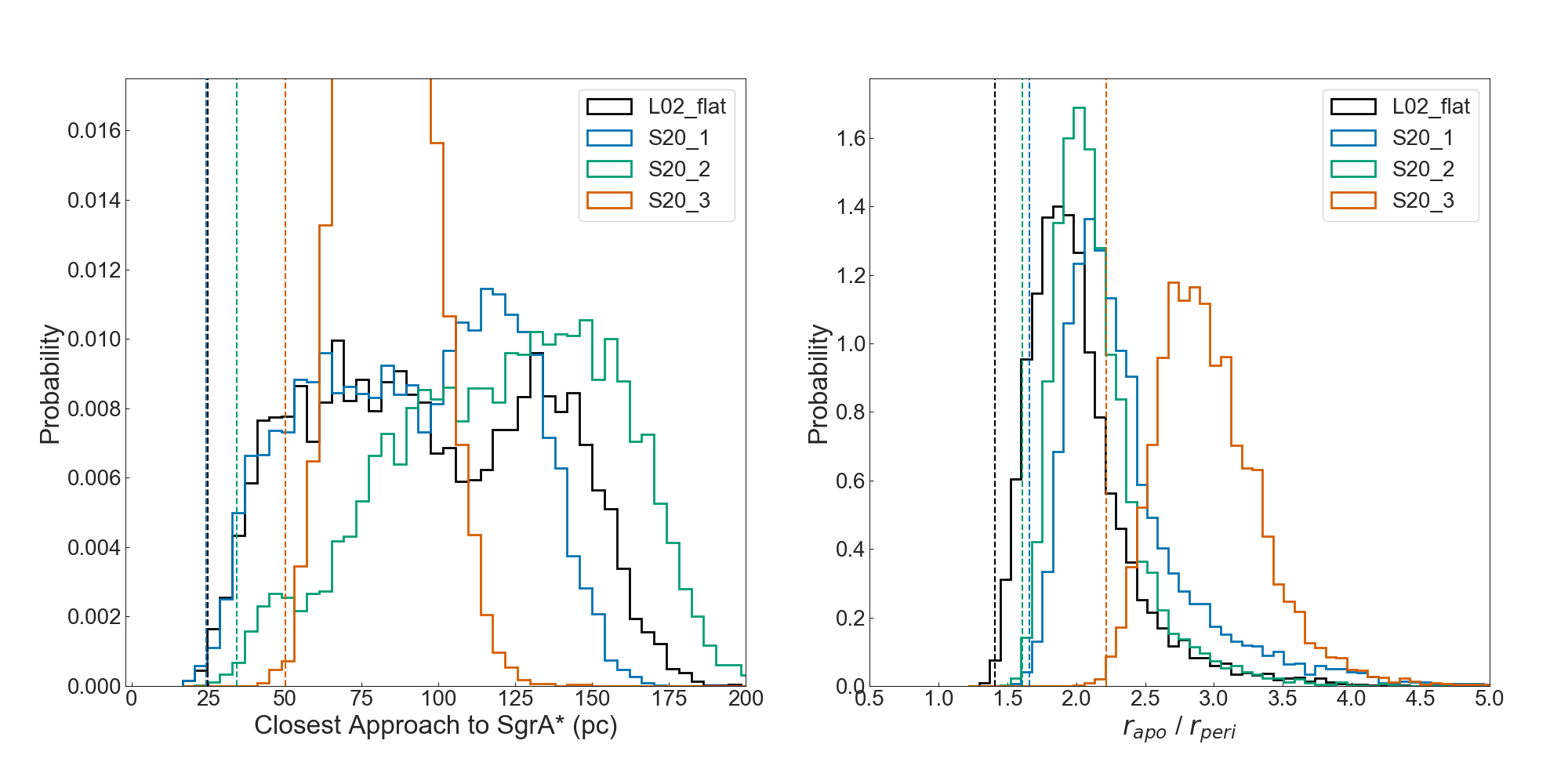}
\caption{The probability distributions for the closest approach to SgrA* (left) and $r_{apo}$ / $r_{peri}$ (right) for
prograde orbits of the Arches cluster using different gravitational potentials for the GC.
The black histogram
shows the results for the L02\_flat potential that is adopted in the main text of the paper,
while the different color histograms
show the results for the S20\_1, S20\_2, and S20\_3 potentials. The 3$\sigma$ lower limits of these properties for each
potential is shown by the vertical dotted line with the same color as the corresponding histogram.
The S20 potentials produce orbits that do not extend as close to SgrA*
and have larger $r_{apo}$ / $r_{peri}$ ratios compared to the L02\_flat potential.
These differences do not change the conclusions drawn in this paper.}
\label{fig:arches_gpot_comp}
\end{center}
\end{figure*}

\begin{figure*}
\begin{center}
\includegraphics[scale=0.3]{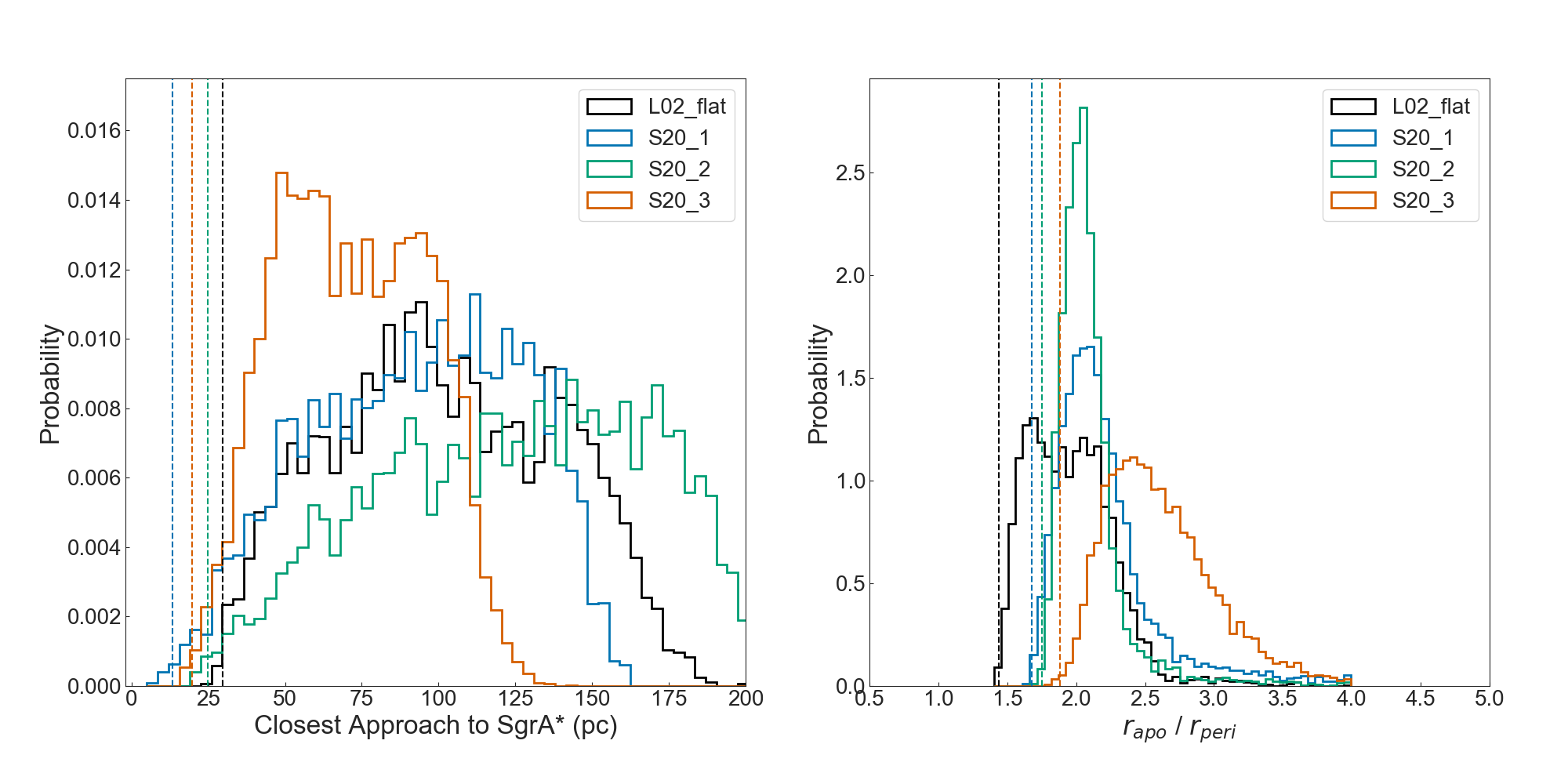}
\caption{The probability distributions for the closest approach to SgrA* (left) and $r_{apo}$ / $r_{peri}$ (right) for
prograde orbits of the Quintuplet cluster using different gravitational potentials for the GC, constructed in the same manner
as Figure \ref{fig:arches_gpot_comp}. The S20 potentials produce orbits that extend closer to SgrA* and have higher
eccentricities compared to the L02\_flat potential. These differences do not change the conclusions drawn in this paper.}
\label{fig:quint_gpot_comp}
\end{center}
\end{figure*}

The distributions of the closest approach distances and
$r_{apo}$ / $r_{peri}$ ratios for the prograde cluster orbits using the
different potentials is shown in
Figure \ref{fig:arches_gpot_comp} and \ref{fig:quint_gpot_comp} for the
Arches and Quintuplet clusters, respectively.
A summary of the limits of these properties for the different
potentials is provided in Table \ref{tab:other_gpot}.
For the Arches, the 3$\sigma$ lower limits for both
the closest approach distance and $r_{apo}$ / $r_{peri}$
for the S20 potentials are similar to or larger than
the limits for the L02\_flat potential.
For the Quintuplet, the 3$\sigma$ limits on the closest
approach distance for the S20 potentials are smaller than for the L02\_flat
potential, but the 3$\sigma$ limits on $r_{apo}$ / $r_{peri}$
are larger.
In general, this indicates that the S20 potentials produce
orbits with generally higher eccentricities
then the L02\_flat potential for both clusters.
Our conclusions that
the Arches and Quintuplet clusters are unlikely
to inspiral into the NSC and that they cannot
be on circular orbits does not change when
these different potentials are used.

We repeat the calculation in Equation \ref{eq:zdist} for the
Arches and Quintuplet orbits in these different potentials
to assess if the clusters could share a common orbit.
Similar to the L02\_flat potential, the Arches always
exhibits larger vertical oscillations in the Galactic Plane,
with $P(b_{max, arch=quint})$ = 0.1\%, 0.2\%, and 0.01\%
for the S20\_1, S20\_2, and S20\_3 potentials,
respectively.
Thus, the choice of gravitational potential
does not change our conclusion that the clusters do
not share a common orbit.

\begin{figure*}
\begin{center}
\includegraphics[scale=0.3]{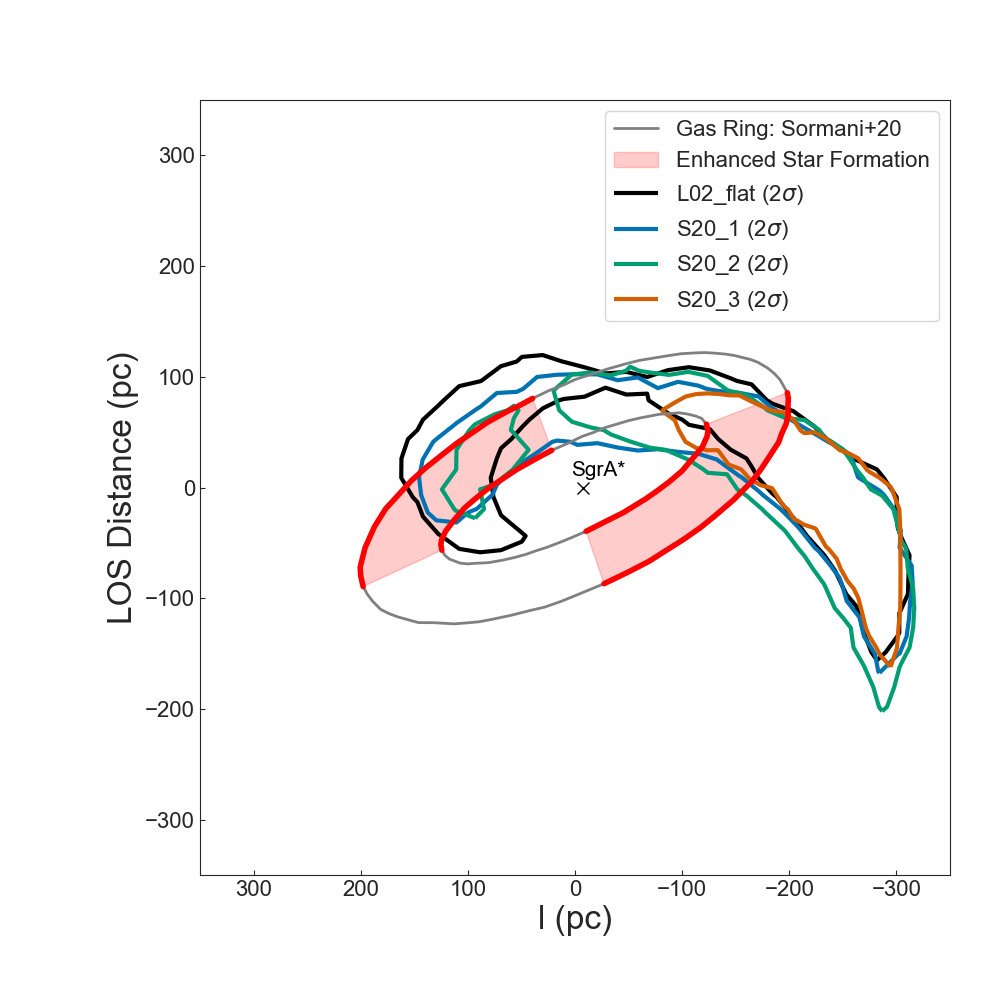}
\includegraphics[scale=0.3]{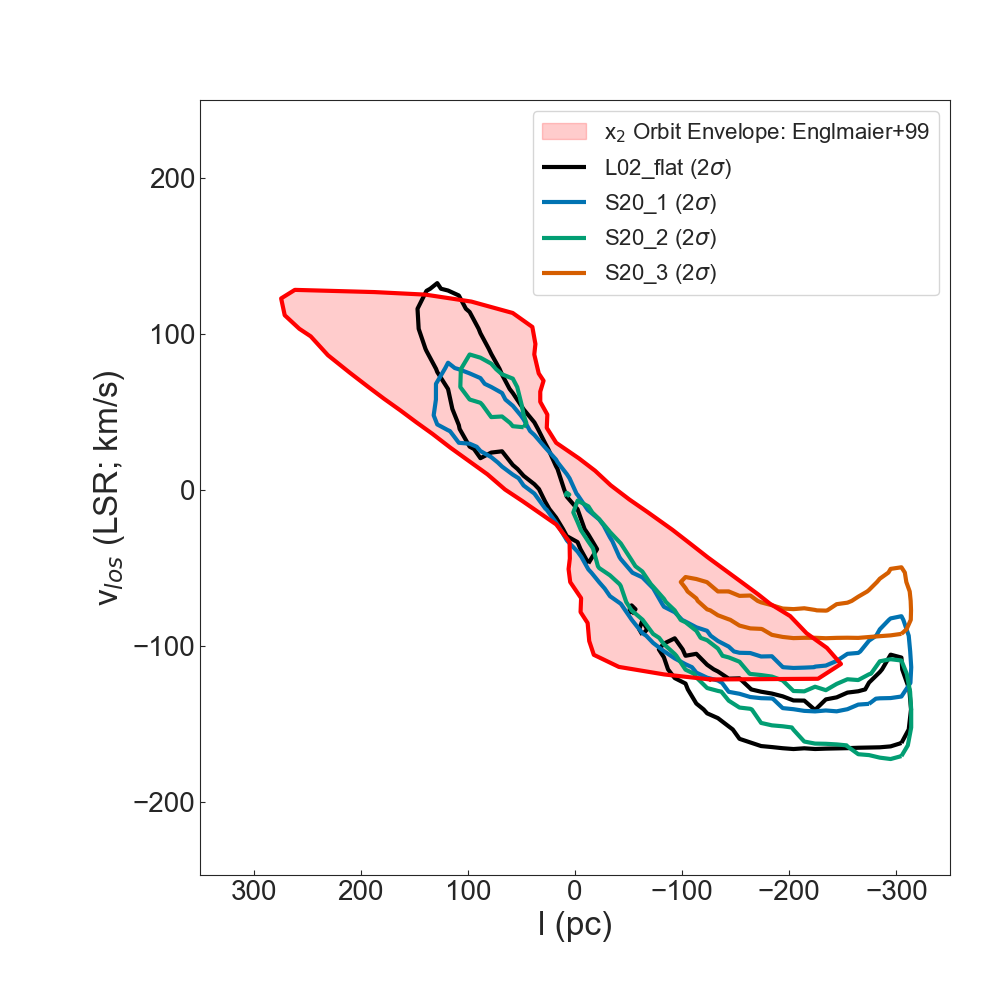}
\caption{The 2$\sigma$ probability contours on the cluster birth location (left) and initial $v_{los}$ (right) for the Arches cluster
assuming different gravitational potentials, plotted similarly to
Figures \ref{fig:x2_pos} and \ref{fig:x2_vlos}.
In the left panel, we find that the Arches cluster has significant probability
of forming in the region of enhanced star formation predicted by the $x_1$ - $x_2$
collision scenario (red shaded region) regardless
of the gravitational potential used.
In the right panel, we find that there is no evidence that the cluster
formed at a higher $v_{los}$ than gas on $x_2$ orbits (red shaded region)
for birth locations within the gas ring ($|l|$ $\lesssim$ 200 pc),
regardless of the gravitational potential used.
Our conclusion that these results are in mild support for the $x_1$ -- $x_2$ collision
scenario is not affected by the choice of gravitational potential.
}
\label{fig:x2_arch_gpot}
\end{center}
\end{figure*}

\begin{figure*}
\begin{center}
\includegraphics[scale=0.3]{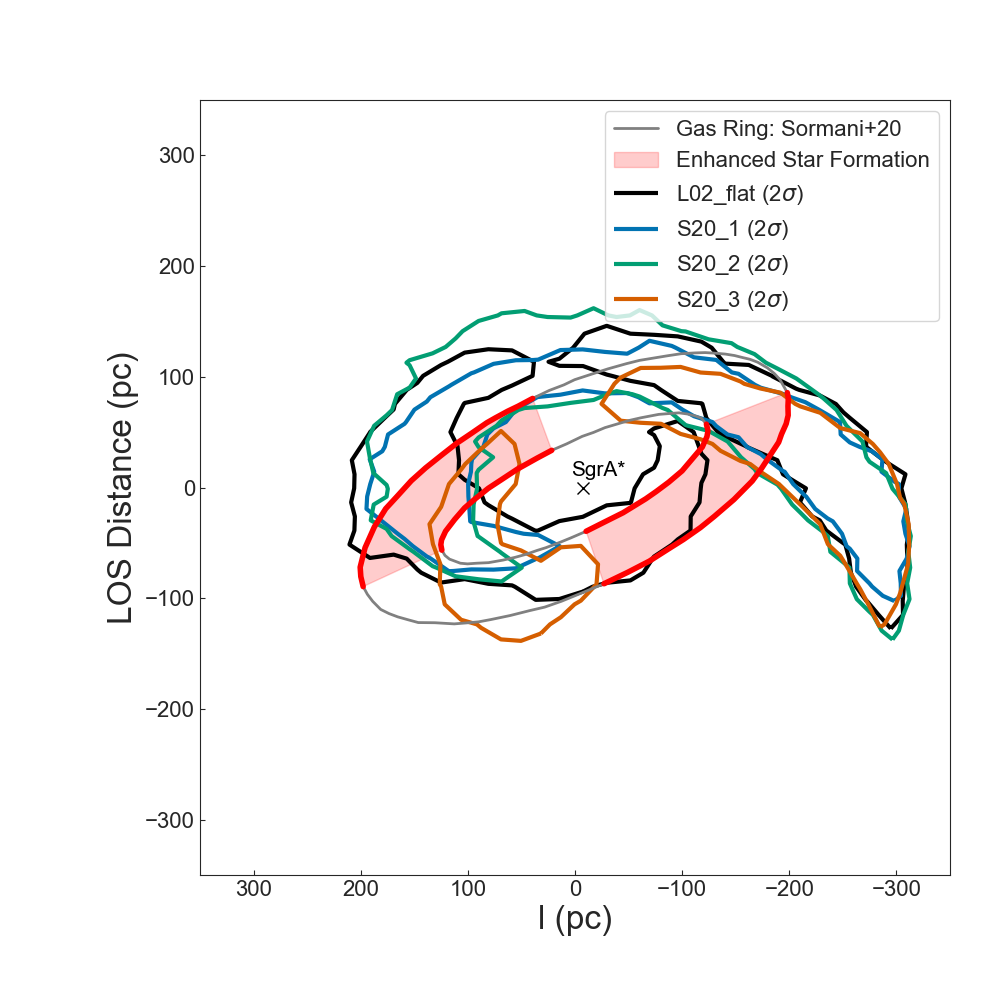}
\includegraphics[scale=0.3]{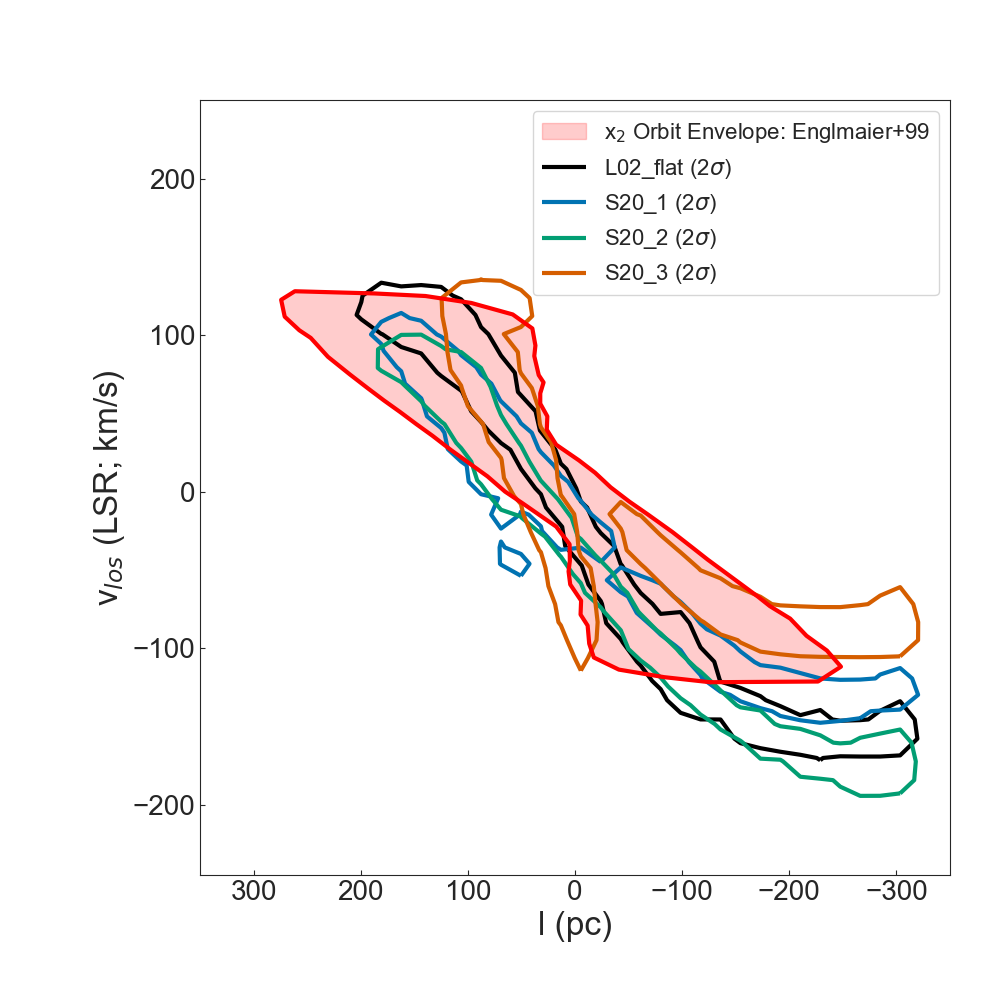}
\caption{The 2$\sigma$ probability contours on the cluster birth location (left) and initial $v_{los}$ (right) for the Quintuplet cluster
assuming different gravitational potentials, plotted similarly to Figure \ref{fig:x2_arch_gpot}.
The cluster birth location remains consistent with the $x_1$ -- $x_2$ formation scenario
for all potentials, and there is some evidence for an enhancement in the
birth $v_{los}$ relative to the $x_2$ orbits if the S20\_2 potential is used.
Our conclusion that these results are in mild support for the $x_1$ -- $x_2$ collision
scenario is not affected by the choice of gravitational potential, and is perhaps
strengthened if the S20\_2 potential is used.
}
\label{fig:x2_quint_gpot}
\end{center}
\end{figure*}

We compare the constraints on the cluster birth locations and
birth $v_{los}$ for different potentials to the predictions of the
$x_1$ -- $x_2$ gas collision scenario in Figures \ref{fig:x2_arch_gpot} and \ref{fig:x2_quint_gpot}.
Changing the gravitational potential has a relatively
minor effect on the cluster birth locations,
as the uncertainty is
primarily driven by the uncertainty in $d_{los}$ rather than
the potential itself.
For all potentials, both the Arches and Quintuplet have
significant probability of forming in the predicted regions of
enhanced star formation in the
$x_1$ -- $x_2$ gas collision scenario.
However, the gravitational potential appears to have
a larger effect on the birth $v_{los}$
of the clusters, especially at negative galactic
longitudes.
That said, we do not find evidence that either
cluster formed with a higher $v_{los}$ than
typical $x_2$ gas velocities for most potentials.
The exception is the $v_{los}$ constraints for the Quintuplet cluster
using the S20\_2 potential,
which produces values that
appear slightly enhanced compared
to the $x_2$ gas orbits
between -200 pc $\lesssim$ $l$ $\lesssim$ -150 pc.
Overall, our conclusion that the clusters are mildly consistent
with the $x_1$ -- $x_2$ formation scenario is not affected by the
choice of gravitational potential, and is perhaps strengthened
for the Quintuplet cluster if the S20\_2 potential is used given
the potential $v_{los}$ enhancement.

\begin{deluxetable}{c c c c c}
\tablecaption{Orbit Properties with Different Potentials}
\label{tab:other_gpot}
\tablehead{
\colhead{Cluster} & \colhead{Potential\tablenotemark{a}} & \colhead{Closest Approach} & \colhead{Min Ratio} & \colhead{Ave Ratio} }
\startdata
Arches & L02\_flat & 24.7 & 1.4 & 1.9 \\
Arches & S20\_1 & 24.4 & 1.7 & 2.2 \\
Arches & S20\_2 & 34.2 & 1.6 & 2.1 \\
Arches & S20\_3 & 50.0 & 2.2 & 2.9 \\
& & & & \\
Quintuplet & L02\_flat & 29.8 & 1.4 & 1.9 \\
Quintuplet & S20\_1 & 13.2 & 1.7 & 2.1 \\
Quintuplet & S20\_2 & 24.7 & 1.8 & 2.0 \\
Quintuplet & S20\_3 & 19.6 & 1.9 & 2.5 \\
\enddata
\tablecomments{Description of columns: \emph{Potential}: potential used, \emph{Closest Approach}: 3$\sigma$ lower limit
on closest approach to SgrA* in pc, \emph{Min Ratio}: 3$\sigma$ lower limit of $r_{apo}$ / $r_{peri}$, \emph{Ave Ratio}: 50th percentile of $r_{apo}$ / $r_{peri}$}
\tablenotetext{a}{Note that the L02\_flat potential is used in the main text of the paper.}
\end{deluxetable}

\end{document}